\documentstyle[11pt,aaspp4,natbib]{article}

\newcommand{\PNM}{P(N|M)}
\newcommand{\Navg}{N_{\mathrm{avg}}}
\newcommand{\NavgM}{N_{\mathrm{avg}}(M)}
\newcommand{\PNNavg}{P(N|N_{\mathrm{avg}})}
\newcommand{\Mmin}{M_{\mathrm{min}}}
\newcommand{\Mcrit}{M_{\mathrm{crit}}}
\newcommand{\Dg}{\Delta\gamma}
\newcommand{\ng}{\bar{n}_g}
\newcommand{\vh}{{\bf v}_{\mathrm{h}}}
\newcommand{\vg}{{\bf v}_{\mathrm{g}}}
\newcommand{\vm}{{\bf v}_{\mathrm{m}}}
\newcommand{\vv}{{\bf v}}
\newcommand{\rr}{{\bf r}}
\newcommand{\k}{{\bf k}}
\newcommand{\Rmax}{R_{\mathrm{max}}}
\newcommand{\Rvir}{R_{\mathrm{vir}}}
\newcommand{\Mvir}{M_{\mathrm{vir}}}
\newcommand{\hvol}{h^{3}{\mathrm{Mpc}}^{-3}}
\newcommand{\hmpc}{h^{-1}{\mathrm Mpc}}
\newcommand{\hkpc}{h^{-1}{\mathrm kpc}}
\newcommand{\hMsun}{h^{-1}M_{\odot}}
\newcommand{\Omegam}{\Omega_{m}}
\newcommand{\Omegal}{\Omega_{\Lambda}}
\newcommand{\ngrpN}{n_{\mathrm{grp}}(>N)}
\newcommand{\nhM}{n_h(>M)}
\newcommand{\xig}{\xi_g(r)}
\newcommand{\xim }{\xi_m(r)}
\newcommand{\xigm}{\xi_{gm}(r)}
\newcommand{\xih}{\xi_{1\mathrm{h}}(r)}
\newcommand{\N}{\left<N\right>}
\newcommand{\NN}{\left<N(N-1)\right>}
\newcommand{\NNN}{\left<N(N-1)(N-2)\right>}
\newcommand{\Qeq}{Q_{\mathrm{eq}}(k)}
\newcommand{\Beq}{B_{\mathrm{eq}}(k)}

\bibpunct[,]{(}{)}{;}{a}{}{,}
\begin{document}

\title{The Halo Occupation Distribution: Towards an Empirical Determination of
the Relation Between Galaxies and Mass}

\author{
Andreas A. Berlind, and David H. Weinberg
}
\affil{Department of Astronomy, The Ohio State University, Columbus,
OH 43210;}
\affil{Email: aberlind,dhw@astronomy.ohio-state.edu}

\begin{abstract}

We investigate galaxy bias in the framework of the ``Halo Occupation
Distribution'' (HOD), which defines the bias of a population of galaxies by the
conditional probability $\PNM$ that a dark matter halo of virial mass $M$
contains $N$ galaxies, together with prescriptions that specify the relative
spatial and velocity distributions of galaxies and dark matter within halos. By
populating the halos of a cosmological N-body simulation using a variety of HOD
models, we examine the sensitivity of different galaxy clustering statistics to
properties of the HOD. The galaxy correlation function responds to different
aspects of $\PNM$ on different scales. Obtaining the observed power-law form of
$\xi_g(r)$ requires rather specific combinations of HOD parameters, implying a
strong constraint on the physics of galaxy formation; the success of numerical
and semi-analytic models in reproducing this form is entirely non-trivial.
Other clustering statistics such as the galaxy-mass correlation function, the
bispectrum, the void probability function, the pairwise velocity dispersion,
and the group multiplicity function are sensitive to different combinations of
HOD parameters and thus provide complementary information about galaxy bias. We
outline a strategy for determining the HOD empirically from redshift survey
data. This method starts from an assumed cosmological model, but we argue that
cosmological and HOD parameters will have non-degenerate effects on galaxy
clustering, so that a substantially incorrect cosmological model will not
reproduce the observations for any choice of HOD. Empirical determinations
of the HOD as a function of galaxy type from the 2dF and SDSS redshift surveys
will provide a detailed target for theories of galaxy formation, insight into
the origin of galaxy properties, and sharper tests of cosmological models.

\end{abstract}


\section{Introduction}

The relation between the galaxy and dark matter distributions depends on
the physics of galaxy formation, and it is expected that galaxies are, at
least to some degree, biased tracers of the mass distribution.  This
expectation, which is supported by observational evidence that galaxy
clustering varies with luminosity, morphology, and color (\citealt{guzzo97};
\citealt{norberg01}; \citealt{zehavi01}; and references therein),
complicates efforts to test cosmological models against observed galaxy
clustering.  However, the presence of bias also implies that galaxy
clustering can be used to constrain the physics of galaxy formation,
especially as independent observations define the background cosmology with
increasing precision (e.g., \citealt{wang01}).  The 2dF and SDSS galaxy
redshift surveys (\citealt{colless01}; \citealt{york00}),
which can measure the clustering of different galaxy types with
unprecedented detail, are now bringing this goal within reach.
Achieving it requires a language for describing bias that is powerful
enough to capture the information in these measurements and thereby connect
observations of galaxy clustering to the physics of galaxy formation.

In this paper, we examine the influence of bias on galaxy clustering
statistics, using the framework of the ``Halo Occupation Distribution''
(HOD).  This approach describes bias at the level of ``virialized'' dark
matter halos, structures of typical overdensity $\rho/\bar{\rho}\sim200$,
which are expected to be in approximate dynamical equilibrium.  Gas
dynamics, radiative cooling, and star formation can strongly influence the
distribution of galaxies {\it within} such halos (e.g., the numbers,
masses, and locations of galaxies), but the masses and spatial distribution
of halos themselves should be determined mainly by gravitational dynamics
of the dark matter.  In the HOD framework, the bias of any particular class
of galaxies is fully defined by the probability distribution $\PNM$ that a
halo of virial mass $M$ contains $N$ galaxies, along with the relations
between the galaxy and dark matter spatial and velocity distributions
within halos.  While the history of the galaxy population is necessarily
entwined with the background cosmology, the HOD description suggests a
useful conceptual division between the ``cosmological model'' and the
``theory of galaxy formation'' in predictions of galaxy clustering: the
cosmological model determines the properties of the halo distribution, and
the theory of galaxy formation specifies how those halos are populated with
galaxies.

The most important strength of the HOD formulation of bias is its
completeness.  For a given cosmological model, the HOD tells us everything
a theory of galaxy formation has to say about the statistics of galaxy
clustering, in real space and redshift space, and on small, intermediate,
and large scales \footnote{We discuss some caveats to this assertion at the
end of \S~2}.  Conversely, if we can determine the HOD empirically, we
will learn everything that observed galaxy clustering has to tell us about
the physics of galaxy formation.  Moreover, the HOD provides a physically
informative basis for interpreting discrepancies between predicted and
observed galaxy clustering, or between predictions of different galaxy
formation theories themselves.  It would be more illuminating to learn,
for example, that a given theory predicts too many red galaxies in
halos of mass $10^{13}-10^{14}M_{\odot}$ than to learn that it predicts
the wrong 3-point correlation function of such galaxies.  Finally, since the
HOD describes bias at the level of systems near dynamical equilibrium,
empirical determinations of the HOD can take advantage of mass estimation
methods that are inapplicable on large scales.  For example, traditional
virial methods and X-ray mass estimates of clusters can provide fairly direct
constraints on $\PNM$ at high $M$ (see \S~5).

The halo occupation framework has a long history, initially in analytic
models that described galaxy clustering as a superposition of randomly
distributed clusters with specified profiles and a range of masses
(\citealt{neyman52}; \citealt{mcclelland77}; \citealt{peebles74}).  The
explosion of recent activity in this field has been fueled partly by the
recognition that a combination of this approach with recently developed
tools for predicting the spatial clustering of halos \citep{mo96} provides a
powerful formalism for analytic calculations of dark matter clustering, which
can be naturally extended to biased galaxy populations (e.g.,
\citealt{seljak00}; \citealt{ma00}; \citealt{peacock00};
\citealt{scoccimarro00}; \citealt{white01a}; and numerous other papers
referred to in subsequent sections).  Our own interest was sparked largely
by the paper of \citeauthor{benson00} (\citeyear{benson00}; see
\citealt{kauffmann97} and \citealt{kauffmann99} for similar analyses),
who demonstrated that they could reproduce
the clustering of galaxies in their semi-analytic models by populating N-body
halos according to a predicted $\PNM$.  Furthermore, they showed that the
predicted clustering depends not only on the complex mass dependence of the
mean occupation, but also on finer details of sub-Poisson fluctuations about the
mean.  This result illustrates the power of the HOD to test detailed
predictions of galaxy formation theories.  Models of $\PNM$ based on
semi-analytic calculations of \citet{benson00} and \citet{kauffmann99} have
been incorporated into several of the papers cited above, and some recent
papers have presented predictions of hydrodynamic simulations for $\PNM$
(\citealt{white01b}; \citealt{yoshikawa01}).  We will compare predictions of
$\PNM$ from hydrodynamic simulations and semi-analytic calculations in a
future study (Berlind et al., in preparation).

The HOD description can be contrasted with another widely used approach that
characterizes bias in terms of the correlation between galaxy density and
properties of the large-scale environment, such as mass density, temperature,
and geometry.  This ``environmental bias'' approach has been used to study
the effects of generic biasing models on galaxy clustering statistics
(\citealt{weinberg95}; \citealt{mann98}; \citealt{dekel99}; Narayanan,
Berlind, \& Weinberg 2000; Berlind, Narayanan, \& Weinberg 2001) and to
encapsulate predictions of hydrodynamic simulations and semi-analytic galaxy
formation models (\citealt{blanton99}; \citealt{cen00}; \citealt{somerville01};
\citealt{yoshikawa01}).  It has also led to valuable analytic results
concerning the shape of the galaxy power spectrum and the influence of bias
on higher-order clustering on large scales (\citealt{coles93};
\citealt{fry93}; \citealt{fry94}; \citealt{juszkiewicz95};
\citealt{scherrer98}; \citealt{coles99}).  However, this formulation cannot
effectively describe bias on scales smaller than the smoothing length used to
define the environment, and the choice of smoothing scale is, itself, rather
arbitrary.  In the HOD framework, there is some range of reasonable methods
for defining halos, but the choice of $\rho/\bar{\rho}\sim200$ for a typical
halo boundary is well motivated by the division between the infall and
dynamical equilibrium regimes.  Also, as already noted, this choice allows
use of virial mass estimates to constrain $\PNM$ empirically, while the
large-scale matter density, which plays a fundamental role in environmental
formulations of bias, is generally unobservable.

Our methodology in this paper is to define generic models of the HOD, apply
them to an N-body simulation of the inflationary cold dark matter scenario
(see \S~2), and investigate the dependence of galaxy clustering
statistics on the HOD parameters.  This numerical approach complements
earlier analytic work on halo bias by considering a wider range of clustering
statistics and HOD models, some of them not readily amenable to analytic
calculations.  We will interpret our results in light of the analytic
formalism developed in other papers and in terms of some heuristic analytic
models presented here as a guide for understanding.  The next section
defines our HOD prescriptions more formally.  We then devote considerable
attention, in \S~3, to the two-point correlation function $\xi(r)$,
because of its intrinsic importance and in order to illustrate some general
features of the way the HOD influences galaxy clustering.  We consider other
clustering statistics in \S~4, and properties of galaxy groups in \S~5.  A
general theme emerging from these results is that different statistics respond
to different features of the HOD, implying that precise measurements of galaxy
clustering can in principle yield an empirical determination of the HOD.
Achieving this goal in practice requires a scheme for getting a good first
guess at the HOD that will reproduce the observed galaxy clustering.  We
outline such a scheme and present an illustrative test in \S~6.  In its
present form, this scheme assumes that the underlying cosmology is known, but
we speculate on prospects for breaking the degeneracy between bias and
cosmology.  We summarize our results in \S~7.

\section{Models of the Halo Occupation Distribution}

In the HOD framework, the relation between the galaxy and matter
distributions is fully defined by
\newline\noindent
(1) the probability distribution $\PNM$ that a halo of virial mass $M$
contains $N$ galaxies,
\newline\noindent
(2) the relation between the galaxy and dark matter spatial distributions
within halos, and
\newline\noindent
(3) the relation between the galaxy and dark matter velocity distributions
within halos.
\newline\noindent
We use the term ``Halo Occupation Distribution'' (or HOD) to refer to all
three of these aspects.  Each individual class of galaxies (defined, for
example, by luminosity and color ranges or by morphological type) has its
own HOD.

For this study we have used the high resolution GIF N-body simulations
carried out by the Virgo consortium \citep{jenkins98}.  The particular
model we have focused on is the flat $\Lambda$CDM model with $\Omegam=0.3$,
$\Omegal=0.7$, $H_0=70~\mathrm{km~s}^{-1}~Mpc^{-1}$, and a spectral shape
parameter $\Gamma=0.21$ (in the parameterization of \citealt{efstathiou92}).
The simulation follows the evolution of $256^3$ particles, each of mass
$1.4\times10^{10}\hMsun$, in a comoving box of size $141.3\hmpc$.  The rms
mass fluctuations on a scale of $8\hmpc$ are $\sigma_8=0.9$, in agreement
with the observed abundance of clusters \citep{eke96}.  Gravitational
forces are significantly softer than $1/r^2$ on scales $\lesssim 30\hkpc$.

We identify halos in the dark matter distribution using a
``friends-of-friends'' (FoF) algorithm \citep{davis85} with a linking
length of $0.2$ times the mean inter-particle separation, and we only
consider halos consisting of 10 or more particles.  This means that the
smallest halos we can resolve have a mass of $1.4\times10^{11}\hMsun$.

For our present purposes, an HOD prescription amounts to a recipe for
selecting a subset of the dark matter particles in these halos to represent
the galaxy population of the simulation.  Most of the models we show in
this paper are tuned to produce a galaxy population with a space density
of $\ng=0.01 \hvol$.  This space density corresponds to galaxies of
luminosity $L\gtrsim 0.5L_*$, assuming no further cuts in color or
morphology.  Our HOD prescriptions consist of the following features:

1. {\it $\NavgM$} --  We consider two types of models for this relation,
power laws and broken power laws.  In the power-law models, the mean
number of galaxies that populate dark matter halos of mass $M$ is
\begin{equation}
\Navg = \left\{ \begin{array}{ll}
		0 & {\mathrm if\ \ } M < \Mmin \\
		(M/M_1)^\alpha & \mathrm{otherwise},
		\end{array}
	\right.
\label{eqn:models1}
\end{equation}
where $\alpha$ is the power-law index, $\Mmin$ is the cutoff halo mass
below which halos cannot contain galaxies, and $M_1$ sets the amplitude of
the relation and corresponds to the mass of halos that contain, on average,
one galaxy.  In the broken power-law models,
\begin{equation}
\Navg = \left\{ \begin{array}{ll}
		0 & {\mathrm if\ \ } M < \Mmin \\
		(M/M_1)^\alpha & {\mathrm if\ \ } \Mmin\leq M\leq \Mcrit \\
		(M/M_1')^\beta & \mathrm{otherwise},
		\end{array}
	\right.
\label{eqn:models2}
\end{equation}
where $\alpha$ and $\beta$ are the low and high mass power-law indices,
$\Mcrit$ is the halo mass at which the power-law slope breaks, and $M_1'$
is required by continuity to be
$M_1' = M_1^{(\alpha/\beta)}\Mcrit^{(1-\alpha/\beta)}$.  For a given HOD
model, the value of $M_1$ is chosen to produce a galaxy population of the
desired space density.

The parameters $\Mmin$, $\alpha$, $\beta$, and $\Mcrit$ are directly related
to the efficiency of galaxy formation as a function of halo mass.  For
example, bright galaxies cannot form in halos below a certain mass because
these halos do not contain enough cold gas, hence the need for $\Mmin$.
The simplest form of $\NavgM$ would have $N$ proportional to $M$ ($\alpha=1$)
for $M>\Mmin$.  However, many physical mechanisms can alter the efficiency
of galaxy formation as a function of halo mass.  For example, the typical
cooling time for gas increases with halo mass, and this suggests that
$\alpha<1$.  This effect could be counteracted by an earlier average
formation time of galaxies that end up in high mass halos.   Galaxy mergers
may alter galaxy numbers preferentially in intermediate mass halos, where
mergers are most frequent.  Any given physical process can affect $\NavgM$
differently for different galaxy classes.  For example, mergers decrease the
number of low luminosity galaxies but increase the number of high luminosity
galaxies.  Morphological transformations, likewise, increase the numbers of
one galaxy type while decreasing those of another.  We will, therefore, be
able to learn much about the processes that determine galaxy properties by
comparing the HOD for different galaxy classes.

\begin{figure}
\centerline{
\epsfxsize=5.0truein
\epsfbox[90 160 460 675]{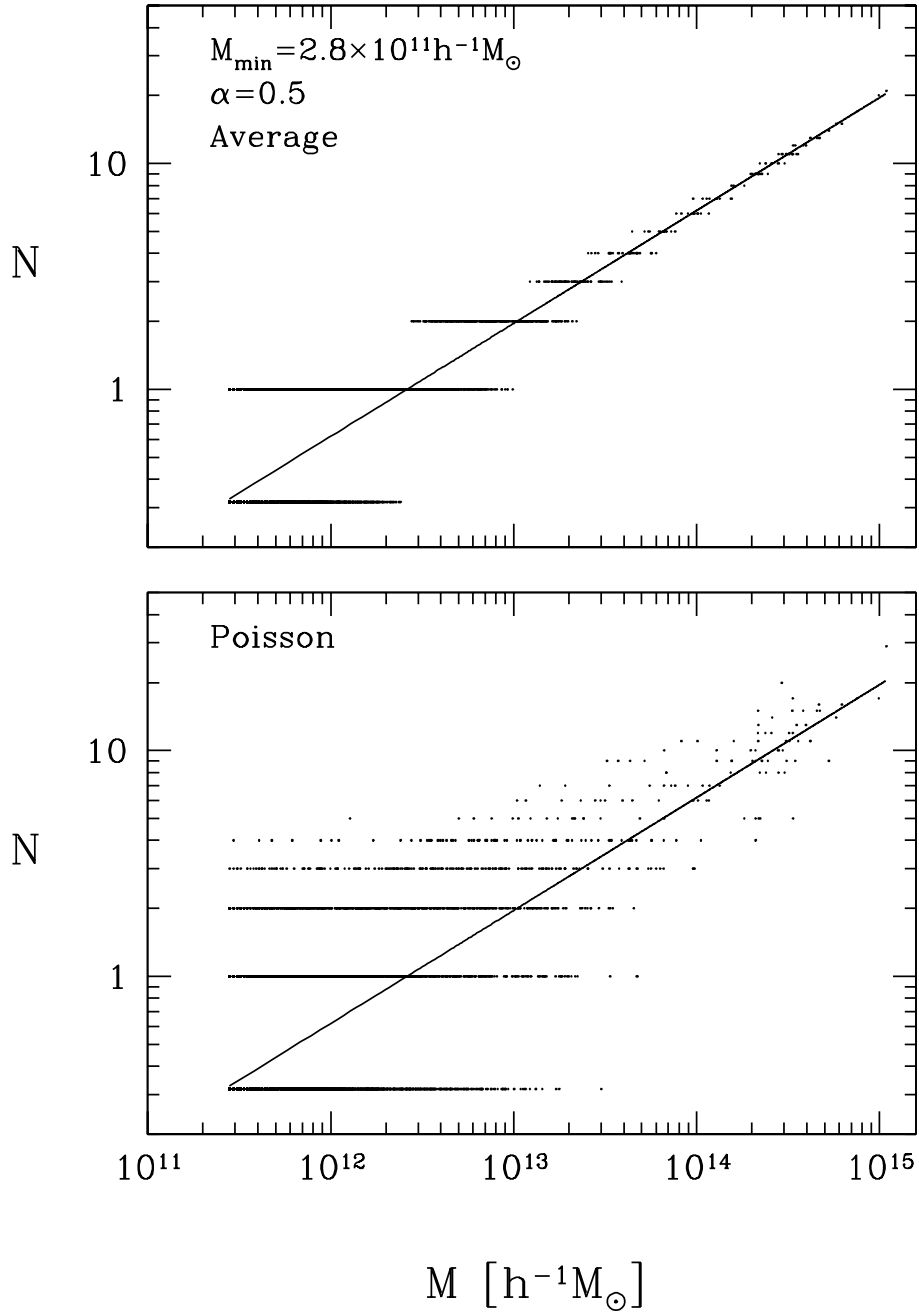}
}
\caption{$\PNM$ relation for two particular HOD models.  Each point
represents the number of galaxies that occupy a single halo in the
N-body simulation.  Points for halos that contain no galaxies are 
arbitrarily placed at log$N=-0.5$.  The HOD prescriptions shown have
a power-law $\NavgM$ with $\alpha=0.5$ and $\Mmin=2.8\times10^{11}\hMsun$
(see definitions in \S~2), as indicated by the solid lines.  The $\PNNavg$ 
distribution are Average (top panel) and Poisson (bottom panel).
} 
\label{fig:1}
\end{figure}

2. {\it $\PNNavg$} -- Once $\Navg$ is determined, the actual number of
galaxies that occupies any given halo is drawn from a probability
distribution $\PNNavg$.  We consider three such probability distributions:
(a) a Poisson distribution; (b) a very narrow distribution, which we call
``Average'', where the actual number of galaxies is the integer either
above or below $\Navg$, with relative frequencies needed to give the
required mean (this is identical to the ``Average'' distribution used by
\citealt{benson00}); (c) a Negative Binomial distribution, which is
substantially wider than Poisson.  Although it is possible for the form of
$\PNNavg$ to vary with $M$, we do not consider such models in this paper.
Figure~\ref{fig:1} shows $\PNM$ for a particular HOD prescription (power-law
$\NavgM$ with $\Mmin=2.8\times10^{11}\hMsun$, $\alpha=0.5$).  Each point
represents the number of galaxies chosen to occupy a particular halo in the
dark matter distribution.  The two panels show the difference between
assuming an Average and a Poisson $\PNNavg$.

In the high halo mass regime, $\PNNavg$ should depend on the
statistics of halo merger histories \citep{lacey93}.  The distribution of
merger histories for halos of a given mass will produce a resulting
distribution of galaxy numbers for those halos.  In the low mass regime,
each halo is expected to contain only one galaxy, but whether that galaxy
would pass a given luminosity threshold depends on the gas cooling and star
formation history of that halo.  Therefore, $\PNNavg$ depends on the
regularity of galaxy formation in halos of a given final mass.

3. {\it Central galaxy} -- Once the actual number of galaxies $N$ that
occupy each halo is determined, we must specify how these galaxies are
distributed within halos.  The first step in this process is to specify
whether or not there must be a galaxy at the center of each halo for which
$N>0$.  We investigate both these cases.  If we force a galaxy to sit at
the halo center, we place it at the halo center of mass and assign it
the mean halo velocity.

4. {\it Galaxy concentration} -- We allow for the possibility that galaxies
are more or less spatially concentrated than the dark matter within halos.
We implement such models by selecting ``galaxy'' particles with probability
$P \propto r^{\Dg}$, so that, on average,
\begin{equation}
\rho_g(r)/\rho_m(r) = r^{\Dg}.
\label{eqn:models3}
\end{equation}
This prescription imposes a difference $\Dg$ in the logarithmic slopes
of galaxy and dark matter profiles without imposing any specific form or
symmetry on the galaxy distribution in the halos; the galaxy distribution
will inherit the geometry of the dark matter.  A non-zero $\Dg$
can be applied together with the central galaxy prescription or on its own.

5. {\it Velocity bias} -- Finally, we allow for the possibility of velocity
bias within halos.  The mean velocity of galaxies in a halo should not differ
from that of the dark matter, since both components are responding to the
same large-scale gravitational field.  However, the galaxies in a halo might
have a higher or lower velocity dispersion than the dark matter particles
at the same locations.  We define a velocity bias factor $\alpha_v$ through
the relation
\begin{equation}
\vg = \vh + \alpha_v (\vm - \vh),
\label{eqn:models4}
\end{equation}
where $\vg$ and $\vm$ are the velocities of the galaxies and the dark
matter particles that they are assigned to and $\vh$ is the mean
center-of-mass halo velocity.  For example, if $\alpha_v=0$ all the
galaxies within a halo have the mean halo velocity, while if $\alpha_v=1$
galaxy velocities trace the dark matter velocities.  Our central galaxy and
galaxy concentration prescriptions impose some degree of velocity bias even
if $\alpha_v=1$, the first because the central galaxy is assumed to move at
$\vh$, and the second because the typical dark matter velocities depend on
radius if the halo is not isothermal.

Prescriptions~3, 4, and 5 allow us to represent a number of physical
processes that could affect the galaxy distribution in important ways.
For example, dynamical friction could cause galaxies to sink to the center
of a halo and end up with a spatial distribution that is more centrally
concentrated and a velocity distribution that is colder than that of dark
matter.  In addition, if galaxies form near the centers of their original
parent halos, they can inherit spatial and velocity bias as a result of
incomplete relaxation during the merging of these halos into a larger common
halo \citep{evrard87}. On the other hand, galaxy mergers happening at the
centers of massive halos could reduce galaxy numbers in those regions,
thus causing galaxies to be less centrally concentrated than dark matter
($\Dg>0$).  Also, an $\alpha_v>1$ velocity bias could arise as a result of
preferential destruction or merging of galaxies that have a low velocity.

Following the above steps, we have created a large number of galaxy
distributions spanning a wide space of HOD parameters.  All of these galaxy
distributions come from the same dark halo population and differ only in the
HOD.  We have calculated a variety of clustering statistics for each of
these galaxy distributions in order to test the sensitivity of each statistic
to features of the HOD.  It would be impractical to show all statistics for
all of our HOD models, so in each of the following sections we focus on a
subset of models that illustrate the sensitivity of the statistic under
examination.
Our approach of populating N-body simulations according to an
HOD with power-law $\NavgM$ is similar to that used by
\cite{jing98b}, \cite{jing98c}, and \cite{jing02}
in their modeling of measurements from the LCRS and PSCz redshift surveys.
However, the implementations are different, and our HOD model
is substantially more general: we allow variations of $\Mmin$ and
the form of $\PNNavg$ in addition to variations in $\alpha$,
and we allow the possibility of spatial and velocity biases within halos.

We asserted in \S~1 that the HOD formulation can provide a complete statistical
description of the bias between galaxies and mass.  Underlying this assertion
is an implicit assumption that the galaxy content of a halo of virial mass
$M$ is statistically independent of that halo's larger scale environment.
This assumption is supported by the N-body simulation results of
\citet{lemson99}, who find, in agreement with predictions of the excursion
set model \citep{bond91}, that halos of fixed mass in different environments
have similar properties and formation histories, although the halo mass
function is itself systematically shifted towards higher mass halos in high
density regions.  However, the alleged independence of halo histories and
large scale environment merits more detailed theoretical investigation.  The
hypothesis that the HOD formulation is complete will ultimately be tested
empirically, by seeing whether an HOD model can reproduce all facets of
observed galaxy clustering when applied to a cosmological model consistent
with other observational data.  Additional implicit assumptions, that all
galaxies reside in virialized dark matter halos and that the halo population
itself is minimally affected by baryonic physics, seem well justified, though
the latter deserves more thorough testing with hydrodynamic simulations
\footnote{The second assumption would not hold if we defined halos at a much
higher overdensity, since dissipative collapse of baryon clumps increases
their ability to retain surrounding dark matter concentrations within groups
and clusters.  However, from our point of view, high density halos surrounding
individual galaxies are substructure within overdensity $200$ halos, so they
are described statistically by the HOD itself.}.  The specific parameterizations
adopted here may not capture all of the important features of the true HOD,
though they are flexible enough to produce a wide range of results and can
easily be generalized to include, e.g., mass dependence of $\PNNavg$ or $\Dg$.

\section{The Galaxy Correlation Function}

We begin our analysis with the two-point correlation function $\xig$,
which plays a fundamental role in understanding galaxy clustering because
it has been thoroughly studied as a function of galaxy type, color, and
luminosity (see \citealt{norberg01}, \citeyear{norberg02},
\citealt{zehavi01}, and numerous
references therein) and because its observed form is remarkably simple.
For typical optically selected samples, $\xig$ is a power-law
$(r/r_0)^{-\gamma}$ for separations $r \lesssim 5\hmpc$, with
$r_0 \approx 5-6\hmpc$ and $\gamma \approx 1.8$.  More luminous galaxies
have a larger $r_0$ and similar $\gamma$, while redder or early-type
galaxies have a larger $\gamma$ and a higher clustering amplitude on small
scales.

Cosmological N-body simulations show that CDM models do not predict a
power-law matter correlation function (e.g., \citealt{jenkins98}), and
analytic theory (\citealt{hamilton91}; \citealt{peacock96}) implies that
the linear theory power spectrum would have to contain a specially tuned
feature in order to yield a power-law $\xi(r)$ on non-linear scales.  If the
primordial power spectrum is a smooth function, as expected on theoretical
grounds, it appears that scale-dependent bias must transform the curved
matter correlation function into a power-law galaxy correlation function.
Remarkably, galaxy distributions predicted by hydrodynamic simulations, by
``sub-halo'' analyses of high resolution N-body simulations, and by
semi-analytic models applied to N-body halos all yield power-law galaxy
correlation functions, at least for some reasonable choices of cosmology
and galaxy definition parameters (\citealt{pearce99}; \citealt{dave00};
\citealt{cen00}; \citealt{yoshikawa01}; \citealt{colin99};
\citealt{kauffmann99}; \citealt{benson00}; \citealt{somerville01}).

In the context of HOD bias, we would like to know whether a power-law
$\xig$ follows from some simple and generic property of the HOD, such as a
particular galaxy profile in high multiplicity halos, or whether it demands
finely tuned parameter choices.   We also want to know more generally how
the amplitude and shape of $\xig$ depend on parameters of the HOD.

\subsection{Analytic Discussion}

Recent papers (\citealt{seljak00}; \citealt{ma00}; \citealt{scoccimarro00};
\citealt{sheth01}) present a fairly complete analytic theory of the galaxy
correlation function in the halo bias model.  The basic idea is to add the
``1-halo'' term representing galaxy pairs within a single halo to the ``2-halo''
term representing pairs in separate, spatially correlated halos (as done in
a different context by \citealt{scherrer91}).  The full analytic theory
becomes simple in the Fourier domain, where convolutions of the halo profile
transform into multiplications.  As a guide to interpreting our numerical
results, we begin with a complementary, more approximate discussion of
correlations in real space.

For separations larger than the virial diameter of the largest halos, all
pairs must come from galaxies in separate halos.  \citet{mo96} derived an
analytic approximation for the bias factor of halos $b_h(M)$ as a function
of halo mass $M$ using the \citet{press74} formalism.  Above the
characteristic mass $M_{*}$ in the Press-Schechter mass function, halo
formation is enhanced in regions of high background density (and suppressed
in underdense regions), so $b_h(M)$ exceeds unity and increases rapidly with
increasing mass.  Halos with $M<M_{*}$ are weakly anti-biased because
they merge into more massive systems in overdense regions.  \citet{jing98a}
numerical results yield $b_h(M) \approx 0.7-0.8$ for $M \ll M_{*}$.  We
assume that the cross-correlation between halos of mass $M_1$ and $M_2$ is
\begin{equation}
\xi_{12}(r) = b_h(M_1)b_h(M_2)\xim ,
\label{eqn:xi1}
\end{equation}
where $\xim $ is the mass correlation function.

Consider a halo of mass $M_1$.  The mean number of excess galaxies in a
volume $dV$ at distance $r$ from the halo is obtained by integrating over
the differential halo mass function $dn/dM$ (which has units of number density
per unit mass), weighting each halo by the
mean number of galaxies $\NavgM$ and by the biased correlation factor:
\begin{equation}
N_{\mathrm{excess}} = \int_{\Mmin}^{\infty}dM_2\frac{dn}{dM_2}\Navg(M_2)
                      b_h(M_1)b_h(M_2)\xim dV.
\label{eqn:xi2}
\end{equation}
If the number density of galaxies is $\ng$, then the number density of
correlated galaxy pairs is
\begin{eqnarray}
\frac{1}{2}\ng^2 \xig dV & = & \frac{1}{2}\int_{\Mmin}^{\infty}dM_1\frac{dn}{dM_1}
                        \Navg(M_1) N_{\mathrm{excess}} \nonumber \\
              & = & \frac{1}{2}\int_{\Mmin}^{\infty}dM_1\frac{dn}{dM_1}\Navg(M_1)
                    \int_{\Mmin}^{\infty}dM_2\frac{dn}{dM_2}\Navg(M_2)
                         b_h(M_1)b_h(M_2)\xim dV \nonumber \\
              & = & \frac{1}{2}\xim dV\int_{\Mmin}^{\infty}dM_1\frac{dn}{dM_1}
                        \Navg(M_1)b_h(M_1)\int_{\Mmin}^{\infty}dM_2
                        \frac{dn}{dM_2}\Navg(M_2) b_h(M_2),
\label{eqn:xi3}
\end{eqnarray}
where the factor of $1/2$ corrects for double counting of each pair.
Thus, at large separations we expect
\begin{equation}
\xig = b^2\xim ,\qquad  b= \ng^{-1}\int_{\Mmin}^{\infty}dM
                                    \frac{dn}{dM}\NavgM b_h(M),
\label{eqn:xi4}
\end{equation}
and the galaxy bias is just the weighted value of the halo bias.

On small scales, the correlation function is dominated by the 1-halo term
representing galaxy pairs that reside in the same halo.  The total number
density of such pairs is
\begin{equation}
n_{\mathrm{pair,1h}} =  \int_{\Mmin}^{\infty}dM\frac{dn}{dM}
                        \frac{\NN_M}{2} \approx
                        \frac{\ng^2}{2}\int_{0}^{2\Rmax}\xig4\pi r^2dr,
\label{eqn:xi5}
\end{equation}
where $\NN_M = \int_{0}^{\infty}dN\PNM N(N-1)$ and the
second equality would be exact if {\it all} correlated pairs out to
the maximum halo virial diameter $2\Rmax$ came from the 1-halo contribution.
(\citealt{bullock02} apply a similar argument to Lyman-break galaxy
clustering.)  More generally, we can write
\begin{equation}
\frac{\ng^2}{2}\int_{0}^{r}\xih4\pi r^2dr = \int_{\Mmin}^{\infty}dM
\frac{dn}{dM}\frac{\NN_M}{2} F\left(\frac{r}{2\Rvir}\right),
\label{eqn:xi6}
\end{equation}
where $\xih$ refers specifically to the correlation function
associated with galaxy pairs in the same halo and the function $F(r/2\Rvir)$
represents the average fraction of galaxy pairs in halos of mass $M$ that
have separation less than $r$ (a fraction $r/2\Rvir$ of the virial
diameter).  Since the only $r$ dependence on the r.h.s. of
equation~(\ref{eqn:xi6}) appears in $F(r/2\Rvir)$, we can differentiate
with respect to $r$ to find
\begin{equation}
2\pi r^2\ng^2\xih = \int_{\Mmin}^{\infty}dM\frac{dn}{dM}
                           \frac{\NN_M}{2}
                        \frac{1}{2\Rvir(M)}F'\left(\frac{r}{2\Rvir}\right).
\label{eqn:xi7}
\end{equation}

While the large scale bias factor (eq.~[\ref{eqn:xi4}]) depends only on
$\NavgM$, the correlation function in the 1-halo regime depends on the 2nd
factorial moment $\NN_M$, and hence on the form of
$\PNNavg$.  A Poisson distribution of mean $\left<N\right>$ has variance
$\left<N^2\right>-\left<N\right>^2=\left<N\right>$, so
$\NN_M = \left<N^2\right>_M - \left<N\right>_M =
\left<N\right>_M^2 = \Navg^2(M)$.  Our Average distribution, on the other
hand, has $\NN_M = \Navg^2(M) - \NavgM$ (even when $\NavgM$
is not an integer), which can be substantially lower than the Poisson value
when $\NavgM$ is small.  The small-scale galaxy correlation function also
depends on the halo profile through $F(r/2\Rvir)$ --- more concentrated halos
have a larger fraction of pairs at small $r$, boosting the small-scale
correlations at the expense of slightly larger separations.

\begin{figure}
\centerline{
\epsfxsize=4.0truein
\epsfbox[135 180 430 710]{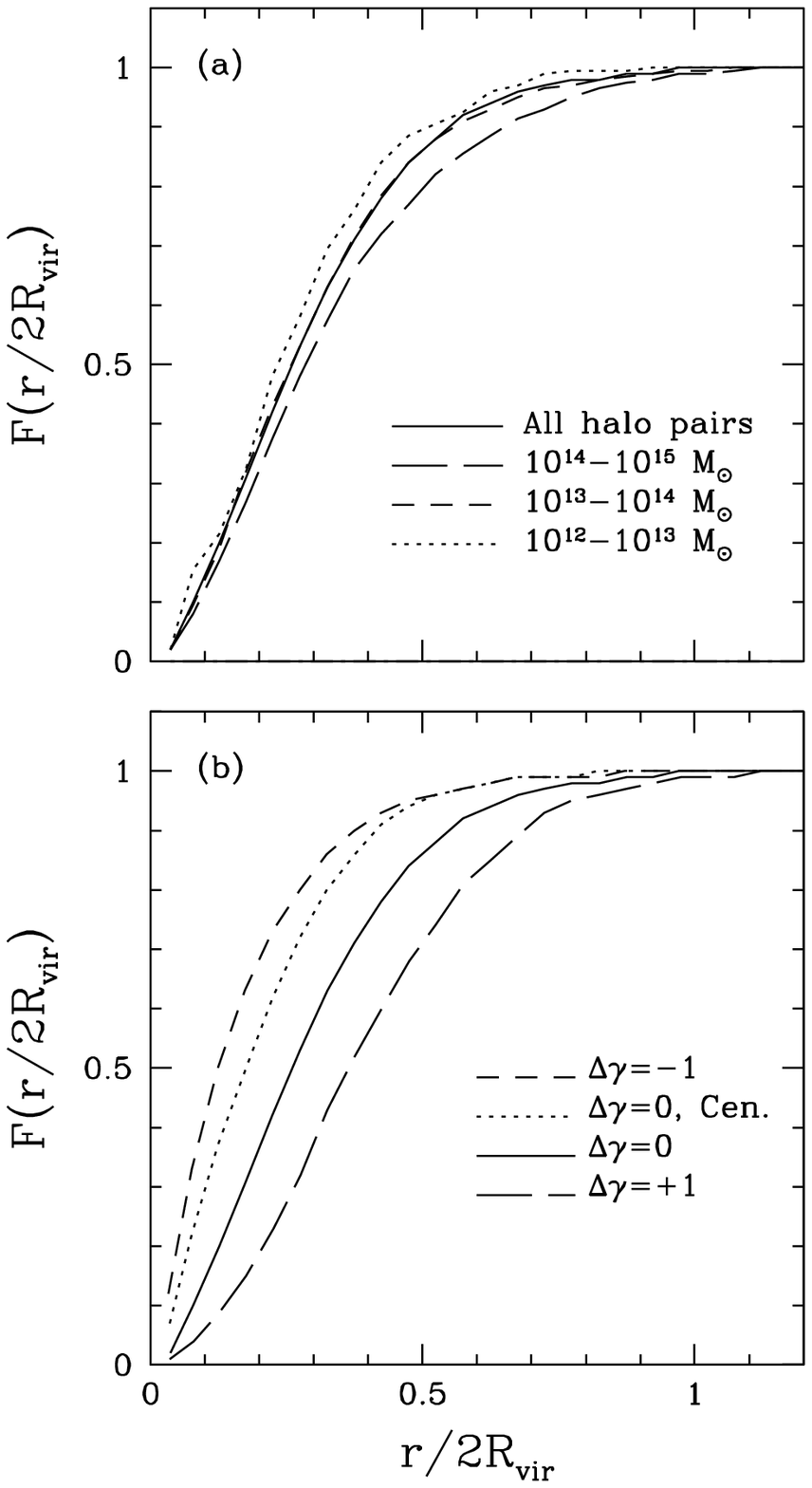}
}
\caption{The cumulative fraction of galaxy pairs within halos as a function
of separation, $F(r/2\Rvir)$, as defined in equation~(\ref{eqn:xi6}).
Panel~(a) shows $F(r/2\Rvir)$ for halos of different mass ranges in the case
of an HOD model with $\alpha=0.5$, $\Mmin=2.8\times10^{11}\hMsun$, and an 
Average $\PNNavg$. 
Panel~(b) shows $F(r/2\Rvir)$ averaged over the halo population for HOD 
models with the same $\PNM$ as panel~(a), but with variations of the galaxy
spatial distribution within halos.  There are three HOD prescriptions with 
different values of $\Dg$ and one where a galaxy is forced to lie at the 
center of each halo for which $N>0$.
} 
\label{fig:2}
\end{figure}

Figure~\ref{fig:2}a shows the cumulative pair fraction $F(r/2\Rvir)$ for
one of our HOD models.  We take the virial radius of each halo to be
\begin{equation}
\Rvir = \left(\frac{3M}{800\pi \bar{\rho}_m}\right)^{1/3},
\label{eqn:xi8}
\end{equation}
where $\bar{\rho}_m$ is the mean mass density and $M$ is the halo mass.
We measure the separation distribution from the numerically realized
galaxy distribution and average over all halos (solid line), as well as
halos in narrower mass bins.  With separations scaled to virial diameters,
the function $F(r/2\Rvir)$ is nearly independent of halo mass, though there
is a slight trend of higher concentration at lower mass, as expected from
N-body studies of halo profiles (\citealt{navarro97}; \citealt{bullock01}).
All pairs have separations less than $2\Rvir$, and about half the pairs have
separations less than $\Rvir/2$.

The halo-averaged form of $F(r/2\Rvir)$ should be virtually independent of
$\PNM$, since the cumulative pair distribution is insensitive to halo mass.
However, $F(r/2\Rvir)$ is sensitive to the relative distribution of galaxies
and dark matter within halos.  Figure~\ref{fig:2}b shows $F(r/2\Rvir)$
averaged over all halos for HOD models in which the galaxies are more
centrally concentrated ($\Dg=-1$; short-dashed curve) or less centrally
concentrated ($\Dg=+1$; long-dashed curve) than the dark matter.  As expected,
$F(r/2\Rvir)$ rises faster for more central concentration, since most pairs
lie close to the halo center.  The dotted curve in Figure~\ref{fig:2}b shows
a model with $\Dg=0$ but a central galaxy in every halo (for which $N>0$),
which yields a result intermediate between the $\Dg=0$ and $\Dg=-1$ models.
In this case, unlike all the others, we expect $F(r/2\Rvir)$ to depend
strongly on halo mass, since the central galaxy contributes to a significant
fraction of pairs in low multiplicity halos but not in high multiplicity
halos.  For central galaxy models, it is more informative to rewrite
equation~(\ref{eqn:xi6}) in the form
\begin{equation}
\frac{\ng^2}{2}\int_{0}^{r}\xih 4\pi r^2dr = \int_{\Mmin}^{\infty}dM
        \frac{dn}{dM}\left[\frac{\left<(N-1)(N-2)\right>_M}{2}
F\left(\frac{r}{2\Rvir}\right)
      + \left<N-1\right>_M F_c\left(\frac{r}{2\Rvir}\right)\right],
\label{eqn:xi9}
\end{equation}
where $F(r/2\Rvir)$ now refers only to pairs that do {\it not}
involve the central galaxy and $F_c(r/2\Rvir)$, which is simply the
cumulative halo profile itself, refers only to pairs that include the
central galaxy.  Here $F(r/2\Rvir)$ and $F_c(r/2\Rvir)$ should both be
approximately independent of halo mass, but $F_c$ rises faster than $F$,
and the relative weight of the two terms depends on $N$.  We see from
equation~(\ref{eqn:xi9}) that a central galaxy should be important only on
scales where low multiplicity halos make a significant contribution to
the correlation function.

\subsection{Numerical Results}

\begin{figure}
\centerline{
\epsfxsize=6.0truein
\epsfbox[60 190 530 660]{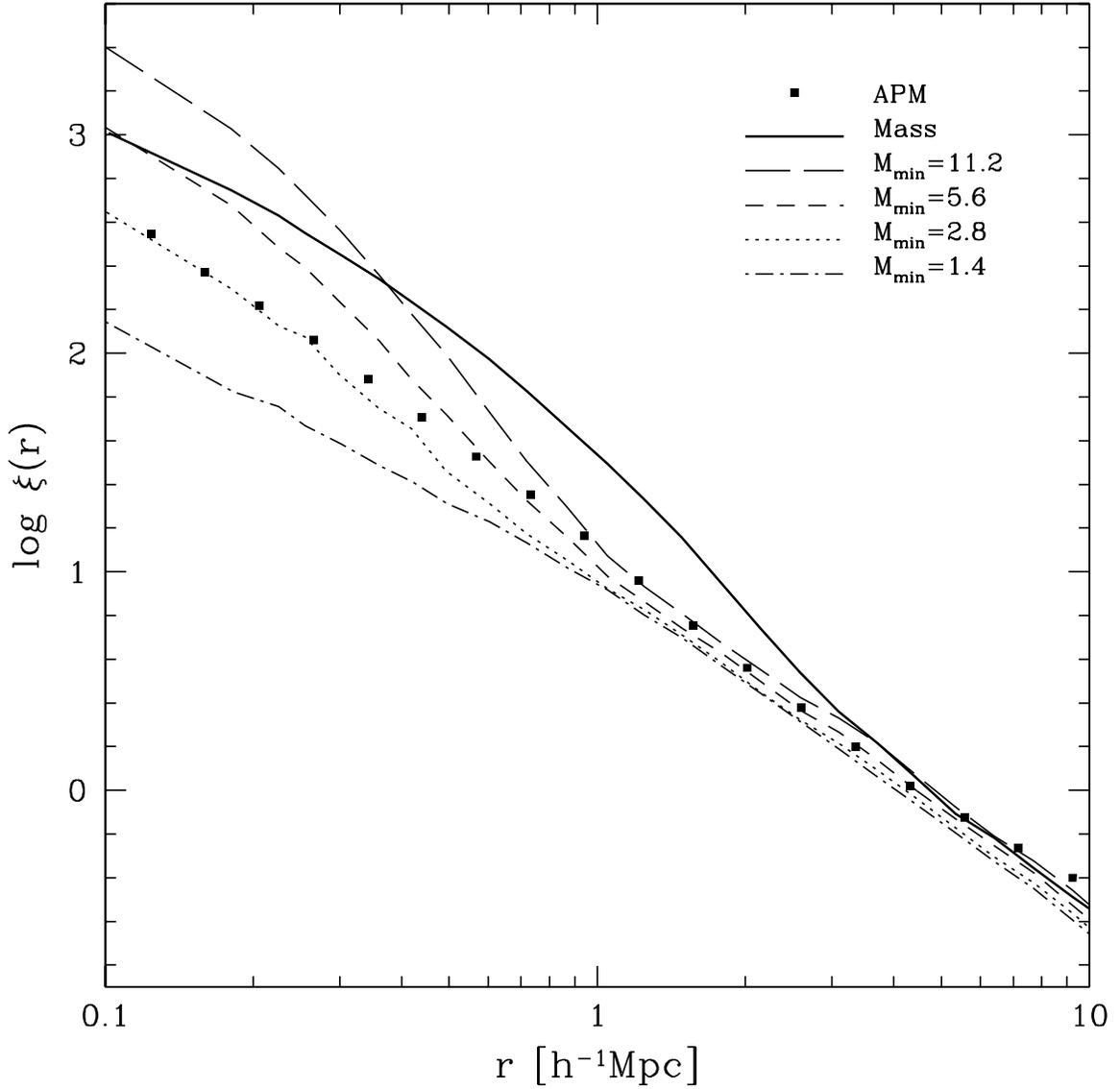}
}
\caption{Influence of $\Mmin$ on the galaxy correlation function.  Curves show
galaxy correlation functions for HOD models with a power-law $\NavgM$,
$\alpha=0.5$, Average $\PNNavg$, and different values of $\Mmin$, which
are listed in the legend in units of $10^{11}\hMsun$.  The solid curve shows
the mass correlation function, and points show the correlation function 
measured from the APM galaxy survey \citep{baugh96}.
} 
\label{fig:3}
\end{figure}

\begin{figure}
\centerline{
\epsfxsize=6.0truein
\epsfbox[60 190 530 660]{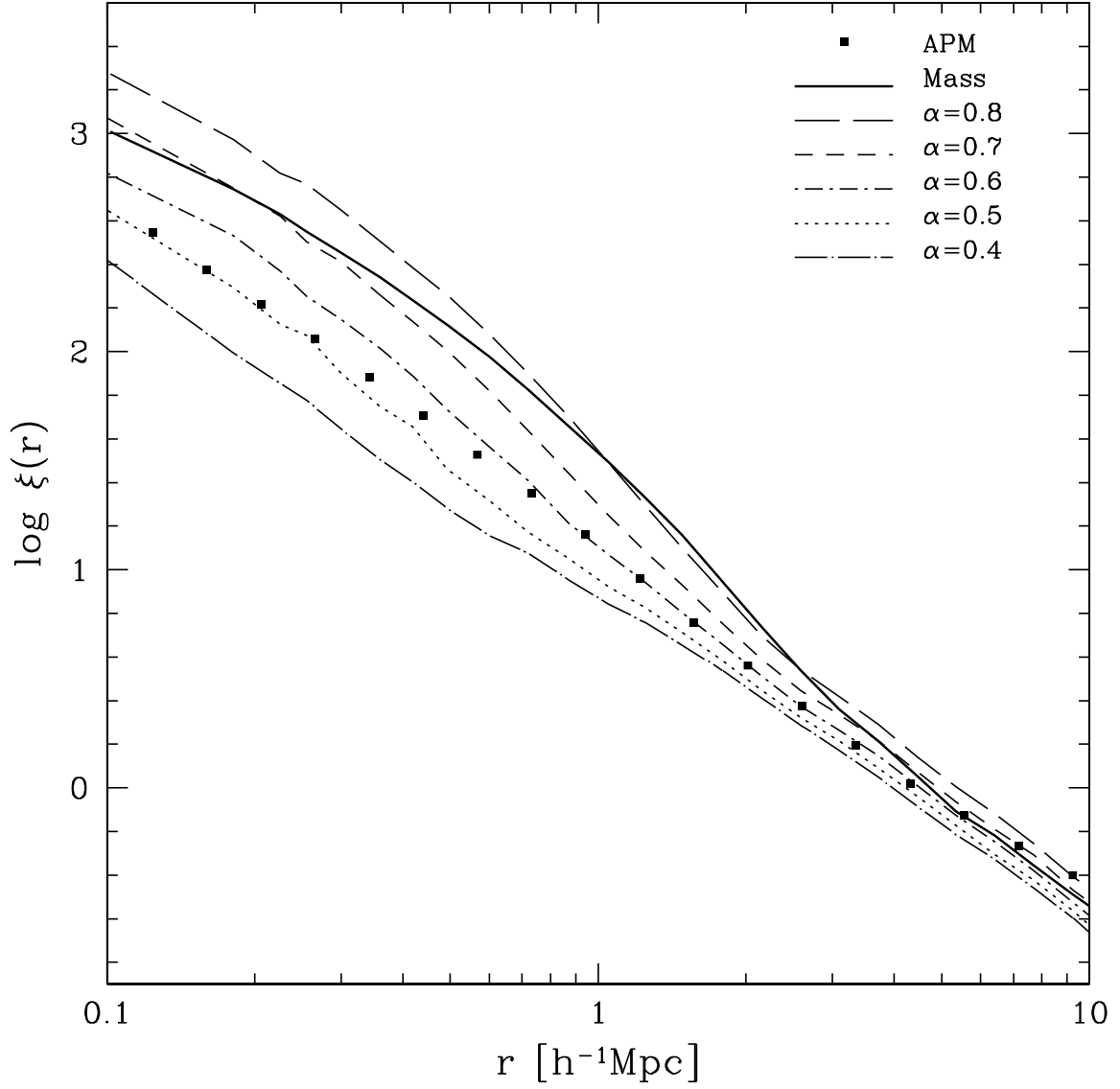}
}
\caption{Influence of $\alpha$ on the galaxy correlation function.  Curves show
galaxy correlation functions for HOD models with a power-law $\NavgM$,
$\Mmin=2.8\times 10^{11}\hMsun$, Average $\PNNavg$, and different values of 
$\alpha$, which are listed in the legend.
} 
\label{fig:4}
\end{figure}

\begin{figure}
\centerline{
\epsfxsize=6.0truein
\epsfbox[60 190 530 660]{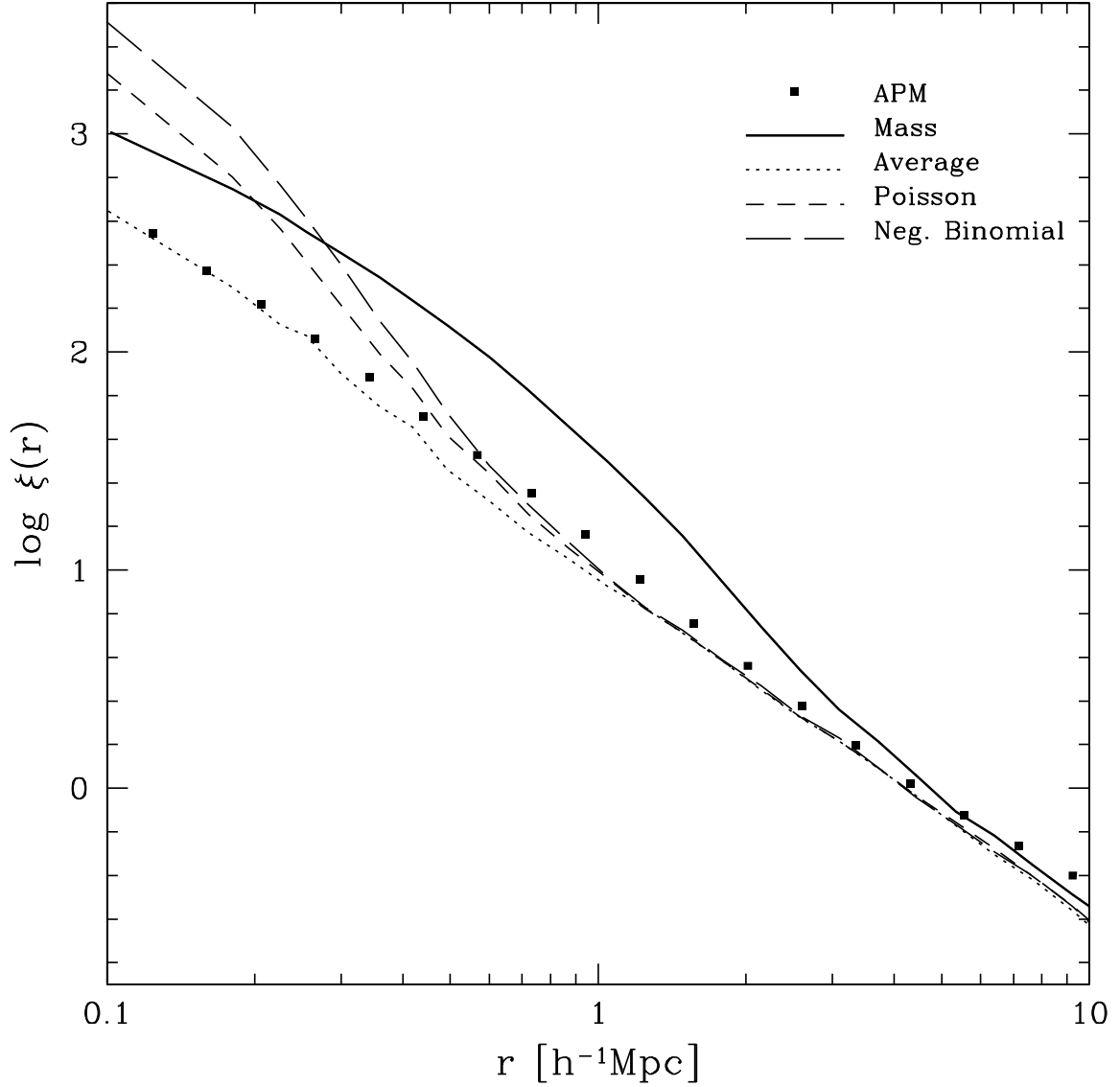}
}
\caption{Influence of $\PNNavg$ on the galaxy correlation function.  Curves show
galaxy correlation functions for HOD models with a power-law $\NavgM$,
$\Mmin=2.8\times 10^{11}\hMsun$, $\alpha=0.5$, and different forms of 
$\PNNavg$, which are listed in the legend.
} 
\label{fig:5}
\end{figure}

\begin{figure}
\centerline{
\epsfxsize=6.0truein
\epsfbox[60 190 530 660]{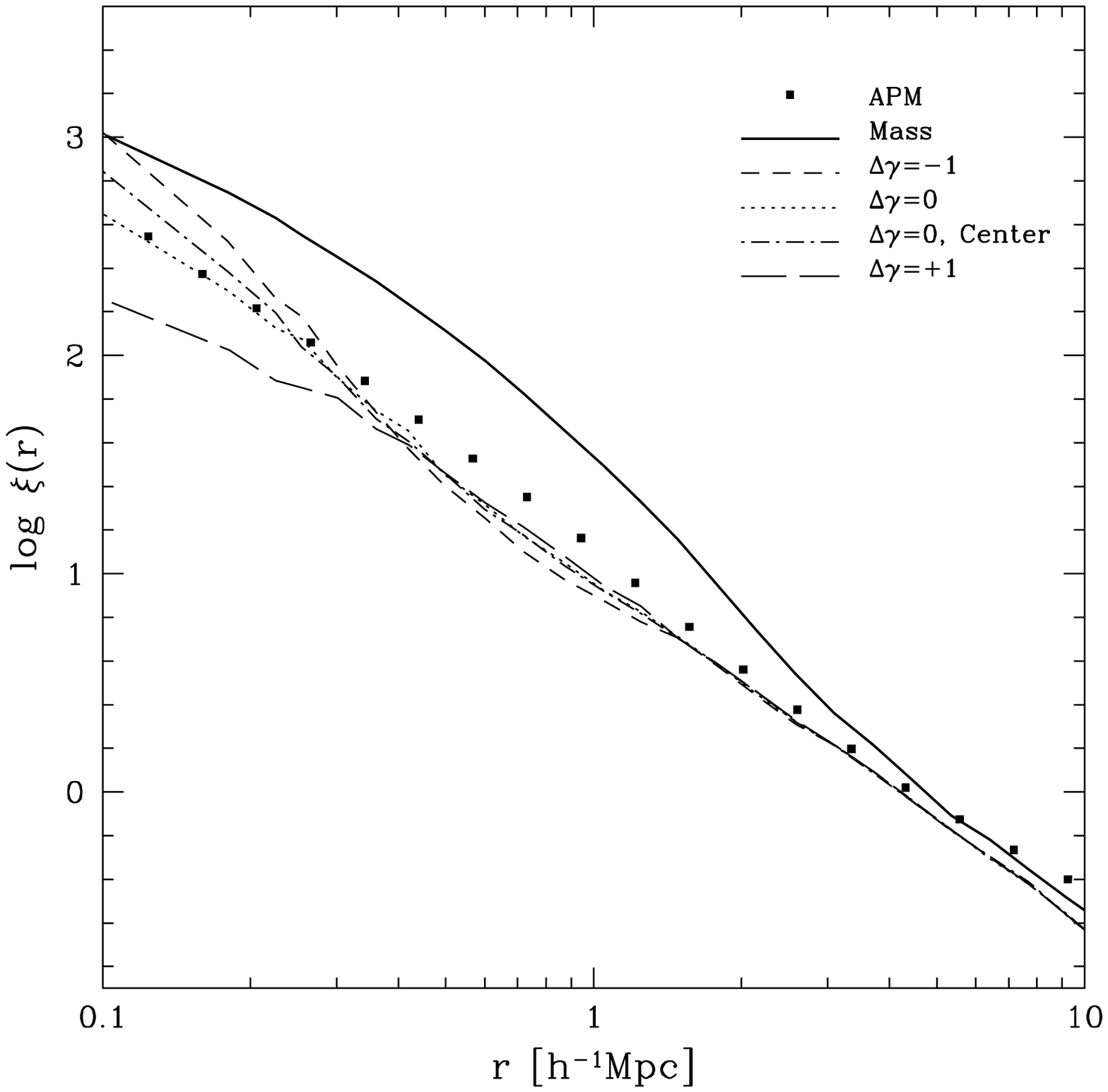}
}
\caption{Influence of the galaxy profiles within halos on the galaxy 
correlation function.  As in Figs.~\ref{fig:3}-\ref{fig:5}, the dotted
curve shows $\xig$ for a model with $\Mmin=2.8\times 10^{11}\hMsun$,
$\alpha=0.5$, Average $\PNNavg$, and galaxies tracing dark matter within
halos.  Short-dashed and long-dashed curves show results for models
in which galaxy profiles are respectively more concentrated or less
concentrated than dark matter profiles ($\Dg=-1$ or $\Dg=+1$, see
eq.~[\ref{eqn:models3}]).  The dot-dashed curve shows a model in which the first
galaxy of each halo lies at the halo center and subsequent galaxies have the
same profile as the dark matter.
} 
\label{fig:6}
\end{figure}

Figures~\ref{fig:3}-\ref{fig:6} show correlation functions for galaxy
distributions created using a variety of HOD models that all have a power-law
$\NavgM$ (as defined in eq.~[\ref{eqn:models1}]).  Each figure demonstrates
the dependence of $\xig$ on a particular feature of the HOD by displaying
models that differ only in that feature.  We note that models with different
values of $\Mmin$ or $\alpha$ also have different values of $M_1$, since this
parameter is used to fix the mean number density of galaxies to $\ng=0.01\hvol$.
For purpose of comparison, each figure also shows the dark matter correlation
function and the real space $\xig$ inferred by \citet{baugh96} from angular
clustering in the APM galaxy survey \citep{maddox90}.


Figure~\ref{fig:3} shows the effect on $\xig$ of varying $\Mmin$.  On large
scales, only the amplitude of the correlation function is affected, with
higher values of $\Mmin$ having a slightly larger amplitude.  Since the
number density of galaxies remains fixed, the result of increasing $\Mmin$
is to remove galaxies from low mass halos and redistribute them into halos of
mass $M>\Mmin$.  In terms of equation~(\ref{eqn:xi4}), this means that weight
is taken away from the lower mass halos that have a smaller halo bias factor
$b_h$, so the galaxy bias $b$ increases.  On scales $r \lesssim 1 \hmpc$,
both the shape and amplitude of the correlation function are very sensitive to
$\Mmin$, with higher values of $\Mmin$ producing a steeper slope and higher
amplitude of $\xig$.  Figure~\ref{fig:4} shows the effect of varying $\alpha$.
Qualitatively, increasing $\alpha$ has an effect similar to that of increasing
$\Mmin$, since both changes boost the fraction of galaxies in high mass,
positively biased halos.  However, the changes to the shape of $\xig$ are
different in detail.  We will return to a discussion of these effects shortly.

Figure~\ref{fig:5} shows the effect of varying $\PNNavg$ while keeping
$\NavgM$ fixed.  It is clear that the impact on large scales is negligible,
while on small scales the amplitude of $\xig$ increases with increasing
width of the $\PNNavg$ distribution.  This behavior is expected because the
large-scale bias factor (eq.~[\ref{eqn:xi4}]) depends only on $\NavgM$,
while $\xih$ depends on $\NN_M$ (eq.~[\ref{eqn:xi7}]), which is larger for
wider distributions that have the same $\Navg$.  The impact of $\PNNavg$
grows at smaller scales, where low mass (and hence low multiplicity) halos
contribute to $\xih$.

Figure~\ref{fig:6} shows the effect of varying the distribution of galaxies
within halos, while keeping $\PNM$ fixed.  The plot shows four models: one
where galaxies are more centrally concentrated than the dark matter within
halos ($\Dg=-1$), one where galaxies are less concentrated within halos
($\Dg=+1$), one where galaxies trace the dark matter within halos ($\Dg=0$),
and a model where a galaxy is forced to lie at the center of every halo for
which $N>0$.  The effects of varying the spatial distribution of galaxies
within halos are confined to small scales.  As expected, the model where
galaxies are more centrally concentrated within halos has a correlation
function that is amplified on very small scales and depressed on scales
corresponding to the virial size of halos.  The model where galaxies are
less centrally concentrated exhibits the opposite behavior.  Forcing halos
to have central galaxies only affects $\xig$ by increasing it slightly on
the smallest scales, as we expect from equation~(\ref{eqn:xi9}).  Relative
to changes in $\PNM$, the effect of changing the distribution of galaxies
within halos is small, at least on the scales $r>0.1\hmpc$ considered here.

\begin{figure}
\centerline{
\epsfxsize=5.0truein
\epsfbox[50 35 555 765]{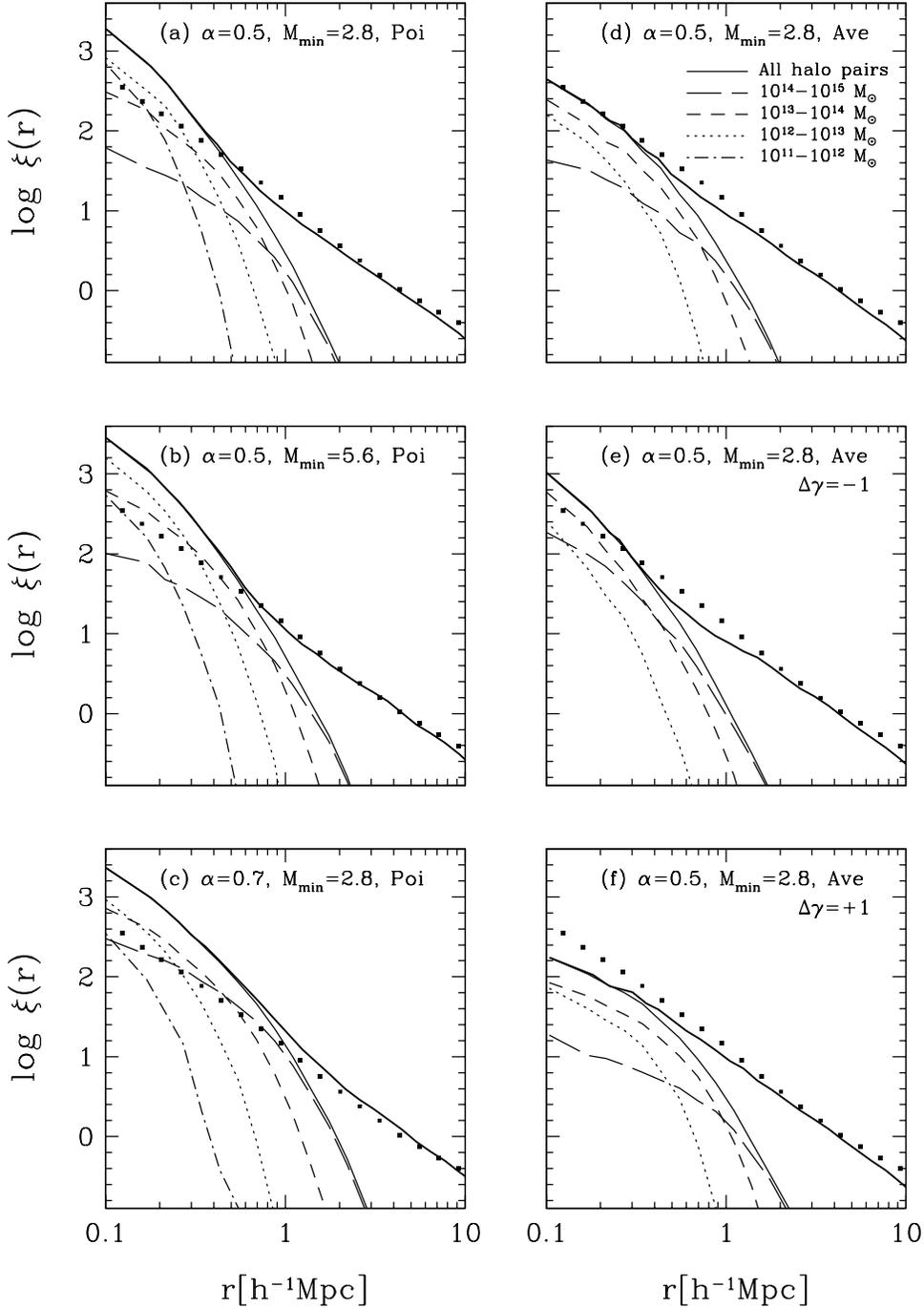}
}
\caption{Contributions to the galaxy correlation function from pairs
within halos of different mass ranges.  Each panel represents a particular 
HOD model.  The models are specified at the top of each panel, where
$\Mmin$ is given in units of $10^{11}\hMsun$, and Poi and Ave 
represent Poisson and Average $\PNNavg$ distributions, respectively.
Each panel shows the full galaxy correlation function (thick solid curve), 
the correlation function including only galaxy pairs that lie within the 
same halos (thin solid curve), and the contribution to $\xig$ from pairs 
that lie only within halos of a certain mass range (remaining 4 curves).  
The mass ranges are indicated in panel~(d). Also shown, for comparison, 
is the APM galaxy correlation function (squares).
} 
\label{fig:7}
\end{figure}

To better understand the behavior in Figures~\ref{fig:3}-\ref{fig:6}, it is
helpful to decompose $\xih$ into contributions from halos in different mass
ranges.  Figure~\ref{fig:7} presents such a decomposition for several HOD
models (see figure~2 of \citealt{seljak00} for a similar decomposition in
the Fourier domain).  Figure~\ref{fig:7}a shows a power-law $\NavgM$ model
with $\alpha=0.5$ and $\Mmin=2.8\times10^{11}\hMsun$ that has a Poisson
$\PNNavg$.  The thick solid curve shows the correlation function for this
galaxy distribution (also seen in Figures~\ref{fig:3}-\ref{fig:6}), and the
thin solid curve represents the contribution from galaxy pairs within the
same halo (the 1-halo term).  As expected, the 1-halo term dominates on
scales up to the virial size of typical halos and drops off quickly at
larger scales.  The remaining four curves show the contribution to the
1-halo term from halos with $M = 10^{11}-10^{12}\hMsun$ (dot-dashed curve),
$10^{12}-10^{13}\hMsun$ (dotted curve), $10^{13}-10^{14}\hMsun$ (short
dashed curve), and $10^{14}-10^{15}\hMsun$ (long dashed curve).  Each curve
is highest at the smallest scales and drops off at larger scales.  However,
the curves for high mass halos start at lower amplitude and extend to larger
$r$, since pairs are spread over a larger virial volume.  Consequently,
galaxy pairs in low mass halos dominate $\xig$ at very small scales, while
galaxy pairs in high mass halos dominate $\xig$ on scales corresponding to
their virial radii.

Figure~\ref{fig:7}b shows the effect of doubling $\Mmin$ while keeping all
other HOD parameters fixed.  Since $\ng$ remains constant, galaxies that were
previously in low mass halos are redistributed to halos above the new value
of $\Mmin$.  In Figure~\ref{fig:7}b, the contributions to $\xig$ from galaxy
pairs in the three highest halo mass bins uniformly increase.  The lowest
mass bin ($10^{11}-10^{12}\hMsun$) behaves differently because $\Mmin$ lies
within it, causing some of the halos in this bin to lose pairs and others to
gain pairs, resulting in no net change to the contribution from that bin.
Doubling $\Mmin$ slightly increases the large scale bias factor, but the
impact on $\xih$ is much greater, so the overall effect is to steepen $\xig$.

Figure~\ref{fig:7}c shows the effect of increasing $\alpha$ while keeping all
other HOD parameters fixed.  The contribution of the high mass halos increases
dramatically, whereas the contribution of the lowest mass halos drops.  We
can thus understand why increasing $\alpha$ has a ``smoother'' effect on
$\xig$ than increasing $\Mmin$, producing less distortion in the overall
shape.  First, increasing $\alpha$ has a larger impact on the large scale bias
factor, so the 1-halo and 2-halo terms rise more nearly in step.  Second,
increasing $\alpha$ flattens the shape of $\xih$ as it increases its
amplitude, by redistributing pairs to halos with larger virial radii.  The
shape of $\xih$ therefore stays closer to the extrapolated shape of $\xig$
from large scales.

Figure~\ref{fig:7}d shows the effect of changing the $\PNNavg$ distribution
from Poisson to Average.  The narrower distribution yields a smaller value of
$\NN_M$ for the same $\NavgM$, especially in low multiplicity halos.  As a
result, the 1-halo contribution to $\xig$ drops dramatically at small scales,
$r\sim 0.1-0.4\hmpc$.  The contribution of halos with $M=10^{11}-10^{12}\hMsun$
disappears completely because $M_1$ exceeds $10^{12}\hMsun$, so with an
Average $\PNNavg$ no halo in this mass range can have more than one galaxy.
The suppression of pairs in low mass halos allows $\xig$ to continue as a
power law down to $r=0.1\hmpc$.

The last two panels of Figure~\ref{fig:7} show the effect of changing the
spatial distribution of galaxies within halos.  The HOD models have the same
$\PNM$ as the model shown in Figure~\ref{fig:7}d, so the number of pairs
in halos of each mass range is the same as before.  However, the radial
separations of these pairs are squeezed towards smaller $r$ for $\Dg=-1$
(Fig.~\ref{fig:7}e) and stretched towards larger $r$ for $\Dg=+1$
(Fig.~\ref{fig:7}f), making the correlation function steeper or shallower at
small scales.  The shapes of these curves are directly related to the
function $F(r/2\Rvir)$, defined in equation~(\ref{eqn:xi6}) and plotted in
Figure~\ref{fig:2}, which depends on the radial profile of galaxies within
halos.

\subsection{Understanding the Observed Correlation Function}

In light of these results, what do we make of the observed power-law form of
$\xig$?  One obvious conclusion is that a power-law $\xig$ is not
a generic prediction of HOD models applied to a $\Lambda$CDM cosmology; most
of the models in Figures~\ref{fig:3}-\ref{fig:6} show clear departures from
a power law.  We can understand this behavior by considering Figure~\ref{fig:7}
and the analytic discussion in \S~3.1.  On large scales, the shape of $\xig$
is the same as the shape of $\xim$, and the amplitude of $\xig$ is determined
by the $\xim$ amplitude and the bias factor $b$ (eq.~[\ref{eqn:xi4}]).  On
small scales, typically $r\lesssim0.5\hmpc$ (Fig.~\ref{fig:7}), the 1-halo
term dominates.  In this regime, the shape and amplitude of $\xig$ are
governed by equation~(\ref{eqn:xi7}), which involves $\NN_M$ rather than
$\NavgM$ and does not involve $\xim$ explicitly at all (though $\xi_m$ and
$dn/dM$ are connected indirectly).  Achieving a power-law $\xig$ requires that
the amplitude of $\xih$ place it on the extension of $b^2\xim $, and it
requires that the distribution of pair counts as a function of halo mass yield
a power law of the same slope in the 1-halo regime.

Suppose we have a model that achieves this somewhat delicate balancing act,
such as the $\alpha=0.5$, $\Mmin=2.8\times10^{11}\hMsun$, Average model.
Changing $\PNNavg$ has no effect at large scales, since $b$ depends only on
$\NavgM$, but it changes the amplitude and shape of $\xih$, destroying
the power-law behavior.  Changing $\Mmin$, with a compensating change in
$M_1$ to keep $\ng$ fixed, has different effects in the 1-halo and 2-halo
regimes.  If $\Mmin \ll M_{*}$, then the impact on $b$ will generally be
modest.  However, raising $\Mmin$ redistributes pairs that were previously
in low mass halos to high mass halos, substantially increasing the 1-halo
contribution of these halos and destroying the previous balance.  Changing
$\alpha$ has a more complicated effect than changing $\Mmin$ because it can
significantly alter $b$ and because it changes both the amplitude and shape of
$\xih$.  However, these three changes generally do not work in a way that
preserves a power law.  Finally, Figure~\ref{fig:6} shows that changes to
the halo profile have relatively little leverage for modifying a non-power-law
$\xig$ into a power law.  Even the rather drastic profile changes represented
by $\Dg=\pm 1$ only alter $\xig$ on fairly small scales.  A central galaxy
restriction affects still smaller scales, since central galaxy pairs only
matter in low-$N$ halos with small virial radii.  Central galaxies probably
have an important influence on $\xig$ at $r<100\hkpc$, but they are not
fundamental to understanding the $0.1-0.5\hmpc$ regime.  More generally,
adopting the ``right'' halo profile can never compensate for having the
``wrong'' $\PNM$.

\begin{figure}
\centerline{
\epsfxsize=6.0truein
\epsfbox[30 150 585 710]{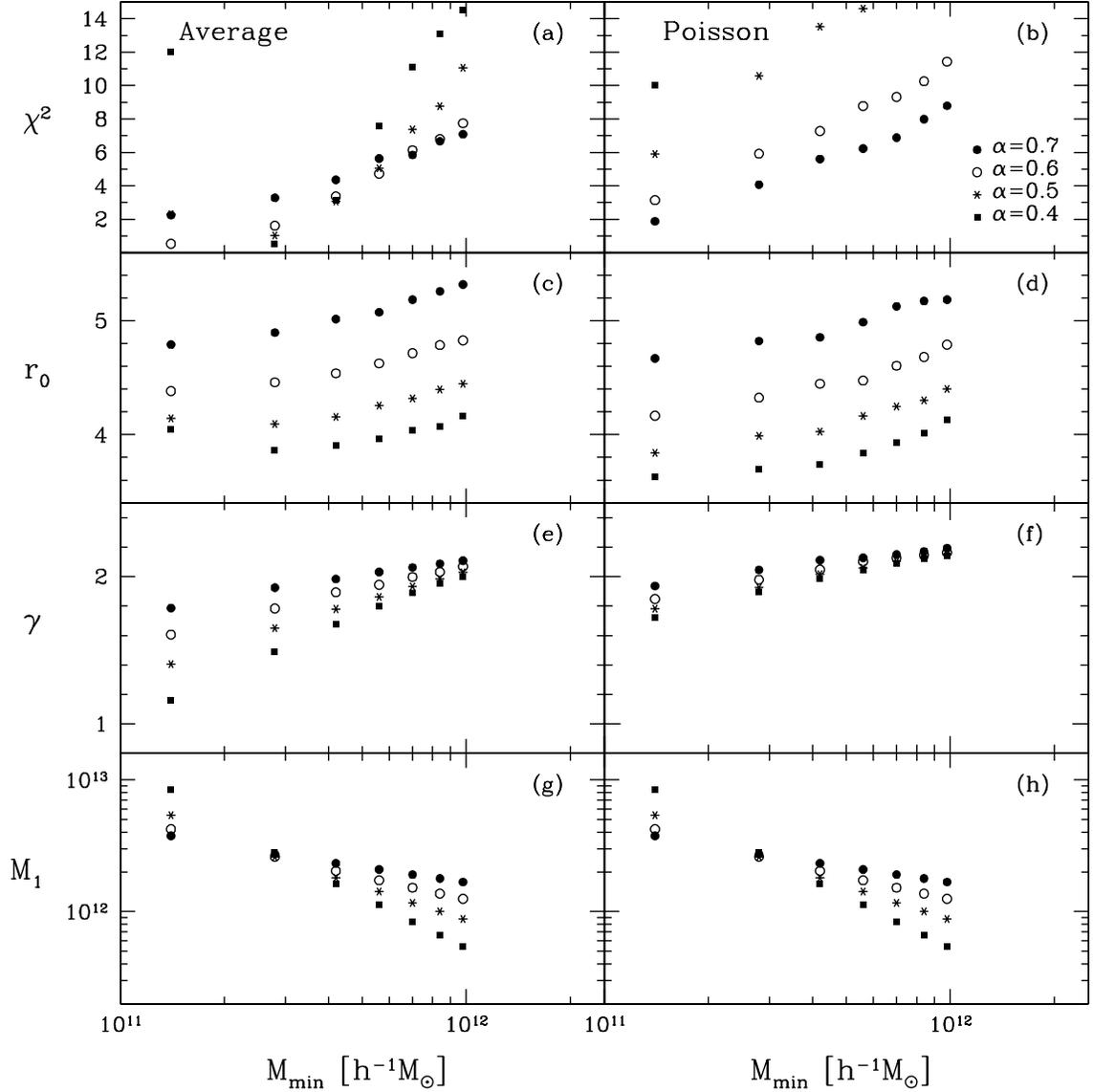}
}
\caption{Power-law fits to a sample of HOD model galaxy correlation 
functions.  Each point corresponds to a specific $\PNM$, where the x-axis 
shows $\Mmin$ and the type of point represents $\alpha$ as shown in 
panel~(b).  The left-hand-side and right-hand-side panels contain models 
with an Average and Poisson $\PNNavg$, respectively.  Panels~(a) and (b) 
show values of $\chi^2$ from these fits.  Since the fits are unweighted, the values of 
$\chi^2$ have been scaled so that models with visually acceptable power-law 
correlation functions have $\chi^2<1$.  Panels~(c) and (d) show the 
best-fit values of the correlation length $r_0$.  Panels~(e) and (f) show 
the best-fit values of the slope $\gamma$.  Panels~(g) and (h) show the 
value of $M_1$ for each model, chosen to yield a galaxy distribution of 
fixed number density $\ng=0.01\hvol$.     
} 
\label{fig:8}
\end{figure}

To roughly quantify the ``difficulty'' of obtaining a power-law $\xig$, we
have carried out a systematic survey of HOD models with power-law $\NavgM$,
Average or Poisson $\PNNavg$, and $M_1$ chosen in all cases to yield
$\ng=0.01\hvol$.  Figure~\ref{fig:8} summarizes the results for a
representative sample of models.  Using an unweighted least-squares fit in
the range $0.1\hmpc<r<10\hmpc$, we determine the best-fit power-law parameters
$r_0$ and $\gamma$ for each model correlation function.  As expected, the slope
$\gamma$ steepens with increasing $\Mmin$ or $\alpha$, since these changes
amplify the 1-halo term relative to the 2-halo term.  The influence of
$\Mmin$ is stronger in Average models than in Poisson models (compare
Figs.~\ref{fig:8}e and \ref{fig:8}f) because $\NN$ cuts off more sharply
near $\Mmin$ in an Average model.  The correlation length $r_0$ generally
increases with increasing $\Mmin$ or $\alpha$ (Figs.~\ref{fig:8}c and
\ref{fig:8}d).  However, the dependence on $\Mmin$ is weak when $\Mmin$ is
small, because the bias factor integral (eq.~[\ref{eqn:xi4}]) remains
dominated by low mass halos.  The dependence becomes stronger when $\Mmin$
becomes large enough to significantly increase the fraction of galaxies in
halos with $M>M_*$ (and thus $b_h>1$).  Increasing $\alpha$, by contrast,
always increases the relative numbers of galaxies in more biased halos, and
the dependence of $r_0$ on $\alpha$ is fairly steady.

The top panels of Figure~\ref{fig:8} show the $\chi^2$ of the power-law fit.
Since we used an unweighted fit, we scale the values of $\chi^2$ by a constant
factor so that models with visually acceptable power-law correlation functions
have $\chi^2\leq1$.  (If we were trying to match specific observational data with
known error bars, we would follow a more rigorous methodology, but that is
not our purpose here.)  As one would guess from the sampling of models shown
in Figures~\ref{fig:3}-\ref{fig:6}, only a small region of this HOD parameter
space --- Average models with $\Mmin\lesssim 3\times10^{11}\hmpc$ and
$\alpha\approx 0.4-0.6$ --- yields power-law correlation functions.  In more
general terms, achieving a power-law $\xig$ requires that a large fraction
of galaxies reside in low mass (and low multiplicity) halos; otherwise the
1-halo portion of $\xig$ is too high relative to the 2-halo portion.
Average models are more successful than Poisson models because they suppress
$\NN$ in low mass halos and thus prevent a steepening of $\xig$ at small $r$.
Values of $\Mmin$ smaller than our resolution threshold
($1.4\times10^{11}\hMsun$) might allow a Poisson model to work, but we are
modeling the population of galaxies with $L>0.5L_*$, and a lower value of
$\Mmin$ would force some of these galaxies into very low mass halos,
contradicting constraints from the \citet{tully77} relation.  In addition,
the bottom panels of Figure~\ref{fig:8} show that a small $\Mmin$ implies a
large $M_1$ ($M_1 \geq 3\times10^{12}\hMsun$ for
$\Mmin \leq 2.8\times10^{11}\hMsun$).  Since halos of mass $\sim 2M_1$ still
have a significant probability of hosting a single galaxy, a high $M_1$ may
force some low luminosity galaxies into fairly massive halos, again running
afoul of Tully-Fisher constraints.

\begin{figure}
\centerline{
\epsfxsize=6.0truein
\epsfbox[60 190 530 660]{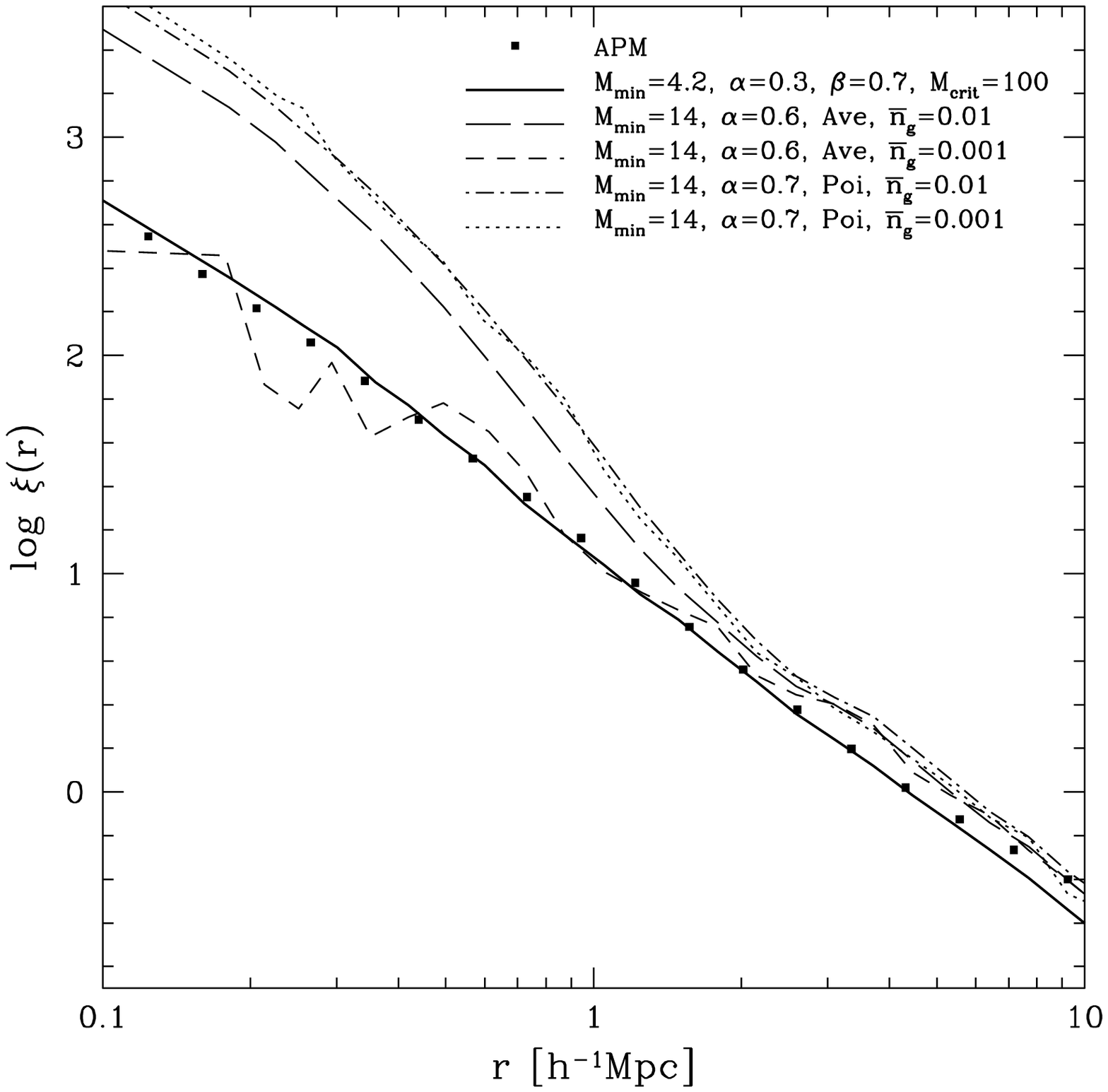}
}
\caption{Real-space correlation functions of galaxy distributions for
different HOD prescriptions.  As indicated in the legend, there is one 
HOD model with a broken power-law $\NavgM$, and four power-law $\NavgM$
models with two values of the galaxy space density $\ng$ (see text for further
discussion).  Values of $\Mmin$ and $\Mcrit$ are given in units of 
$10^{11}\hMsun$, and values of $\ng$ are in units of $\hvol$.  
} 
\label{fig:9}
\end{figure}

This discussion suggests that the power-law $\NavgM$ parameter space may
itself be too restrictive, since it ties the fraction of galaxies in low mass
halos directly to the relative distribution of galaxies among high mass halos
(which affects the large scale bias and the larger separation end of the
1-halo regime).  We have also examined the family of broken power-law models
defined by equation~(\ref{eqn:models2}), which (for $\alpha<\beta$) allow
more galaxies in low mass halos for a given slope in the high mass regime.
With a break point at $\Mcrit=10^{13}\hmpc$, we find acceptable correlation
functions for some Average models with $\Mmin\sim3-5\times10^{11}\hMsun$,
$\alpha\sim0.2-0.3$, $\beta\sim0.6-0.8$, and for some Poisson models with
$\Mmin\sim1-2\times10^{11}\hMsun$, $\alpha\sim0.4-0.5$, $\beta\sim0.8-0.9$.
The solid line in Figure~\ref{fig:9} shows the correlation function of one of
these models, which matches the APM results quite well.  Even with broken
power-law $\NavgM$, acceptable Poisson models require low values of $\Mmin$.
To the extent that such low values are implausible, we concur with arguments in
previous papers (\citealt{benson00}; \citealt{seljak00}; \citealt{peacock00};
\citealt{scoccimarro00}) that sub-Poisson number fluctuations in low
multiplicity halos are  important in explaining the observed form of $\xig$.

At scales $r\lesssim5\hmpc$, peculiar velocities severely distort the shape
of the correlation function in redshift space.  Observational evidence for a
power-law $\xig$ therefore rests on measurements of the angular correlation
function $w(\theta)$ or the projected redshift-space correlation function
$w_p(r_p)$ (see, e.g., \citealt{baugh96}; \citealt{norberg01};
\citealt{zehavi01}).  While a power-law $\xig$ projects into a power-law
$w(\theta)$ or $w_p(r_p)$ (\citealt{limber54}; \citealt{davis83}), one might
worry that projection could wash out departures from a power-law $\xig$.  We
have checked that this is not the case by computing $w_p(r_p)$ for many of
our HOD models.  We find that any models that predict visually evident
departures from a power-law $\xig$ also predict visually evident departures
from a power-law $w_p(r_p)$.

At higher luminosities, the observed $\xig$ has a higher amplitude and a
similar power-law slope.  The higher amplitude is not surprising, since
$\Mmin$ should be higher for more luminous galaxies.  Retaining the power-law
form of $\xig$, however, requires that other HOD parameters adjust in the
right way.  A higher luminosity threshold also implies a lower galaxy space
density $\ng$, so in general we expect both $\Mmin$ and $M_1$ to be higher
for more luminous samples.  The influence of $M_1$ on $\xig$ depends
crucially on the form of the probability distribution $\PNNavg$.  For a
Poisson model, raising $M_1$ is equivalent to randomly sampling the galaxy
population, which does not alter the correlation function.  For an Average
model, on the other hand, raising $M_1$ reduces $\NN$ by a factor larger
than $\ng^2$, thus reducing the 1-halo contributions to $\xig$.
We demonstrate this point with an explicit example in Figure~\ref{fig:9}.  The
long-dashed and dot-dashed lines show $\xig$ for two models, one Average and
one Poisson, with space densities $\ng=0.01\hvol$.  The two correlation
functions show similar departures from a power-law at $r\lesssim2\hmpc$.
Raising $M_1$ to reduce the number density to $\ng=0.001\hvol$ does not
change the correlation function of the Poisson model (dotted line).  However,
the correlation function of the Average model drops dramatically on small
scales and stays the same on large scales (short-dashed-line), yielding a
(noisy) power-law $\xig$.  To raise the amplitude of $\xig$ on large scales,
one would want to increase $\Mmin$ and/or $\alpha$ as well as $M_1$, changes
that tend to boost the 1-halo term relative to the 2-halo term.  The generic
difficulty of finding HOD models that yield a high bias factor and a power-law
$\xig$ (see Fig.~\ref{fig:8}) suggests that suppression of 1-halo clustering by
sub-Poisson fluctuations in low multiplicity halos is especially important for
explaining the correlation function of luminous galaxies.

The quantitative results in this Section (and the rest of the paper) apply to
the specific $\Lambda$CDM cosmological model used for the N-body simulation
described in \S~2.  The galaxy correlation function depends on the HOD and
on the mass and spatial distributions of dark matter halos themselves.  In
terms of the analytic discussion in \S~3.1, the cosmological model influences
$\xig$ by determining $dn/dM$, $b_h(M)$, and $\xim $ (see eqs.~\ref{eqn:xi4}
and \ref{eqn:xi7}).  However, we expect that our conclusions about the
influence of HOD parameters on $\xig$ and the generic requirements for
obtaining a power-law correlation function would continue to hold for a
fairly wide range of cosmological models.

In HOD models, a power-law correlation function emerges from a balance of
several competing effects and therefore requires rather specific combinations
of parameters.  This complex explanation of a simple result may at first seem
unreasonably contrived, but we see no physically realistic
alternative.  A model with $\NavgM \propto M$ above a mass threshold $\Mmin$
predicts a strongly scale-dependent bias in the non-linear regime if $\Mmin$
is large enough to exclude a significant fraction of the mass.  Thus, even if
the non-linear matter correlation function were a pure power law, that
simplicity would be lost as soon as one demanded that luminous galaxies
inhabit halos above some minimum mass.  Eliminating the mass in low mass
halos will in general change the 1-halo and 2-halo regimes of the correlation
function by different factors, so there is no reason to think that a power-law
dark matter correlation function would produce a power-law galaxy correlation
function if a significant fraction of the mass resided in halos that were too
small to host galaxies above the luminosity threshold, even if the galaxies
traced the mass in larger halos.  Furthermore, obtaining a power-law
$\xim$ requires features in the primordial matter power spectrum that have no
obvious physical motivation \citep{peacock96}.

We conclude instead that the observed power-law form of $\xig$ provides a
strong constraint on the physics of galaxy formation, and that the
success of semi-analytic models (\citealt{kauffmann99}; \citealt{benson00}),
hydrodynamic simulations (\citealt{pearce99}; \citealt{cen00};
\citealt{dave00}; \citealt{yoshikawa01}), and high resolution N-body
simulations \citep{colin99} in reproducing this form is entirely non-trivial.
In a subsequent paper (Berlind et al., in preparation), we will show that
semi-analytic calculations and SPH simulations of the $\Lambda$CDM model
predict $\PNM$ distributions that are similar to those of the broken
power-law Average model shown in Figure~\ref{fig:9}, which does indeed yield a
power-law $\xig$.  Improving observations of galaxy clustering already show
evidence for statistically significant departures from a power-law
correlation function on small scales at low \citep{connolly01} and high
\citep{porciani01} redshift, in addition to the ``shoulder'' in $\xig$ at
$r\approx r_0$ (\citealt{gaztanaga01} and references therein).  Such
departures are theoretically expected at some level, and they may provide
further important clues to the properties of the HOD.

\section{Other Statistics}

\subsection{Galaxy-Mass Cross-Correlation Function}

Advances in wide-field digital imaging have recently begun to provide a
new and direct probe of the relation between galaxies and dark matter,
the galaxy-mass cross-correlation function $\xigm$.  This statistic can be
measured using galaxy-galaxy lensing (\citealt{brainerd96};
\citealt{griffiths96}; \citealt{dellantonio96}; \citealt{hudson98};
\citealt{fischer00}; \citealt{smith01}) or by cross-correlating
cosmic shear maps with galaxy light maps \citep{wilson01}.  Since the
lensing signal depends on the surface density in units of the critical density
for lensing, rather than the cosmic mean density, these techniques measure
a combination of $\Omegam$ and $\xigm$ rather than $\xigm$ alone.  Recent
analyses of galaxy-galaxy lensing in the SDSS \citep{mckay01} constrain
$\xigm$ out to $r \sim 1\hmpc$ and demonstrate its dependence on galaxy
luminosity, color, and environment.

Theoretical calculations of $\xigm$ have been presented in the context of
halo occupation models \citep{seljak00}, semi-analytic galaxy formation models
\citep{guzik01}, and hydrodynamic simulations \citep{white01c}.  Much of our
analytic discussion of $\xig$ in \S~3.1 carries over to $\xigm$, except that
we are now considering the correlation between the galaxies and a second
``population'' of objects that trace the mass.  On large scales, we expect
\begin{equation}
\xigm = b \xim,
\label{eqn:xigm1}
\end{equation}
with the same bias factor given in equation~(\ref{eqn:xi4}).  On small scales,
we expect a ``1-halo'' term analogous to equation~(\ref{eqn:xi6}), but with
halos weighted by $M \NavgM$ instead of $\NN_M$.  An important consequence of
this change is that the shape of $\PNNavg$ should not influence $\xigm$ even
in the 1-halo regime.

\begin{figure}
\centerline{
\epsfxsize=6.0truein
\epsfbox[40 185 550 675]{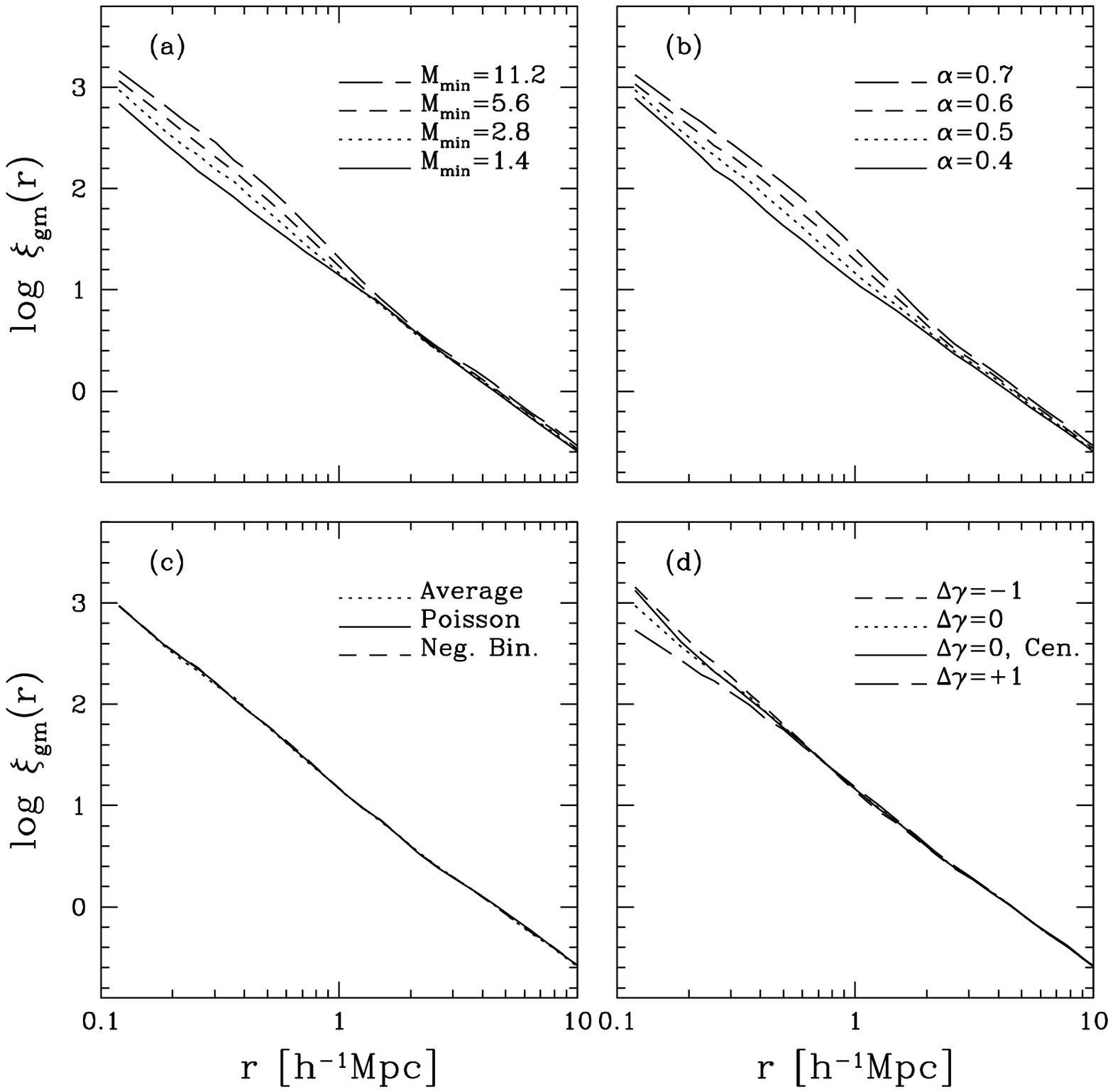}
}
\caption{Galaxy-mass cross-correlation functions for different HOD 
prescriptions.  In each panel, the dotted curve shows $\xigm$ for a
model with power-law $\NavgM$, $\Mmin=2.8\times 10^{11}\hMsun$, $\alpha=0.5$, 
Average $\PNNavg$, and $\Dg=0$.
Other curves show $\xigm$ for models with different $\Mmin$ (panel~a),
different $\alpha$ (panel~b), different $\PNNavg$ (panel~c), or different
galaxy distributions within halos (panel~d).
}
\label{fig:10}
\end{figure}

Figure~\ref{fig:10} shows the galaxy-mass cross-correlation function for a
variety of HOD models, with each panel concentrating on a specific feature of
the HOD.  Figure~\ref{fig:10}a shows the effect of varying $\Mmin$.  On large
scales, models with higher $\Mmin$ have a slightly higher bias factor, in
agreement with our results for $\xig$ in Figure~\ref{fig:3}.  On small scales,
higher $\Mmin$ models have a significantly larger $\xigm$, since a larger
fraction of their galaxies live in high mass halos.  Increasing $\alpha$
raises $\xigm$ for the same reasons, as shown in Figure~\ref{fig:10}b.  The
influence of $\alpha$ extends over a wider range of scales, as it did for
$\xig$ (see the discussion of Fig.~\ref{fig:7}c in \S~3.2).
Figure~\ref{fig:10}c confirms that $\xigm$ does not depend on the shape of
$\PNNavg$, since the 1-halo and 2-halo contributions are both controlled by
$\NavgM$.  For $\Dg=-1$, galaxies are more concentrated in the central, high
density regions of halos, boosting $\xigm$ on small scales
(Fig.~\ref{fig:10}d).  A central galaxy restriction has a similar effect, but
the impact is confined to smaller $r$.  Conversely, setting $\Dg=+1$
suppresses $\xigm$ by pushing galaxies to lower density regions of their
parent halos.

For $r\lesssim 200\hkpc$, our approach of selecting N-body particles to
represent galaxies necessarily underestimates $\xigm$ for a given HOD because
it cannot account for the ability of condensed baryons to concentrate dark
matter around them.  One could do somewhat better by locating galaxies at
density peaks of halo substructure, but this method still would not capture the
effects of baryonic dissipation on the dark matter distribution.  We expect
our approach to be accurate at larger scales (note, for example, that the
effect of a central galaxy restriction is negligible beyond $200\hkpc$) and to
illustrate the correct qualitative dependence on HOD parameters even at small
$r$.

\begin{figure}
\centerline{
\epsfxsize=6.0truein
\epsfbox[25 345 590 560]{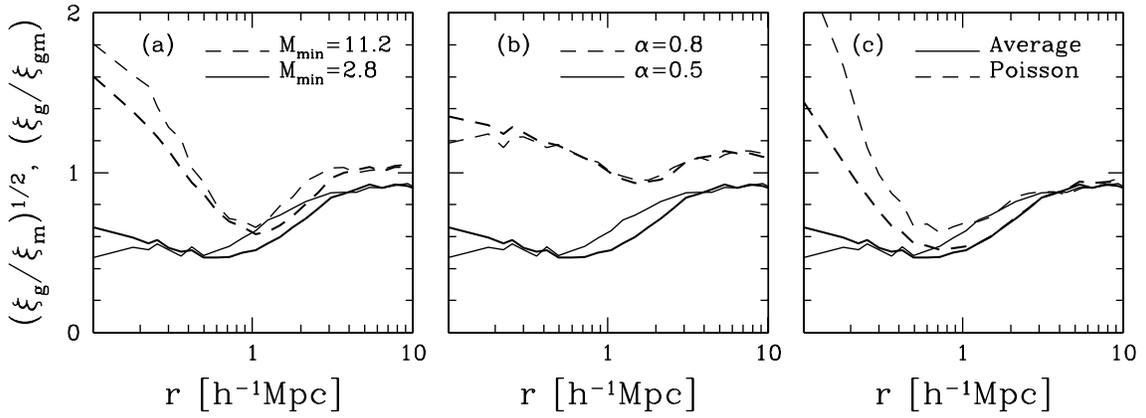}
}
\caption{Effective bias functions defined by ratios of correlation functions
for different HOD prescriptions.  Thick curves represent 
$(\xi_{g}/\xi_{m})^{1/2}$, where $\xi_{g}$ and $\xi_{m}$ are the galaxy 
and mass autocorrelation functions, respectively.  Thin curves represent
$(\xi_{g}/\xi_{gm})^{1/2}$, where $\xi_{gm}$ is the galaxy-mass 
cross-correlation function.  In each panel, solid curves represent an HOD
model with power-law $\NavgM$, $\Mmin=2.8\times 10^{11}\hMsun$, $\alpha=0.5$, 
and Average $\PNNavg$.  Dashed curves represent models with different $\Mmin$
(panel~a), different $\alpha$ (panel~b), or different $\PNNavg$ (panel~c).
} 
\label{fig:11}
\end{figure}

With observations of $\xig$ and $\xigm$, one can define a bias function
$b(r) = \xig/\xigm$.  At least on large scales, we expect this ratio to equal
the bias between the galaxy and matter correlation functions
$[\xig/\xim]^{1/2}$, allowing $\xim$ to be inferred indirectly.
Figure~\ref{fig:11} confirms this expectation for several HOD models.  In each
panel, thick and thin solid curves represent the two bias functions for an HOD
model with $\Mmin=2.8\times10^{11}\hMsun$, $\alpha=0.5$, and an Average $\PNNavg$.
Thick and thin dashed curves show corresponding results for a model with
higher $\Mmin$ (\ref{fig:11}a), higher $\alpha$ (\ref{fig:11}b), or Poisson
$\PNNavg$ (\ref{fig:11}c).  All of the curves go flat at $r\gtrsim 4\hmpc$,
indicating that $\xig$ and $\xigm$ have the same shape as $\xim$.  In this
regime, the two definitions of $b$ are in essentially perfect agreement.
At smaller $r$, the bias functions are scale-dependent and do not track each
other perfectly, though for a given model, they have similar shapes even in
this regime.  From Figure~\ref{fig:11} we see that a similarity of shape
between $\xig$ and $\xigm$ should not be taken as evidence that galaxies
trace mass (Wilson et al. 2001).  However, if measurements of $\xigm$ can be
pushed to $r\gtrsim 4\hmpc$, they should yield an accurate estimate of
$\xig/\xim$, and even on smaller scales they can indicate whether galaxies
are more or less clustered than the matter.  Since the weak lensing signal
depends on $\Omegam$ as well as $\xigm$, these measurements will generally
constrain a combination of $\Omegam$ and $b$ rather than $b$ alone.

\clearpage

If one's goal is to determine the HOD given a known cosmology, then the
principal virtue of $\xigm$ is its sensitivity to $\NavgM$ in a regime
where $\xig$ is sensitive to $\NN_M$.  By combining the two statistics, one
can discriminate between models that have different $\PNNavg$, even if
$\NavgM$ is adjusted so that the models produce similar galaxy correlation
functions.\footnote{The constraint of fixed number density greatly restricts
one's ability to trade off $\NavgM$ against $\PNNavg$, since
$\int dM\frac{dn}{dM}\NavgM$ must remain constant.  However, $\xigm$ provides
more discriminating power because different scales constrain the behavior
of $\NavgM$ in different halo mass ranges.}  However, independent data will
not determine cosmological parameters perfectly, and the more crucial role
of galaxy-galaxy lensing measurements will be to help break the degeneracy
between the HOD and the background cosmology, by providing direct estimates
of the masses of the halos that galaxies inhabit.  We return to this point
in \S~6.

\subsection{Bispectrum}

Higher order clustering statistics complement the information in the
two-point correlation functions $\xig$ and $\xigm$.  Because rare, high
multiplicity halos make a larger contribution to high order statistics, we
expect them to be more sensitive to the occupation of massive halos and to the
high-$N$ tails of $\PNNavg$.  \citet{ma00} and \citet{scoccimarro00} give
fairly comprehensive analytic discussions of the three-point correlation
function and its Fourier transform, the bispectrum, in the halo occupation
framework.  The basic principles are similar to those that govern $\xig$
and $P_g(k)$ --- for example, small scale correlations are dominated by a
1-halo term, in which halos of mass $M$ contribute in proportion to the
mean number of galaxy triples per halo, $\NNN_M$.  However, the mathematics of
three-point correlations is considerably more intricate, in part because
there are three terms (1-, 2-, and 3-halo) contributing to the galaxy
correlations, and in part because the bias of the halos is more complicated.
On large scales, where second-order perturbation theory is valid, the relation
between the halo and mass three-point correlations depends on two bias
factors, one describing the ratio of rms halo and mass fluctuations and one
describing the non-linearity of the relation between halo number and mass
density contrast.

To facilitate comparison with other work, we focus here on the reduced
bispectrum,
\begin{equation}
Q(\k_1,\k_2,\k_3) = \frac{B(\k_1,\k_2,\k_3)}
            {P(k_1)P(k_2)+P(k_2)P(k_3)+P(k_3)P(k_1)}, \qquad k_i = |\k_i|,
\label{eqn:bk1}
\end{equation}
where $B(\k_1,\k_2,\k_3)$ is the bispectrum itself, and $P(k)$ is the power
spectrum.  The three wave-vectors form a triangle $\k_1+\k_2+\k_3=0$, so they
can also be specified by the magnitude $k_1$, the ratio $k_1/k_2$, and the
angle $\theta$ between $\k_1$ and $\k_2$.  In second-order perturbation
theory, the reduced matter bispectrum $Q_m$ is independent of the scale $k_1$,
but it depends on the triangle shape, with higher $Q_m$ at the elongated
configurations ($\theta\approx 0^{\circ}$ and $\theta \approx 180^{\circ}$)
favored by anisotropic gravitational collapse \citep{fry84}.  In bias models
where the galaxy density contrast is a local function of the mass density
contrast, $\delta_g=f(\delta_m)$, a second-order calculation yields
\begin{equation}
Q_g = \frac{Q_m}{b_1} + \frac{b_2}{b_1^2},
\label{eqn:bk2}
\end{equation}
where $b_1=f'(0)$ and $b_2=\frac{1}{2}f''(0)$ (\citealt{fry94}; see also
\citealt{fry93}; \citealt{juszkiewicz95}).  On smaller scales, the value
of $Q_m$ rises and then plateaus at a higher level, and the dependence of
$Q_m$ on triangle shape decreases (\citealt{ma00}; \citealt{scoccimarro00}).

\begin{figure}
\centerline{
\epsfxsize=6.0truein
\epsfbox[40 185 550 675]{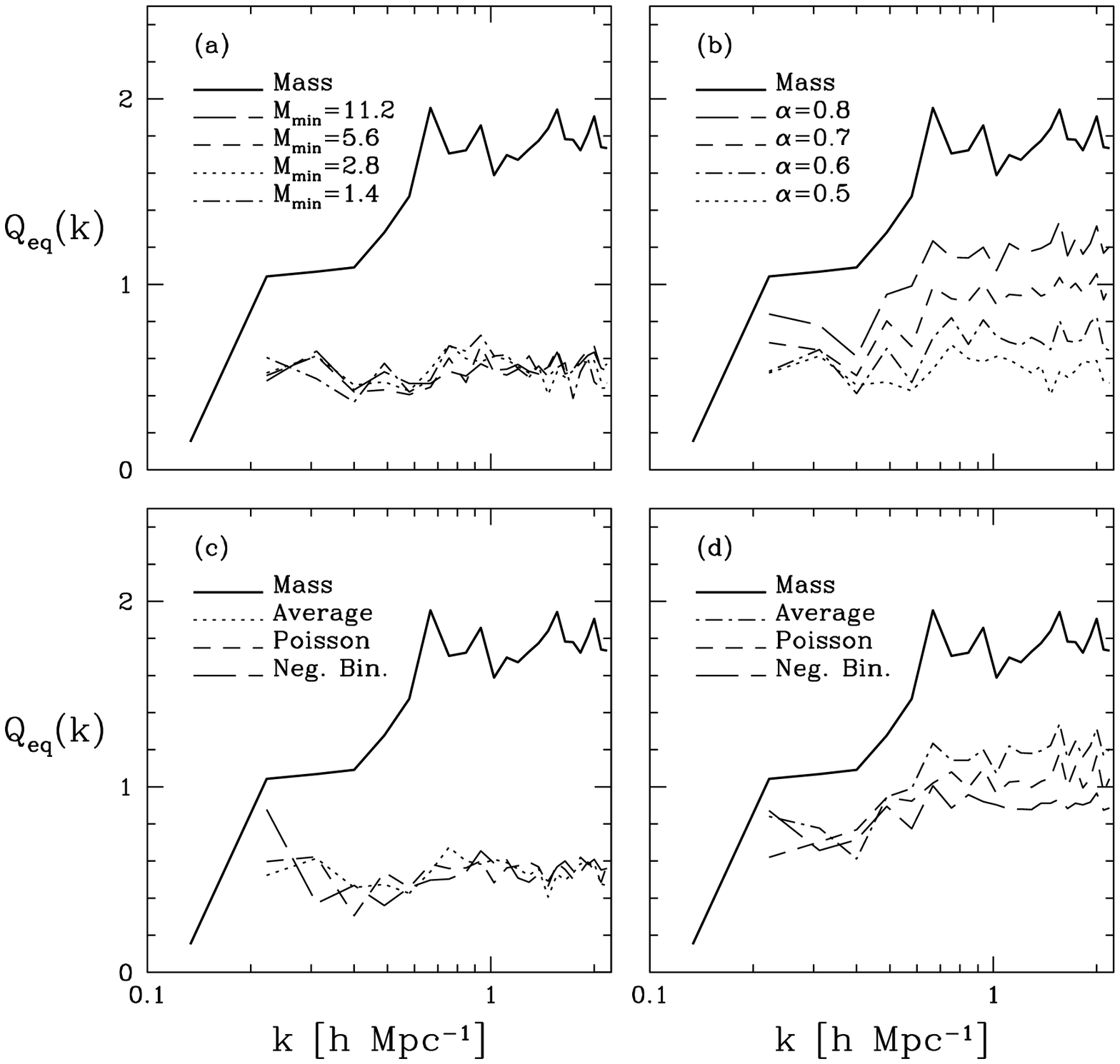}
}
\caption{Reduced bispectrum $\Qeq$ for equilateral configurations
as a function of scale, for different HOD prescriptions.  In panels~(a)-(c), 
the dotted curve shows $\Qeq$ for a model with power-law $\NavgM$, 
$\Mmin=2.8\times 10^{11}\hMsun$, $\alpha=0.5$, and Average $\PNNavg$.
Other curves show $\Qeq$ for models with different $\Mmin$ (panel~a),
different $\alpha$ (panel~b), or different $\PNNavg$ (panel~c).  Panel~(d)
sgows models with different $\PNNavg$ for $\alpha=0.8$.  Thick solid
curves show $\Qeq$ for the unbiased mass distribution.
}
\label{fig:12}
\end{figure}

\begin{figure}
\centerline{
\epsfxsize=6.0truein
\epsfbox[40 185 550 675]{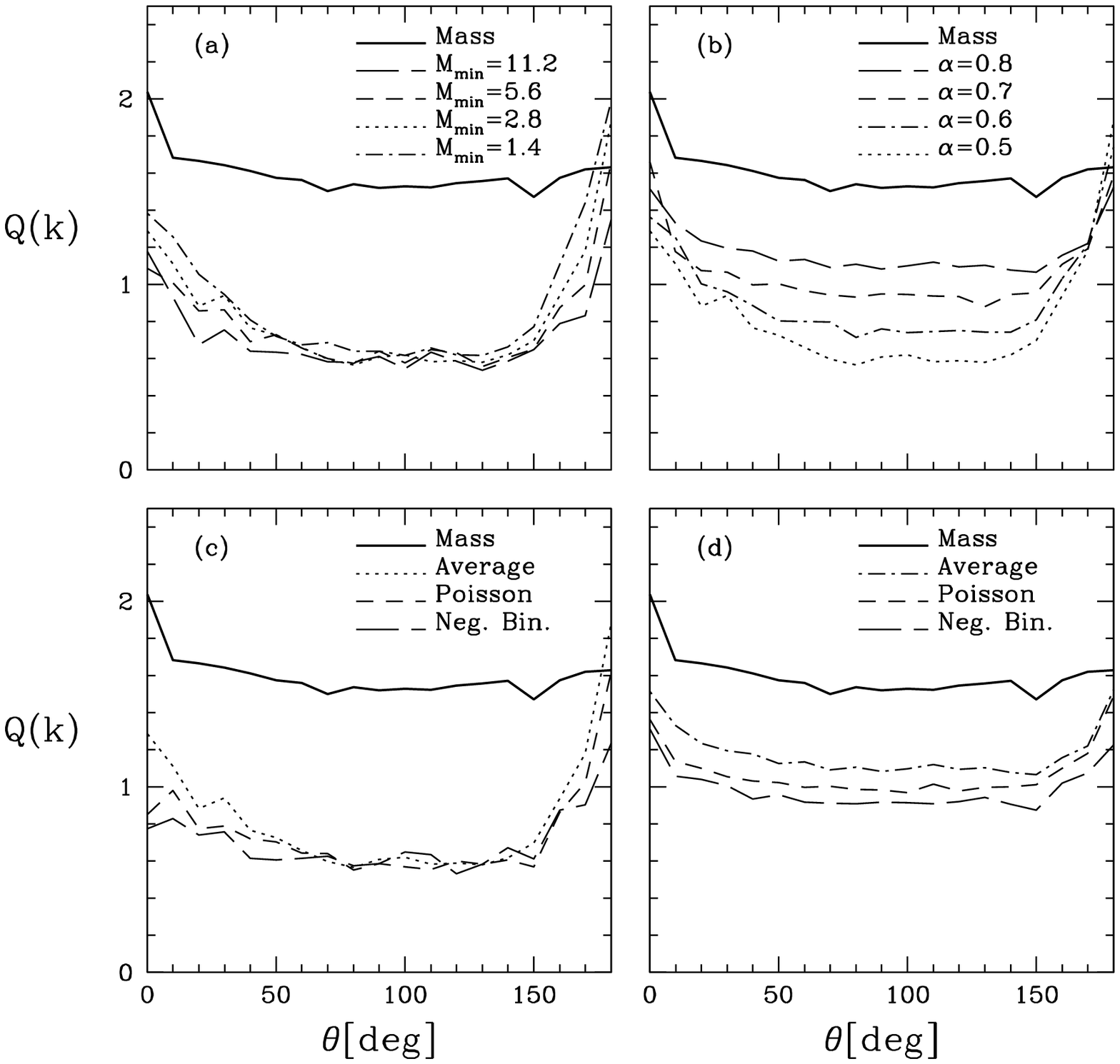}
}
\caption{The reduced bispectrum $Q(\theta)$ as a function of triangle shape,
where $k_1=2k_2=1h\mathrm{Mpc}^{-1}$ ($k=2\pi/\lambda$).  The four panels 
show the same HOD models as in Fig.~\ref{fig:12}.
} 
\label{fig:13}
\end{figure}

To compute the reduced bispectrum, we form continuous density fields by
cloud-in-cell (CIC) binning the discrete mass or galaxy distributions onto
a $200^3$ grid and calculate $P(k)$ and $B(\k_1,\k_2,\k_3)$ using a Fast
Fourier Transform (FFT).  Figure~\ref{fig:12} shows $\Qeq$, the reduced
bispectrum as a function of scale in the equilateral triangle case
($k_1=k_2=k_3$), for a variety of HOD models and for the unbiased mass
distribution.  For the same set of HOD models, Figure~\ref{fig:13} shows
$Q(\theta)$, the reduced bispectrum as a function of triangle shape, in the
case $k_1=2k_2=1h\mathrm{Mpc}^{-1}$.  The fundamental mode of the $141.3\hmpc$
simulation cube is $k_f=0.0445 h\mathrm{Mpc}^{-1}$.  In Figure~\ref{fig:12},
$\Qeq$ curves for the mass distribution begin at $k_1=3k_f$, and in
Figure~\ref{fig:13} $k_1=2k_2=22.5k_f$.

Figures~\ref{fig:12}a and \ref{fig:13}a show models with $\alpha=0.5$,
Average $\PNNavg$, and varying values of $\Mmin$.  Because of the limited
size of the simulation volume, the mass $\Qeq$ does not show the low-$k$
plateau predicted by perturbation theory very clearly, though it does show
the usual rise into the non-linear regime and plateau at high-$k$.
The scale $k_1$ adopted in Figure~\ref{fig:13} is
sufficiently non-linear that most of the shape dependence in
$Q_m(\theta)$ has disappeared, though there is still a slight
enhancement for elongated triangles.  The galaxies have much lower values of
$\Qeq$.  In the context of equation~(\ref{eqn:bk2}), we interpret the
depression of $\Qeq$ as a sign of negative $b_2$, since the galaxies are
{\it anti}-biased in an rms sense (compare the galaxy and mass
correlation functions in Figure~\ref{fig:3}), and this anti-bias on its own
would tend to boost $Q$ ($b_1<1$). \footnote{The association of $b_1$ and
$b_2$ with derivatives of $f(\delta_m)$ is no longer accurate in the
non-linear regime, but one can still define $b_1=[P_g(k)/P_m(k)]^{1/2}$ and
define $b_2$ through equation~(\ref{eqn:bk2}), though $b_2$ may now be
dependent on triangle shape.}  Turning to non-equilateral triangles in
Figure~\ref{fig:13}a, we see that $Q_g$ is strongly depressed relative to
$Q_m$ for $30^{\circ} \lesssim \theta \lesssim 140^{\circ}$ but rises towards
$Q_m$ near $\theta=0^{\circ}$ and, especially, near $\theta=180^{\circ}$.  As a
result, the shape dependence of $Q_g$ is stronger than that of $Q_m$.  We are
unsure how to explain the shape dependence of the suppression of $Q_g$, though
a greater impact of bias on roughly equilateral configurations than on elongated
configurations is not surprising.

In Figure~\ref{fig:12}a, we see that changing $\Mmin$ has no significant
impact on $\Qeq$ in these models.  While $\Beq$ and $P(k)$ both increase for
larger $\Mmin$, the two effects cancel in the ratio that defines $Q$.
Figure~\ref{fig:13}a shows a modest dependence of $Q(\theta)$ on $\Mmin$
for $\theta \gtrsim 140^{\circ}$, perhaps because the signature of elongated
structures is stronger when galaxies are distributed among lower mass halos
(smaller $\Mmin$).  Figure~\ref{fig:12}b shows the dependence of $\Qeq$ on
$\alpha$; while $\Beq$ and $P(k)$ both increase for larger $\alpha$, $\Beq$
grows faster, and $\Qeq$ rises closer to the value of $Q_m$.  Because low
$\alpha$ suppresses $Q(\theta)$ primarily at intermediate values of $\theta$,
the triangle-shape dependence weakens as $\alpha$ increases
(Fig.~\ref{fig:13}b).


Figures~\ref{fig:12}c and \ref{fig:12}d show the dependence of $\Qeq$ on the
form of $\PNNavg$ for $\Mmin=2.8\times10^{11}\hMsun$ and $\alpha=0.5$ and
$0.8$, respectively.  The dependence is minimal in the first case but
significant in the second, with the Average distribution producing the
highest $\Qeq$ at high $k$.  While the narrower width of the Average
distribution leads to a smaller $\Beq$, it causes a still larger reduction in
the denominator of equation~(\ref{eqn:bk1}), thus boosting $Q$.  The greater
dependence on $\PNNavg$ for higher $\alpha$ reflects the greater importance
of the 1-halo and 2-halo terms when massive halos are highly occupied, since
the 3-halo term is independent of $\PNNavg$ but the other two are not.
Figures~\ref{fig:13}c and \ref{fig:13}d show a slight dependence of
$Q(\theta)$ on $\PNNavg$ for $\alpha=0.5$ and a depression of $Q(\theta)$
at all angles for the broader distributions when $\alpha=0.8$.

We have found the influence of HOD parameters on the reduced bispectrum
relatively difficult to interpret cleanly, for several reasons.  First,
HOD changes influence both the individual terms (1-, 2-, and 3-halo)
contributing to the bispectrum and the relative importance of those terms,
and it is not always clear which effect is more significant.  Second, the use
of a Fourier space measure makes it more difficult to separate terms based on
scale --- the 1-halo contribution to the three-point function, for example,
must vanish beyond the virial diameter of the largest halos, but it can still
have an influence on the bispectrum down to fairly low $k$.  Finally, while the
influence of an HOD change on the bispectrum or power spectrum may be easy to
guess, the effect on the ratio $Q$ is often much harder to predict.  The
reduced bispectrum is a natural statistic for describing mass clustering,
especially in the perturbative regime.  However, the unreduced, configuration
space three-point function may prove more useful in unraveling the HOD, at
least on scales of individual halos.

We have also computed the hierarchical amplitudes $S_3$ and $S_4$ (related
to the skewness and kurtosis, respectively), as a function of smoothing scale.
These results are qualitatively similar to those for the reduced bispectrum
shown in Figure~\ref{fig:12}, so we do not show them here.

Despite the complexities, three-point correlations clearly have information
content that is not present in $\xig$, because of their dependence on higher
moments of $\PNNavg$ and their greater weighting of massive halos.  The
dependence of $Q$ on $\Mmin$, $\alpha$, and $\PNNavg$ is significantly
different from the dependencies found for $\xig$ in \S~3.  We therefore expect
three-point correlations to play an important role in constraining the HOD and
breaking degeneracies between bias and cosmology.
The suppression of $Q_g$ relative to $Q_m$ in HOD models with $\alpha < 1$
(seen for all of the cases in Figure~\ref{fig:12}) may help to explain the
low amplitude that \cite{jing98c} find for the configuration-space three-point
function of LCRS galaxies, which is reduced relative to the N-body
prediction for the mass distribution by about a factor of two.

\subsection{Void Probability Function}

We now turn to a statistic which focuses on the lowest density regions, the
voids in the galaxy distribution.  The statistic we examine is the void
probability function (VPF), and it is defined as the probability $P_0(R)$
that a randomly placed sphere of radius $R$ contains no galaxies.  The best
observational study of the VPF to date is the analysis of the CfA2 redshift
survey by \citet{vogeley94}, though 2dFGRS and SDSS should yield results in the
near future.  \citet{little94} showed that the VPF is sensitive to the details
of biasing, using environmental bias models.  In the HOD context, the VPF
should depend on the expected number of halos contained within a given sphere,
as well as the expected number of galaxies per halo.  The VPF does not
distinguish between halos that contain one galaxy and those that contain
multiple galaxies, since one galaxy alone prevents a region from being empty.
Therefore, the VPF should be very sensitive to the halo mass regime
$\Mmin \leq M \leq M_1$, where most halos contain either one galaxy or no
galaxies.  \citet{benson01} has analytically studied count
probabilities in the HOD framework, and \citet{casas01} have studied halo
counts in voids.  These are steps towards an analytic description of the galaxy
VPF under HOD bias.  Here we provide an approximate model that is useful for
understanding our numerical results.

The number of galaxies expected in a volume $V$ that has a density contrast
$\delta$ is
\begin{equation}
\left<N|V,\delta\right> = V\int_0^{\infty}dM
                          \left.\frac{dn}{dM}\right|_{\delta}\NavgM,
\label{eqn:vpf1}
\end{equation}
where the halo mass function is conditional on $\delta$ because halos are, on
average, more abundant and more massive in high density regions than in low
density regions.  The amplitude of $dn/dM$ is enhanced by the same factor
$(1+\delta)$ that affects the mass density.  However, the shape of the
mass function also depends on environment: in low density regions,
gravitational growth of structure is suppressed, and the halo mass function
shifts towards lower masses.  This dependence of the halo mass function on
density is the basis for the \citet{mo96} calculation of halo bias.  For our
present purposes, it is useful to write equation~(\ref{eqn:vpf1}) as
\begin{equation}
\left<N|V,\delta\right> = V(1+\delta)\int_0^{\infty}dM
                          \left.\frac{dn}{dM}\right|_{\delta_L}\NavgM,
\label{eqn:vpf2}
\end{equation}
where $\delta_L$ denotes the Lagrangian density contrast associated with the
evolved non-linear density contrast $\delta$.  In equation~(\ref{eqn:vpf2}),
the dispersal of halos by the expansion of underdense regions is represented by
$(1+\delta)$, while the shift of the mass function is represented by
$\left.\frac{dn}{dM}\right|_{\delta_L}$.

For a Poisson distribution of mean $\N$, the probability of having $N=0$ is
$P(0)=\mathrm{exp}(-\N)$.  The actual probability distribution of galaxy
counts depends both on $\PNNavg$ and on the probability distribution of halo
counts; however, approximating the combined effect as Poisson provides a
useful guide for understanding the VPF.  In the Poisson approximation, we
can write an expression for the VPF by integrating over $\delta$,
\begin{equation}
P(N=0|V) = \int_{-\infty}^{\infty} d\delta P(\delta|V)(1+\delta){\mathrm exp}
  \left[-V\int_0^{\infty}dM\left.\frac{dn}{dM}\right|_{\delta_L}\NavgM\right],
\label{eqn:vpf3}
\end{equation}
where $P(\delta|V)$ is the probability that a sphere of volume $V$ has a mass
density contrast $\delta$.  If we ignored the dependence of $dn/dM$ on
$\delta_L$, the exponential factor would simply be ${\mathrm exp}(-V\ng)$,
and the VPF would thus not depend on the HOD.  For $V\ng \lesssim 1$, the
exponential factor will not be too far from unity, so the VPF will depend
mainly on $P(\delta)$.  However, at large $R$, where $V\ng \gg 1$, the
exponential suppression is usually large, and changes to $\NavgM$ can
potentially have very large effects on the VPF.

\begin{figure}
\centerline{
\epsfxsize=6.0truein
\epsfbox[40 185 550 675]{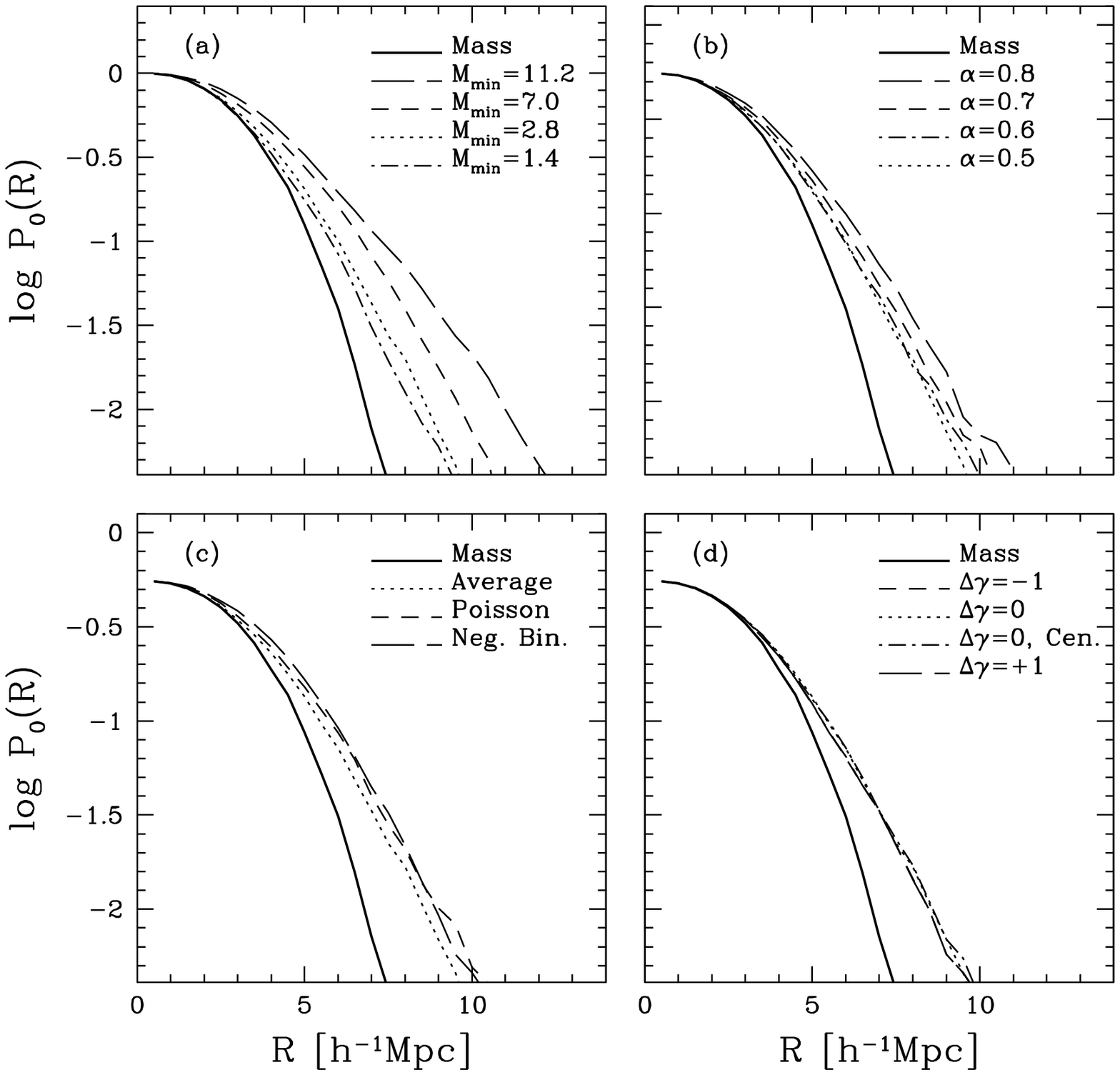}
}
\caption{Void probability functions (VPF) for different HOD 
prescriptions.  In each panel, the dotted curve shows the VPF for a
model with power-law $\NavgM$, $\Mmin=2.8\times 10^{11}\hMsun$, $\alpha=0.5$, 
Average $\PNNavg$, and $\Dg=0$.
Other curves show VPFs for models with different $\Mmin$ (panel~a),
different $\alpha$ (panel~b), different $\PNNavg$ (panel~c), or different
galaxy distributions within halos (panel~d).  Thick solid curves show the VPF
for the unbiased mass distribution.
} 
\label{fig:14}
\end{figure}

Figure~\ref{fig:14} shows VPFs for a variety of HOD models, as well as for the
unbiased mass distribution.  Since the VPF is very sensitive to the number
density of particles in the distribution being studied, we have randomly
sampled the mass distribution to have the same space density of particles
as the galaxy distributions.  All HOD models shown have a power-law
$\NavgM$.  Figure~\ref{fig:14}a shows that changing $\Mmin$ from $1.4$ to
$2.8$ (in units of $10^{11}\hMsun$) has little effect but that raising it
to $7$ and again to $11.2$ has a dramatic impact, especially at large $R$.
This behavior has a natural explanation in the context of
equation~(\ref{eqn:vpf3}).  When $\Mmin$ is low, the integral inside the
exponential factor is never far from $\ng$.  However, when $\Mmin$ is large,
the break in the halo mass function can shift below $\Mmin$ in the lowest
density regions, decreasing the integral and producing an enormous increase
in the exponential factor.  Figure~\ref{fig:14}b shows the effect on the VPF
of varying $\alpha$.  HOD models with higher values of $\alpha$ have somewhat
higher probability of containing voids, since they have lower $\Navg$ at
masses just above $\Mmin$, but the dependence of the VPF on $\alpha$ is
clearly weaker than the dependence on $\Mmin$.  Figure~\ref{fig:14}c shows
the effect of varying $\PNNavg$.  Models with wider $\PNNavg$ distributions
have a slightly higher probability of containing voids because they
sometimes produce empty halos in the halo mass regime where $\Navg\sim1-2$.
Finally, Figure~\ref{fig:14}d shows that the VPF is completely insensitive
to the distribution of galaxies within halos.

The VPF complements correlation statistics by responding sensitively to the
low mass end of $\PNM$ and especially to the cutoff mass $\Mmin$.  The VPF is
a special case of count probabilities or of the closely related probability
distribution function (PDF) of smoothed density fields.  We have also measured
these full PDFs, and they respond roughly as one would expect: models with
more ``bias'' produce wider PDFs.  However, we have been unable to draw simple
associations between specific features of PDFs and specific properties of the
HOD, so we have chosen not to present these results in detail.

\subsection{Pairwise Velocity Dispersion}

The statistics we have examined so far characterize the clustering of
galaxies in real space.  However, there is also important information
stored in galaxy peculiar velocities, which are gravitationally induced by
the underlying mass distribution.  One of the statistics most widely used
to characterize the galaxy velocity distribution is the pairwise radial
velocity dispersion,
\begin{equation}
\sigma^2_v(r) \equiv \left<|(\vv_1-\vv_2)\cdot\rr_{12}|^2\right> -
                     \left<(\vv_1-\vv_2)\cdot\rr_{12}\right>^2,
\label{eqn:vdisp1}
\end{equation}
where $\vv_1$ and $\vv_2$ are the velocities of two galaxies separated by
$\rr_{12}$, and the average is over all galaxy pairs of separation
$r=|\rr_{12}|$.  This quantity can be estimated from the 2-dimensional
correlation function $\xi(r_p,\pi)$ (e.g., \citealt{davis83}), and its
relatively low observed value provided one of the strongest early arguments
against an $\Omegam=1$ cosmology with an unbiased galaxy distribution
\citep{davis85}.  Recent observations give
$\sigma_v=500-600\mathrm{km~s}^{-1}$ at $r\sim 1\hmpc$ for typical optical
samples, with a weak dependence on scale but a strong dependence on galaxy
color or morphological type (\citealt{jing98b}; \citealt{zehavi01}).
\citet{narayanan00} found that
$\sigma^2_v(r)$ is highly sensitive to the details of biasing, and here we
wish to explore this sensitivity in the HOD framework of bias.

On small scales, $\sigma^2_v(r)$ will be dominated by pairs of galaxies
within the same halo.  If we assume that the halo velocity distribution is
isotropic and isothermal, then the mean pairwise radial dispersion of galaxy
pairs in a halo of mass $M$ is $2\alpha_v^2\sigma^2(M)$, where $\sigma(M)$ is
the halo's one-dimensional velocity dispersion and $\alpha_v$ is the galaxy
velocity bias parameter defined in \S~2.  The pairwise dispersion of the
galaxy population is the average value of $2\alpha_v^2\sigma^2(M)$ weighted by
the fraction of galaxy pairs of separation $r$ contributed by halos of mass
$M$.  Reference to equation~(\ref{eqn:xi7}) for $\xih$ shows that this average
is
\begin{equation}
\sigma^2_{v,\mathrm{1h}}(r)= \left[2\pi r^2 \ng^2 \xig\right]^{-1}
   \int_{\Mmin}^{\infty}dM\frac{dn}{dM}\frac{\NN_M}{2}\frac{1}{2\Rvir(M)}
   F'\left(\frac{r}{2\Rvir}\right)2\alpha_v^2 \sigma^2(M),
\label{eqn:vdisp2}
\end{equation}
where we have assumed $\xig = \xih \gg 1$.

At large scales, each member of a galaxy pair resides in a separate halo,
and the velocity of each galaxy is the center-of-mass velocity of its halo
plus a random velocity of 1-d dispersion $\alpha_v^2\sigma^2(M)$.  Since
the random velocities are uncorrelated with the center-of-mass velocities
and with each other, the mean pairwise dispersion of galaxy pairs in halos
of mass $M_1$ and $M_2$ at separation $r$ is
$\sigma_{12}^2(r) + \alpha_v^2\sigma^2(M_1) + \alpha_v^2\sigma^2(M_2)$,
where $\sigma_{12}^2(r)$ is the pairwise dispersion of halos.  In the 2-halo
regime, the pairwise dispersion of the galaxy population is the average of
this quantity weighted by the number of pairs in halos of mass $M_1$, $M_2$,
which is proportional to $(1+\xi_{12}(r))\Navg(M_1)\Navg(M_2)$.  Using
equation~(\ref{eqn:xi3}) we have
\begin{eqnarray}
\sigma^2_{v,\mathrm{2h}}(r) & = & \left[\ng^2(1+\xig)\right]^{-1}
                \int_{\Mmin}^{\infty}dM_1\frac{dn}{dM_1}\Navg(M_1)
                \int_{\Mmin}^{\infty}dM_2\frac{dn}{dM_2}\Navg(M_2) \nonumber \\
     & & \times (1+\xi_{12}(r))
 \left[\alpha_v^2\sigma^2(M_1)+\alpha_v^2\sigma^2(M_2)+\sigma_{12}^2(r)\right] \nonumber \\
          & = & \frac{2\alpha_v^2}{1+\xig}
 \left[\left<\sigma^2(M)\right>_N + \xig\left<\sigma^2(M)\right>_{Nb} \right]
	     +  \left<\sigma^2_{12}(r)\right>,
\label{eqn:vdisp3}
\end{eqnarray}
where
\begin{equation}
\left<\sigma^2(M)\right>_N \equiv
              \ng^{-1}\int_{\Mmin}^{\infty}dM\frac{dn}{dM}\NavgM\sigma^2(M)
\label{eqn:vdisp4}
\end{equation}
is the mean halo dispersion weighted by galaxy number,
\begin{equation}
\left<\sigma^2(M)\right>_{Nb} \equiv
              (b\ng)^{-1}\int_{\Mmin}^{\infty}dM\frac{dn}{dM}\NavgM b_h(M)\sigma^2(M)
\label{eqn:vdisp5}
\end{equation}
is the mean halo dispersion weighted by galaxy number and halo bias factor
(and $b$ is the galaxy bias factor of eq.~[\ref{eqn:xi4}]), and
\begin{eqnarray}
\left<\sigma^2_{12}(r)\right> &\equiv & \left[\ng^2(1+\xig)\right]^{-1}
                \int_{\Mmin}^{\infty}dM_1\frac{dn}{dM_1}\Navg(M_1)
                \int_{\Mmin}^{\infty}dM_2\frac{dn}{dM_2}\Navg(M_2) \nonumber \\
     & & \times (1+\xi_{12}(r))\sigma_{12}^2(r)
\label{eqn:vdisp6}
\end{eqnarray}
is the mean pairwise halo dispersion weighted by number of galaxy pairs.
Since $\sigma_{12}(r)$ can in principle depend on both $M_1$ and $M_2$, it
cannot generally be pulled out of the integral in equation~(\ref{eqn:vdisp6}),
though at large scales it may be close to the value
$\sigma^2_{v,\mathrm{lin}}(r)$ predicted for the mass by linear theory, since
halos obtain their relative velocities from large scale density perturbations.

For a more detailed discussion, we refer the reader to \citet{sheth01},
who give a much more comprehensive analytic treatment of the
pairwise dispersion and mean streaming in the halo occupation framework and
address a number of subtle issues (like non-isothermality) that we have
glossed over here.  For our purposes, the crucial features of
equations~(\ref{eqn:vdisp2}) and (\ref{eqn:vdisp3}) are the following.  In the
1-halo regime, $\sigma_v^2$ depends on $\alpha_v^2$ and $\NN_M$, and different
halo masses contribute on different scales just as they do in $\xih$.  In the
2-halo regime, $\sigma_v^2(r)$ has one contribution that is proportional to
$\alpha_v^2$ and another that is independent of $\alpha_v$.  At large
separations, where $\xig$ is small, the first contribution becomes constant,
so the {\it change} in $\sigma_v^2(r)$ is entirely due to
$\left<\sigma^2_{12}(r)\right>$.

\begin{figure}
\centerline{
\epsfxsize=6.0truein
\epsfbox[40 185 550 675]{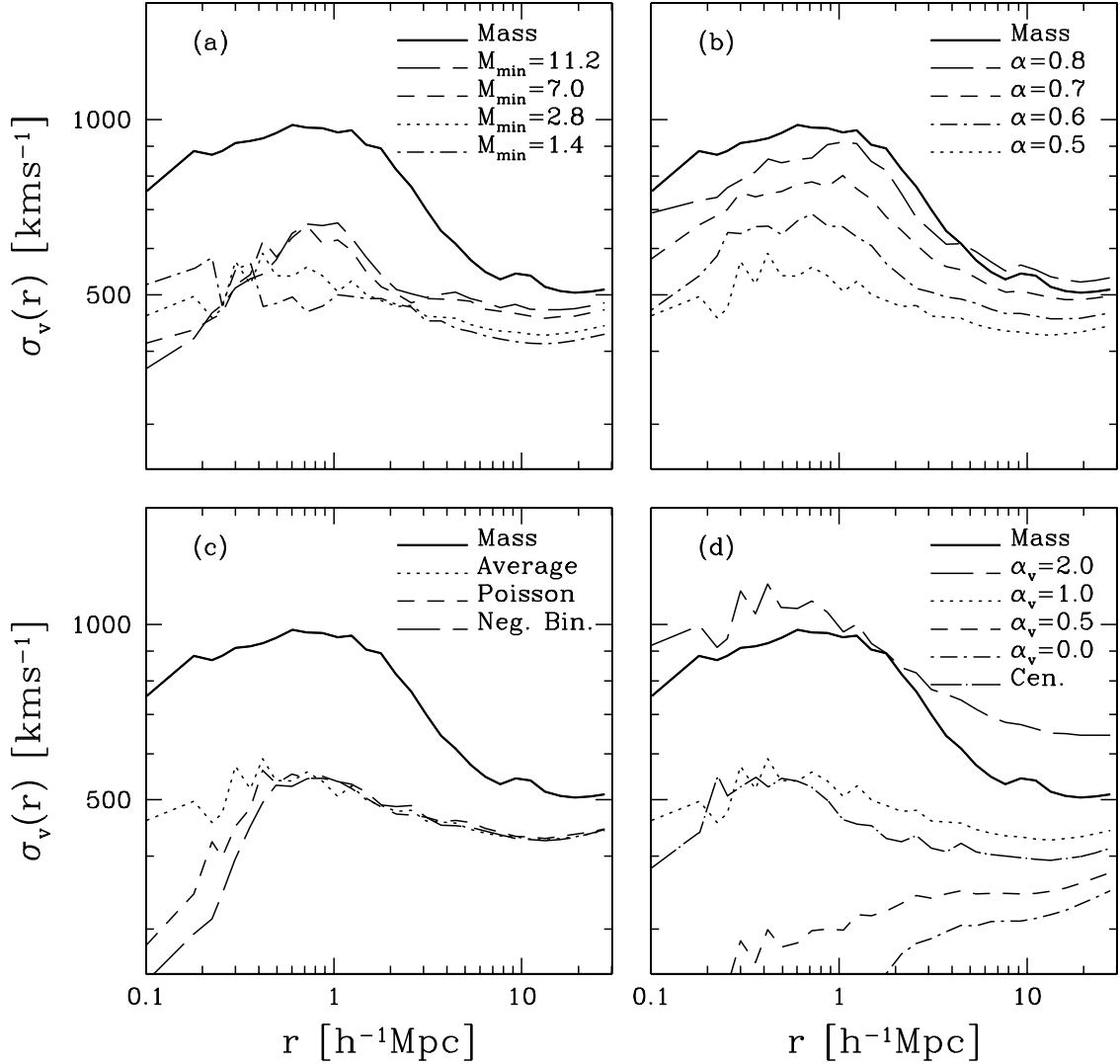}
}
\caption{Mean pairwise radial velocity dispersions $\sigma_v$ as a function 
of the pair separation $r$ for different HOD prescriptions.  In each panel, 
the dotted curve shows $\sigma_v$ for a model with power-law $\NavgM$, 
$\Mmin=2.8\times 10^{11}\hMsun$, $\alpha=0.5$, Average $\PNNavg$, and 
$\alpha_v=1$.
Other curves show $\sigma_v$ for models with different $\Mmin$ (panel~a),
different $\alpha$ (panel~b), different $\PNNavg$ (panel~c), or different
velocity distributions within halos (panel~d).  Thick solid curves show $\sigma_v$
for the unbiased mass distribution.  The dot-long-dash curve in panel~(d)
shows a model with $\alpha_v=1$ but the first, central galaxy in each halo
forced to move at the halo's mean velocity.
} 
\label{fig:15}
\end{figure}

Figure~\ref{fig:15} shows the pairwise velocity dispersion for a variety of
HOD models and for the unbiased mass distribution.  Increasing $\Mmin$
(Fig.~\ref{fig:15}a) slightly increases $\sigma_v^2(r)$ because it forces
galaxies into more massive halos with high velocity dispersions.  However,
increasing $\alpha$ (Fig.~\ref{fig:15}b) has a much more dramatic effect
because it redistributes galaxies preferentially towards the most massive
halos.  For all $\alpha$, $\sigma_v^2(r)$ has a broad maximum at $r\sim 1\hmpc$,
where the highest mass halos contribute a significant number of pairs.  Higher
$\alpha$ models have higher $\sigma_v^2(r)$ even at large scales because of
the continuing contribution of $\left<\sigma^2(M)\right>$
(eq.~[\ref{eqn:vdisp3}]).  Figure~\ref{fig:15}c confirms that $\sigma_v^2(r)$
is independent of $\PNNavg$ on large scales, where the 2-halo term dominates.
However, on small scales $\PNNavg$ has an important impact.  While making
$\PNNavg$ broader boosts the correlation function (Fig.~\ref{fig:5}), it
drives $\sigma_v^2(r)$ {\it down} by allowing $\sigma_{v,1\mathrm{h}}^2(r)$
to be dominated by pairs in low mass halos.  For example, the Average model
never has pairs in halos with $\NavgM<1$, but the Poisson model does.
Consequently, the small scale pairs in the Average model necessarily come from
higher mass halos with larger velocity dispersions than in the Poisson model.

Figure~\ref{fig:15}d shows the influence of $\alpha_v$ on $\sigma_v^2(r)$,
which is, as one might expect, very strong.  The dotted line shows the extreme
case of $\alpha_v=0$, where $\sigma_v^2(r)$ comes only from the relative
velocities of halos themselves.  This contribution goes to zero in the 1-halo
regime.  It is usually sub-dominant even on large scales, but it dominates
the {\it change} in $\sigma_v^2(r)$ because the internal dispersion contribution
becomes constant.  Increasing or reducing galaxy random velocities by a factor
of two ($\alpha_v=2$ or 0.5) changes $\sigma_v^2(r)$ by a large amount at all
scales; note, however, that values of $\alpha_v$ so far from unity are
probably implausible on dynamical grounds.  The dot-long-dash line in
Figure~\ref{fig:15} shows the impact of a central galaxy on a model with
$\alpha_v=1$; the pairwise dispersion drops slightly because the central
galaxy moves with the halo mean velocity.

In the absence of velocity bias, $\alpha_v \approx 1$, the pairwise dispersion
is a good diagnostic of $\alpha$, which controls the fraction of galaxies in
high dispersion halos (\cite{jing98b} reach a similar conclusion).
It also constrains $\PNNavg$, with a rather different
signature from $\xig$.  For the $\Lambda$CDM cosmology and HOD models
considered here, matching the observed amplitude and weak scale-dependence of
$\sigma_v^2(r)$ requires $\alpha\approx 0.5$ to suppress the peak at
$r\sim 1\hmpc$ associated with massive halos, and sub-Poisson fluctuations
at low $M$ to prevent low mass halos from dragging down $\sigma_v^2(r)$ at
small $r$.  These are the same features of $\PNM$ required to match $\xig$.
The pairwise dispersion is also sensitive to $\Omegam$ and to the amplitude of
the mass power spectrum, since these determine the scale of halo velocity
dispersions and of relative halo velocities.

\subsection{Redshift Space Distortions}

The dispersion of galaxy peculiar velocities in collapsed regions smears
structure along the line of sight in redshift space; it is this induced
anisotropy that allows $\sigma_v^2(r)$ to be inferred from redshift maps
without measuring galaxy peculiar velocities individually.  On large scales,
coherent flows into overdense regions and out of underdense regions amplify
the contrast of structure along the line of sight (\citealt{sargent77};
\citealt{kaiser87}; for an excellent and comprehensive review see
\citealt{hamilton98}).  In the linear regime, the linear bias approximation
$\delta_g=b\delta_m$, and the distant observer approximation, the
redshift-space power spectrum $P^S(k)$ is enhanced by a factor
\begin{equation}
\frac{P^S(k)}{P^R(k)} = 1 + \frac{2}{3}\beta + \frac{1}{5}\beta^2
\label{eqn:PSR1}
\end{equation}
over the real-space power spectrum $P^R(k)$, where
$\beta\equiv f(\Omegam)\approx\Omegam^{0.6}/b$ \citep{kaiser87}.  For models
in which galaxy bias is a more complex function of local environment,
equation~(\ref{eqn:PSR1}) still applies on large scales, with $b$
corresponding to the large-scale asymptotic value of the rms bias factor
\citep{berlind01}.  However, non-linear
effects, especially those associated with velocity dispersions in collapsed
regions, influence $P^S(k)$ out to scales $2\pi/k \sim 50-100\hmpc$ or more
\citep{cole94}.  Measurement of $\beta$ with redshift-space distortions
therefore requires very large redshift surveys and accurate modeling of
non-linear effects.  The analysis of the 2dF galaxy redshift survey by
\citet{peacock01} provides the best measurement to date.

\begin{figure}
\centerline{
\epsfxsize=6.0truein
\epsfbox[40 185 550 675]{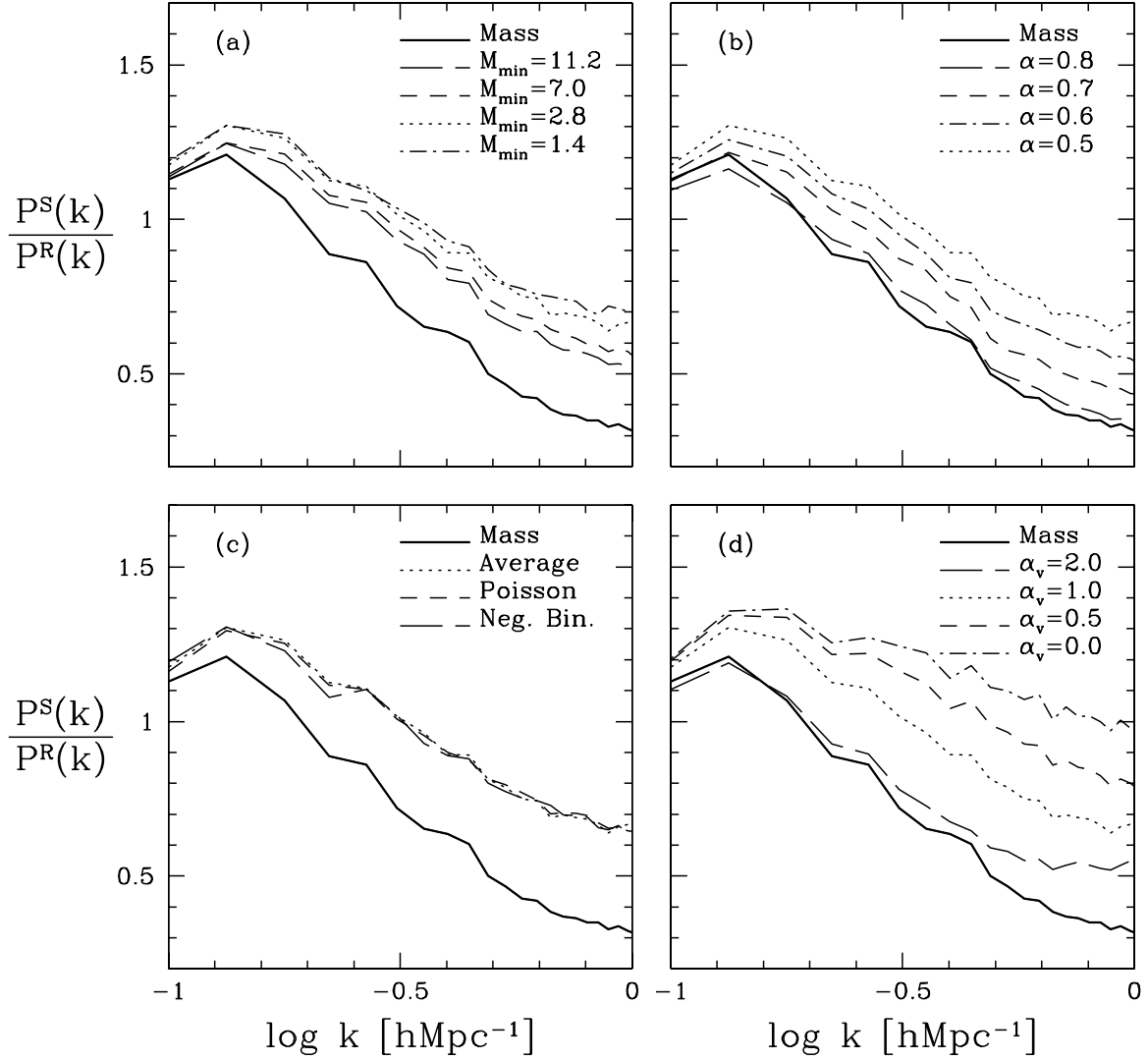}
}
\caption{Ratios of the redshift to real space power spectra $P^S/P^R$ as a
function of wavenumber $k$ for different HOD prescriptions.  In each panel, 
the dotted curve shows $P^S(k)/P^R(k)$ for a model with power-law $\NavgM$, 
$\Mmin=2.8\times 10^{11}\hMsun$, $\alpha=0.5$, Average $\PNNavg$, and $\alpha_v=1$.
Other curves show $P^S(k)/P^R(k)$ for models with different $\Mmin$ (panel~a),
different $\alpha$ (panel~b), different $\PNNavg$ (panel~c), or different
velocity distributions within halos (panel~d).  Thick solid curves show 
$P^S(k)/P^R(k)$ for the unbiased mass distribution.
} 
\label{fig:16}
\end{figure}

We compute power spectra of our mass and galaxy distributions by CIC-binning
them onto a $200^3$ grid and applying an FFT.  For the redshift-space power
spectrum, we take the line-of-sight direction to be a Cartesian axis of the
simulation cube, implicitly assuming that the whole simulation volume is far
enough away to satisfy the distant observer approximation.
Figure~\ref{fig:16} shows $P^S(k)/P^R(k)$ for a variety of HOD models and for
the unbiased mass distribution.  The general behavior of these curves is as
expected: suppression of the redshift-space power spectrum on small scales
(high $k$), where non-linear effects dominate, and enhancement at low $k$.
However, it is clear that the largest scales probed by our $141.3\hmpc$
simulation cube are not yet in the linear regime described by
equation~(\ref{eqn:PSR1}), since there is no clear plateau in $P^S(k)/P^R(k)$
and the maximum ratio for the mass distribution is substantially lower
than the value 1.37 implied for $\beta=0.3^{0.6}=0.48$.

In Figure~\ref{fig:16}a, raising $\Mmin$ depresses $P^S(k)/P^R(k)$ at all $k$.
At small scales, where $P^S(k)/P^R(k) < 1$, this behavior reflects the
increased suppression of line-of-sight structure by velocity dispersions
in high mass halos.  At large scales, one expects higher $\Mmin$ models to
have lower $P^S(k)/P^R(k)$ because they have higher $b$ (lower $\beta$), but
the effect seen in Figure~\ref{fig:16}a may still be dominated by the
increased residual impact of small scale velocity dispersions.  All four HOD
models have $\alpha=0.5$, so relative to the mass distribution the high
dispersion halos are down-weighted, and the large scale bias factor is
slightly less than one.  Thus, all four models have higher $P^S(k)/P^R(k)$
than the mass distribution.  Raising $\alpha$ reduces $P^S(k)/P^R(k)$ by
increasing velocity dispersions and the bias factor, as seen in
Figure~\ref{fig:16}b.  The $\alpha=0.8$ model has nearly the same pairwise
dispersion as the mass (Fig.~\ref{fig:15}b) and a large scale bias factor
slightly larger than one (Fig.~\ref{fig:4}); as one might therefore expect,
its $P^S(k)/P^R(k)$ ratio matches the mass on small scales and is slightly
lower on large scales.  Figure~\ref{fig:16}c shows that the form of
$\PNNavg$ has no significant impact on the power spectrum distortion.

Figure~\ref{fig:16}d shows the impact of velocity bias within halos.  Even
though all four models have the same mean velocity field, their power spectrum
ratios are different on all scales probed by the simulation.  The comparison
shows that the suppression of $P^S(k)/P^R(k)$ on small scales is dominated
by dispersions within virialized halos.  However, even the $\alpha_v=0$ model
shows a drop in $P^S(k)/P^R(k)$ relative to the value at large scales, showing
that the dispersion of the halo velocities themselves also plays a role.
Above all, Figure~\ref{fig:16}d confirms previous indications that small scale
galaxy velocity dispersions influence the redshift-space power spectrum out to
remarkably large scales.

Redshift-space distortions can also be quantified using the ratio of
quadrupole and monopole moments of $P^S(k)$ \citep{cole94}.  We have measured
this ratio as well as $P^S(k)/P^R(k)$ and find similar dependence on HOD
parameters.

Although our simulation is not large enough to show it, we expect on the basis
of \citet{berlind01} that the power spectrum ratio would approach the value
defined by $\beta=\Omegam^{0.6}/b_P$ at sufficiently low $k$, where
$b_P^2$ is the asymptotic value of $P_g(k)/P_m(k)$.  Even with a large
galaxy redshift survey, constraining $\beta$ requires modeling the non-linear
effects on redshift-space distortions, which is most often done using the
\citet{peacock94} model of a linear theory distortion modulated by random
velocities with a single value of the velocity dispersion.  With galaxy
redshift samples of the size expected from the 2dF and SDSS, the limitation of
this model introduces a systematic error that is larger than the statistical
uncertainty (\citealt{cole95}; \citealt{hatton99}).  The HOD seems the natural
framework in which to develop a more realistic and more accurate model of
non-linear redshift-space distortions.  The analytic studies of
\citet{seljak01} and \citet{white01a} provide the basis for such a model,
which should improve constraints on $\Omegam$ and $b$ from future analyses
of larger data sets.

\section{Group Statistics}

The statistics that we have examined so far treat individual galaxies as the
basic units of structure.  A useful complementary strategy is to define groups
of galaxies and measure their properties.  We identify galaxy groups in real
space using the same friends-of-friends algorithm that we used to
find halos in the mass distribution.  We use a linking length of $0.2$ times
the mean inter-galaxy separation, and we only retain groups containing
four or more galaxies.  Since our galaxy distributions have a space density of
$0.01\hvol$, the linking length is approximately $0.9\hmpc$.  With this choice,
most galaxies within the same halo are identified as belonging to the same
group, but galaxies in separate halos are only occasionally linked.  However,
there are situations in which the galaxy populations of two halos are linked
into a single group, and others where the population of one halo is split into
two or more components, so the match between groups and halos is not exact.
For a galaxy redshift survey there are additional complications including the
need for different angular and redshift linking lengths to account for
peculiar velocities, scaling linking lengths with distance to account for
varying galaxy density in a flux-limited survey, and, in surveys that employ
fiber spectrographs, accounting for incompleteness imposed by minimum fiber
separations.  (For discussion of some of these issues, see, e.g.,
\citealt{moore93} and \citealt{nolthenius94}.)  Here we consider the idealized
case and leave the problem of design and performance of group-finding
algorithms to studies that are tailored to specific data sets.

The most basic group statistic is the multiplicity function, the abundance of
galaxy groups as a function of group richness.  This can be measured for groups
identified in a redshift survey (\citealt{maia89}; \citealt{ramella89};
\citealt{ramella99}), or from the galaxy surface density in an imaging survey
(\citealt{gott77}; \citealt{defilippis99}).  We define the cumulative
multiplicity function $\ngrpN$ to be the number density of galaxy groups that
contain $N$ or more galaxies.  To the extent that groups and halos correspond
one-to-one, the differential group multiplicity function is simply a
convolution of the halo mass function with $\PNM$, and the cumulative
multiplicity function is
\begin{equation}
\ngrpN = \sum_N^{\infty} \int_0^{\infty} dM\frac{dn}{dM} \PNM.
\label{eqn:grp1}
\end{equation}
In the limit of a narrow $\PNM$, the space density of groups of multiplicity
$N$ is equal to the space density of halos with $\NavgM=N$.  The direct
connection implied by equation~(\ref{eqn:grp1}) makes the multiplicity
function a promising tool for constraining $\PNM$.

\begin{figure}
\centerline{
\epsfxsize=6.0truein
\epsfbox[40 185 550 675]{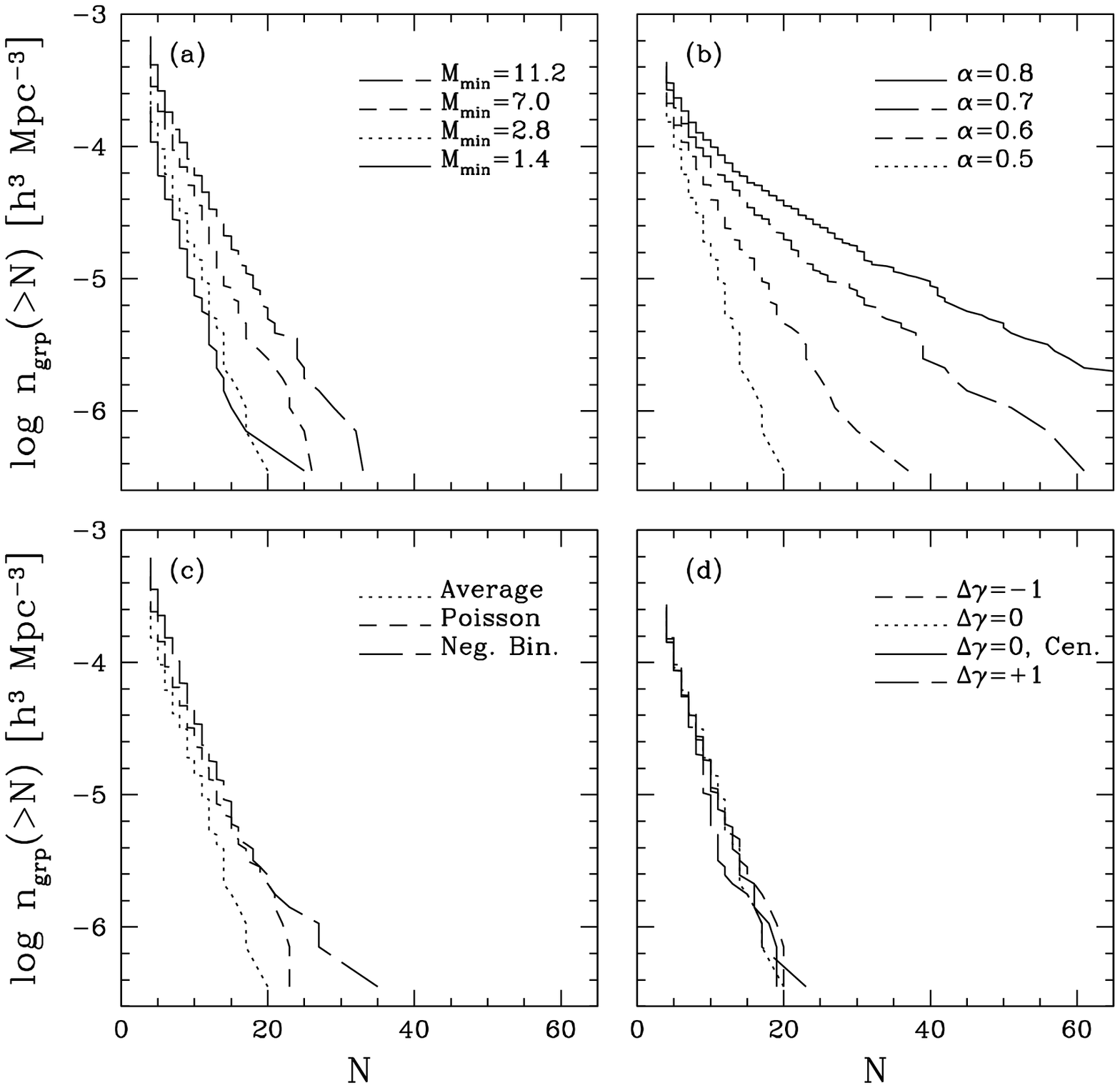}
}
\caption{Cumulative group multiplicity functions for different HOD prescriptions.  
In each panel, the dotted curve shows $\ngrpN$ for a model with power-law 
$\NavgM$, $\Mmin=2.8\times 10^{11}\hMsun$, $\alpha=0.5$, Average $\PNNavg$, 
and $\Dg=0$. Other curves show $\ngrpN$ for models with different $\Mmin$ 
(panel~a), different $\alpha$ (panel~b), different $\PNNavg$ (panel~c), or 
different galaxy distributions within halos (panel~d).  
}
\label{fig:17}
\end{figure}

Figure~\ref{fig:17} plots $\ngrpN$ for a variety of HOD models, starting in
each case at the largest group in the box and moving up to higher densities
as $N$ gets smaller.  The function becomes noisy at high $N$ where the number
of groups is small.  Increasing $\Mmin$ (Fig.~\ref{fig:17}a) shifts $\ngrpN$
towards larger $N$, since the overall galaxy space density remains constant
and the number of galaxies in high mass halos therefore increases.
Increasing $\alpha$ has a much more drastic effect on the multiplicity
function (Fig.~\ref{fig:17}b) because it boosts the multiplicity of higher
mass halos by a larger factor, changing the slope of $\ngrpN$.  Broadening
$\PNNavg$ has a modest impact (Fig.~\ref{fig:17}c); while a given halo's
multiplicity can scatter in either direction, there are more low mass halos to
scatter to high $N$ than vice versa, so the net effect is to boost the
high-$N$ tail of the multiplicity function.
Finally, Figure~\ref{fig:17}d shows that the group multiplicity function is
completely insensitive to the spatial distribution of galaxies within halos,
a reassuring indication that the friends-of-friends algorithm is consistently
identifying the same groups regardless of their detailed internal structure.

From the HOD perspective, the other promising aspect of groups is the
possibility of measuring their masses dynamically and thus obtaining a direct
association between $N$ and $M$.  For each group in our simulated galaxy
population, we apply the projected virial mass estimator given by
\citet{heisler85},
\begin{equation}
\Mvir = \frac{3\pi N}{2G} \frac{\sum_i V^2_{z,i}}{\sum_{i<j}1/R_{\perp,ij}},
\label{eqn:grp2}
\end{equation}
where $V_{z,i}$ is the line-of-sight velocity of galaxy $i$ relative to
the group velocity centroid and $R_{\perp,ij}$ is the relative tangential
separation between galaxies $i$ and $j$.  For this calculation, we treat
one axis of the simulation box as the line-of-sight direction.  (We have
also tried the other mass estimators suggested by Heisler et al. 1985, with
similar results.)

\begin{figure}
\centerline{
\epsfxsize=5.0truein
\epsfbox[50 35 555 765]{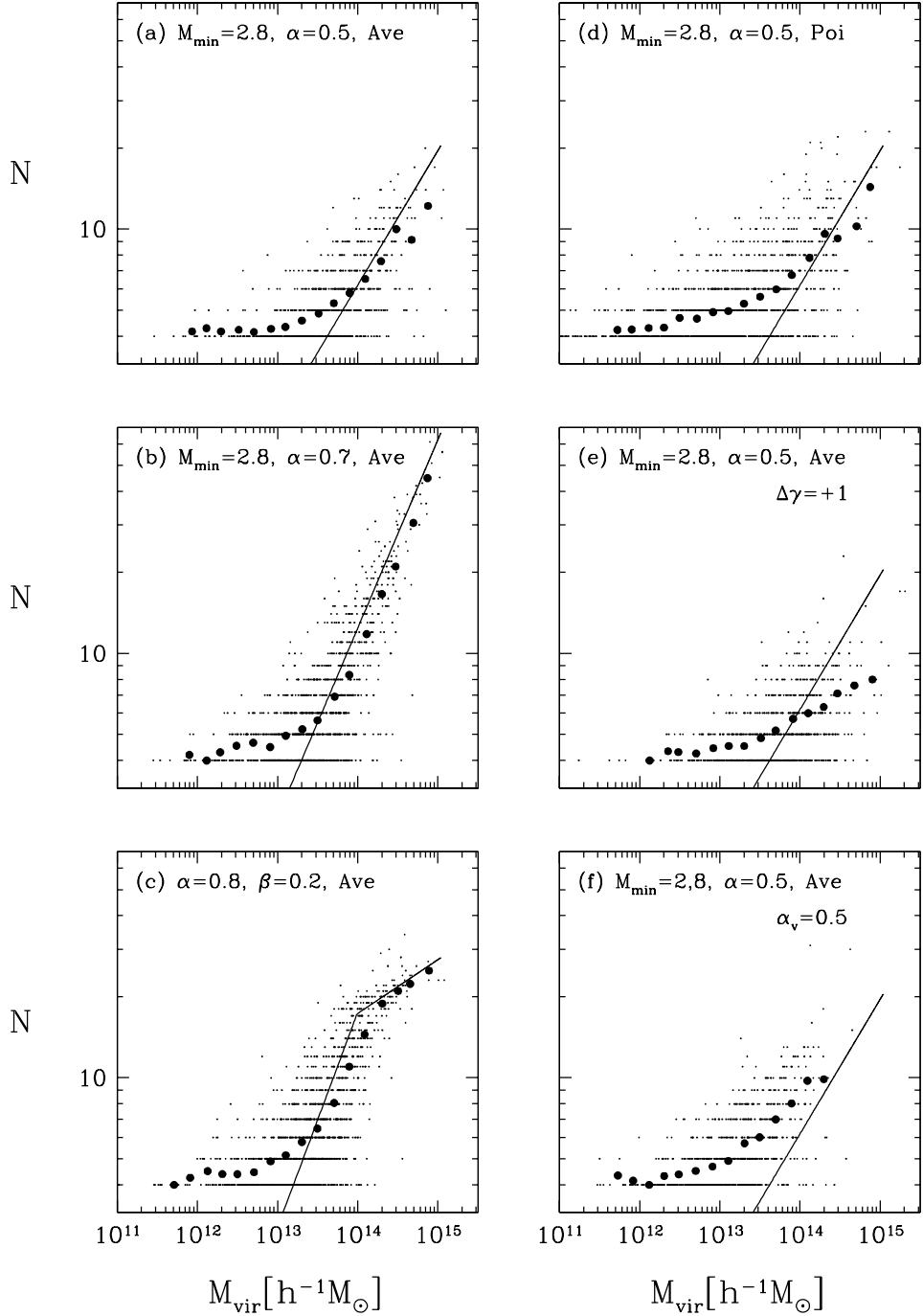}
}
\caption{Richness vs. galaxy group virial mass estimates. Each panel 
represents a particular HOD model.  The models are specified at the top of 
each panel, where $\Mmin$ is given in units of $10^{11}\hMsun$, and Poi 
and Ave represent Poisson and Average $\PNNavg$ distributions, 
respectively.
Galaxy groups were identified with a friends-of-friends algorithm in real
space, and a virial estimator was used to estimate their mass.  Each panel 
also shows the mean group richness in bins of $\Mvir$ (large points) as well
as the input $\NavgM$ relation (line).
} 
\label{fig:18}
\end{figure}

Figure~\ref{fig:18} plots group richness against estimated virial mass for
a variety of HOD models.  Figure~\ref{fig:18}a shows a model with power-law
$\NavgM$, $\alpha=0.5$, $\Mmin=2.8\times10^{11}\hMsun$, and Average $\PNNavg$.
Each small point represents an individual galaxy group, and large points show
the mean richness in bins of $\Mvir$.
If the group-halo match and virial mass estimates were perfect, this figure
would reproduce the top panel of Figure~\ref{fig:1}, which illustrates the
true $\PNM$ used to create this galaxy population.  However, random errors
in $\Mvir$ make the scatter-plot in Figure~\ref{fig:18} much broader.  Perhaps
more seriously, the values of $\NavgM$ recovered from this diagram are biased
with respect to the true relation, shown by the solid line.  At
$M\lesssim 10^{13}\hMsun$, the mean multiplicity is overestimated because the
group catalog excludes halos in this mass range that have $N<4$.  At high
masses, random errors in $\Mvir$ cause a slight underestimate of $\NavgM$,
since the number of low-$N$ groups with overestimated $\Mvir$ exceeds the
number of high-$N$ groups with underestimated $\Mvir$.  Both of these biases
are calculable, so data like those in Figure~\ref{fig:18}a could be corrected
to yield a more accurate estimate of $\NavgM$.  However, such a correction
would require some assumptions about $\PNM$ (such as an extrapolation at the
low mass end) and an accurate understanding of the error distribution of
$\Mvir$.

The situation is considerably more promising for the $\alpha=0.7$ model shown
in Figure~\ref{fig:18}b.  This model has a much larger number of high
multiplicity groups, which yield more accurate $\Mvir$ estimates because they
have more galaxy tracers.  Furthermore, the dynamic range in $N$ is large
enough to allow an accurate estimate of the slope of $\NavgM$ at high $M$,
though the amplitude of the relation is still slightly underestimated.
Figure~\ref{fig:18}c shows a model with a break in $\NavgM$ at
$\Mcrit=10^{14}\hMsun$.  The virial mass estimates reveal the existence of the
break and recover the high-$M$ slope of $\NavgM$ reasonably well.  However,
the mass at which the break occurs and the slope at the low mass end of the
relation are not clearly determined.  A break at lower mass, like the one in
the model illustrated in Figure~\ref{fig:9}, would pass undetected.

Figure~\ref{fig:18}d shows an $\alpha=0.5$ model with a Poisson $\PNNavg$.
Despite the scatter in the mass estimates themselves, the broader $\PNM$ is
clearly detectable in comparison to the Average model of Figure~\ref{fig:18}a.
The larger scatter increases the bias in the recovered $\NavgM$ at low $M$.
Figure~\ref{fig:18}e shows a model with $\Dg=+1$, i.e., galaxies more
extended than dark matter within halos, for which the recovered $\NavgM$ lies
below the true relation at high $M$.  Figure~\ref{fig:18}f shows a model with
a strong velocity bias, $\alpha_v=0.5$, which drives estimated masses down
relative to true values.  For rich groups and clusters, mass estimates from
galaxy dynamics can be tested against masses estimated from X-ray observations
or gravitational lensing, and systematic errors as large as those in
Figures~\ref{fig:18}e and \ref{fig:18}f (which display fairly extreme models)
can probably be ruled out on the basis of existing studies
(see, e.g., \citealt{wu98}).  Such comparisons become harder for low
multiplicity groups, though it may soon be possible to test for any overall
bias by comparing mean dynamical masses to the group-mass correlation function
derived from weak lensing.

Figure~\ref{fig:18} suggests that it will be difficult to recover $\PNM$ or
$\NavgM$ simply by measuring dynamical masses as a function of multiplicity.
Even if there are no internal biases (like $\alpha_v \neq 1$ or $\Dg \neq 0$)
that systematically influence $\Mvir$, the random errors in $\Mvir$ and
absence of low multiplicity halos from the group catalog drive the recovered
$\NavgM$ away from the true value.  However, viewed as another clustering
statistic, the richness-virial mass relation can provide another valuable
constraint on the HOD, especially if the form of $\NavgM$ has been determined
from other clustering measures, since it tests the overall mass scale and
constrains the role of internal spatial and velocity biases.

\begin{figure}
\centerline{
\epsfxsize=5.0truein
\epsfbox[50 35 555 765]{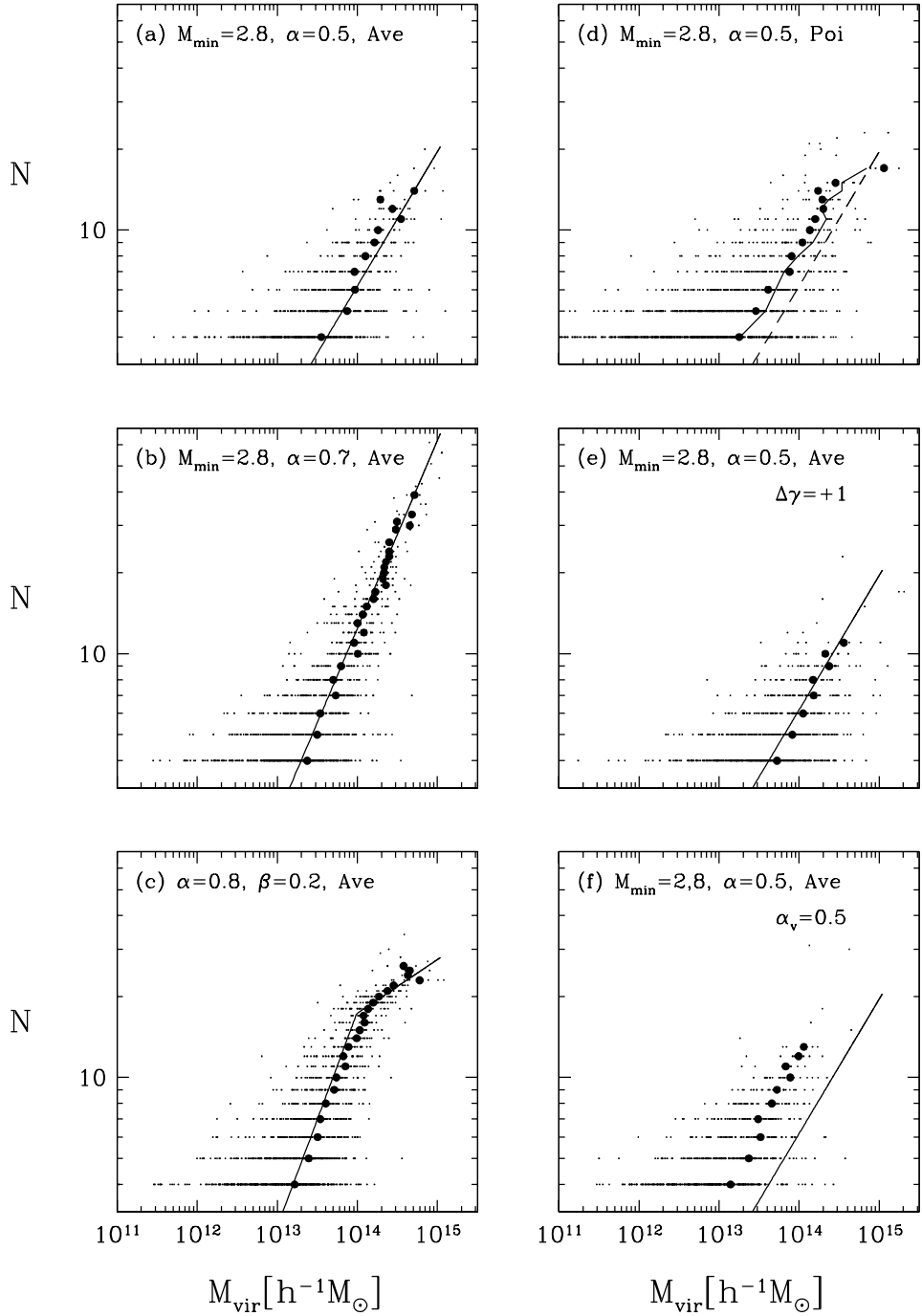}
}
\caption{Same as Fig.~\ref{fig:18}, except that large points represent
the mean estimated virial mass at fixed group richness $N$, solid lines
show the true $M_{\mathrm{avg}}(N)$, and the dashed line in panel~(d)
shows $\NavgM$.
}
\label{fig:19}
\end{figure}

The poor performance of the mean richness as a function of virial mass in
recovering $\NavgM$ leads us to consider an alternative way of interpreting
the richness virial mass relation.  Figure~\ref{fig:19} shows this
relation for the same set of HOD models as Figure~\ref{fig:18}, but large points
now represent the mean virial mass at each multiplicity $N$, and solid lines
show the true $M_{\mathrm{avg}}(N)$ relation.  For a narrow $\PNNavg$, such
as the Average distribution, $M_{\mathrm{avg}}(N)$ and $\NavgM$ contain
identical information, since $\Navg(M_{\mathrm{avg}}(N))=N$.  However, in the
Poisson case these two functions differ, as can be seen in
Figure~\ref{fig:19}d, where $\NavgM$ is shown by the dashed line.
Figure~\ref{fig:19} demonstrates that most of the biases that affected the
recovery of $\NavgM$ do not affect the recovery of $M_{\mathrm{avg}}(N)$.
Only velocity bias seriously hampers the recovery of $M_{\mathrm{avg}}(N)$
(Fig.~\ref{fig:19}f), and, even in that case, only the amplitude of the relation
is affected.  Unfortunately, the $M_{\mathrm{avg}}(N)$ relation cannot be
inverted to recover $\NavgM$ without assuming a form for $\PNNavg$, so the
improved performance of this statistic comes with the loss of model independence.

\section{Measuring the HOD Empirically}

The recurring theme of preceding sections is that different galaxy clustering
statistics are sensitive to different aspects of the halo occupation
distribution.  The encouraging implication is that combinations of clustering
measurements can pin down the properties of the HOD, hemming it in with
complementary constraints.  The 2dF and SDSS redshift surveys are ideally
suited to this task because they can measure clustering with high precision
over a wide dynamic range.  The SDSS also provides detailed photometric
information for all galaxies in the redshift survey, allowing studies of the
HOD for a multitude of galaxy classes, and weak lensing measurements from the
SDSS images can determine the galaxy-mass correlation function for each of
these classes.

In this section, we outline a strategy for determining the HOD from
observations, and although this outline is far from complete, we believe that
it offers a promising route forward.  Since the possible parameter space of
HOD models is extremely broad, a practical strategy requires some procedure
for getting a good first guess at $\PNM$, which can then be refined and tested
using a battery of clustering measures.  Our procedure assumes that the
underlying cosmological model is known in advance based on independent
observations, though the model can be further tested by seeing whether it
reproduces all aspects of observed galaxy clustering for some HOD.  The
ability of large scale structure to constrain cosmology depends in part on
whether two different cosmological models with two different halo occupation
distributions can produce the same galaxy clustering; we consider this
question briefly at the end of the section.

We take as our starting point the tight connection between $\PNM$ and the
group multiplicity function implied by equation~(\ref{eqn:grp1}).  If we
assume that $\PNNavg$ is narrow, then group multiplicity is a monotonic
function of halo mass, and determining $\NavgM$ from the multiplicity function
is simply a matter of matching space densities.  Given the measured $\ngrpN$
and the cumulative halo mass function $\nhM$ computed from the cosmological
model, we choose $\NavgM$ so that $n_{\mathrm{grp}}[>\NavgM]=\nhM$.  More
generally, one can assume a form of $\PNNavg$ and infer $\NavgM$ by reversing
the convolution in equation~(\ref{eqn:grp1}).  The relatively small impact of
$\PNNavg$ on the multiplicity function compared to changes in $\NavgM$ (see
Fig.~\ref{fig:17}) shows that results will not be overly sensitive to the
assumed width of the distribution.

\begin{figure}
\centerline{
\epsfxsize=5.0truein
\epsfbox[85 150 495 718]{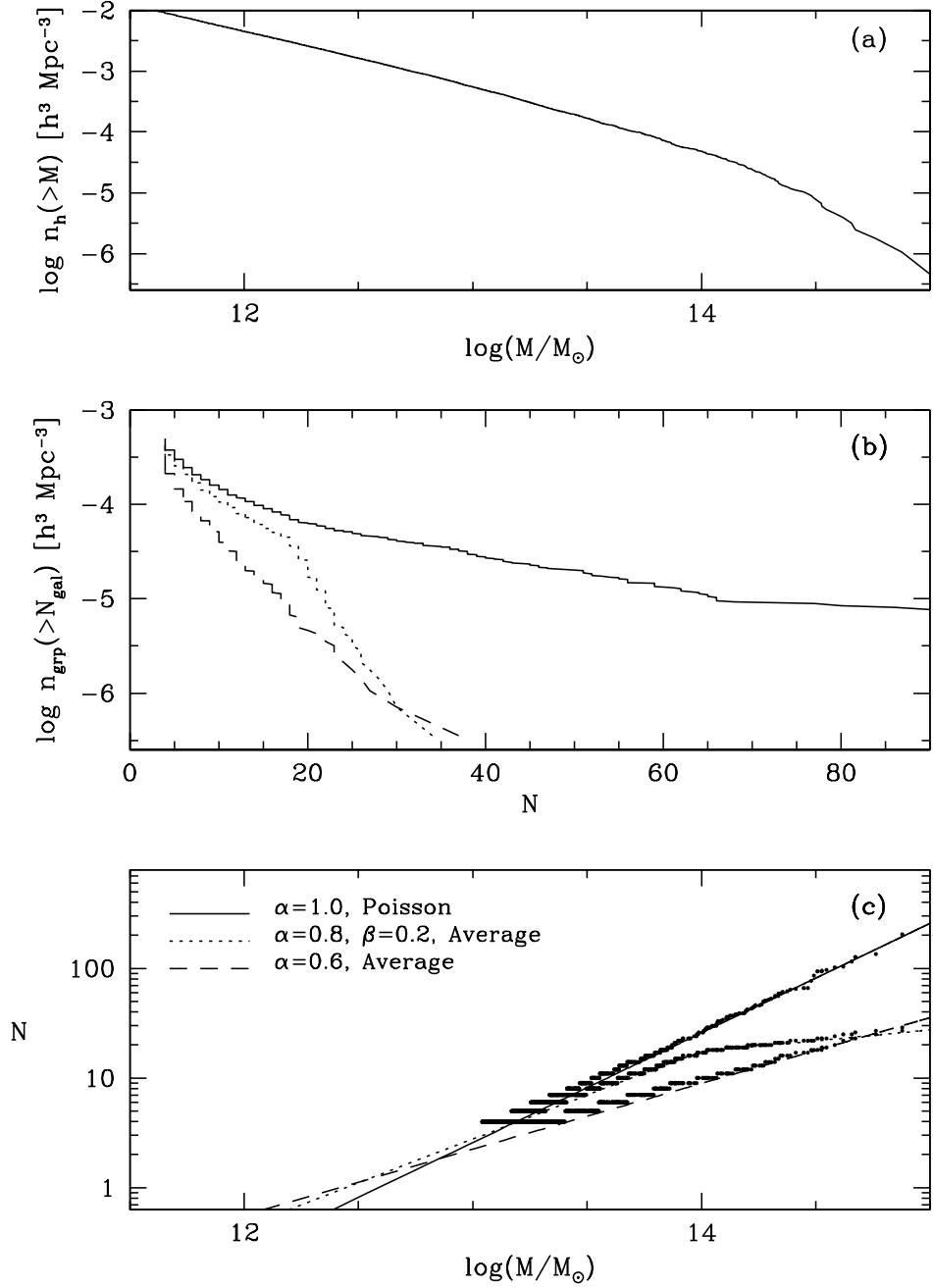}
}
\caption{Recovery of $\NavgM$ from the group multiplicity function. 
Panel~(a) shows the halo mass function $n_{\mathrm{h}}(M)$ derived from the 
GIF N-body simulation, where halos were identified with a FoF algorithm.
Panel~(b) shows group multiplicity functions $\ngrpN$ for three HOD models, 
where galaxy groups were identified with the same FoF algorithm.  Panel~(c) 
shows the resulting $\NavgM$ relations from matching $n_{\mathrm{h}}(M)$ to
$\ngrpN$ (points).  Also shown, for comparison, are the true $\NavgM$ 
functions that were used as inputs for the HOD models (lines).} 
\label{fig:20}
\end{figure}

Figure~\ref{fig:20} illustrates the performance of the monotonic matching
method.  The top panel plots the cumulative halo mass function derived from
the GIF N-body simulation, with halos identified by the FoF algorithm as
usual.  The middle panel shows cumulative group multiplicity functions for
three HOD models, with groups identified by the same FoF algorithm as
described in \S~5.  The bottom panel shows the $\NavgM$
relations derived by matching $\nhM$ to $\ngrpN$ and the true $\NavgM$ used as
inputs to the HOD models.  In all three cases, the method is impressively
successful at recovering $\NavgM$ for $M \gtrsim 10^{13.5}\hMsun$, obtaining
the correct power-law slopes for the $\alpha=0.6$ and $\alpha=1.0$ models and
revealing the break in the broken power-law model.  At low multiplicities, the
discreteness of $N$ causes a step-like behavior in $\NavgM$, but this could be
easily repaired.  The method works well for the $\alpha=1.0$ Poisson model
even though we have not applied any deconvolution correction.

The matching method illustrated in Figure~\ref{fig:20} is a promising way of
determining $\NavgM$ at high $M$, where halos contain large numbers of
galaxies.  For application to redshift survey data, the main challenge is to
develop a good group-finding algorithm and understand any biases associated
with it.  When studying the HOD of galaxy classes, one would start with the
group catalog and associated halo masses derived from the full, high density
galaxy catalog, then measure the content of individual classes.

Group matching provides a partial guess at the HOD, which must be extrapolated
to low masses, refined, and tested using other statistics.  One important
constraint on the extrapolation comes from the mean galaxy number density,
$\ng=\int_0^{\infty}dM\frac{dn}{dM}\NavgM$.  A similar constraint with a
different weighting of halos comes from the large scale bias factor,
$b=\ng^{-1}\int_0^{\infty}dM\frac{dn}{dM}b_h(M)\NavgM$, which can be inferred
from the measured galaxy power spectrum given the assumed cosmological model.
The most valuable constraints on halo occupation at low masses comes from the
1-halo regime of $\xig$, since different separations probe different halo
mass scales.

We can illustrate the constraining power of $\xig$ using an approximation
suggested by Figure~\ref{fig:2}a, which shows that the cumulative pair
separation distribution $F(r/2\Rvir)$ is roughly a linear ramp out to a
maximum separation $x_m \approx 0.6$ of the virial diameter, and is close to
unity at larger separations.  Under this approximation, the 1-halo term $\xih$
only receives contributions from halos with virial diameters $2\Rvir > rx_m^{-1}$,
or, using equation~(\ref{eqn:xi8}), with virial masses
\begin{equation}
M_l(r) = \frac{800\pi \bar{\rho}_m}{3} \left(\frac{r}{2x_m}\right)^3.
\label{eqn:emp1}
\end{equation}
We can therefore rewrite equation~(\ref{eqn:xi7}) by substituting the constant
value $F'(x)=x_m^{-1}$ and changing the lower limit of the integral to
$M_l(r)$, which is now the only part of the integral with an $r$ dependence.
Differentiating  both sides of the equation with respect to $r$ and
rearranging terms yields an expression relating the second factorial moment at
$M_l$ to the 1-halo correlation function and its logarithmic derivative at
$r$:
\begin{equation}
\NN_{M_l} = \frac{4\pi}{3}\ng^2
            \left(\left.\frac{dn}{d\mathrm{ln}M}\right|_{M_l}\right)^{-1}
            r^3\left(2+\frac{d\mathrm{ln}\xih}{d\mathrm{ln}r}\right)\xih.
\label{eqn:emp2}
\end{equation}

\begin{figure}
\centerline{
\epsfxsize=6.0truein
\epsfbox[85 220 530 650]{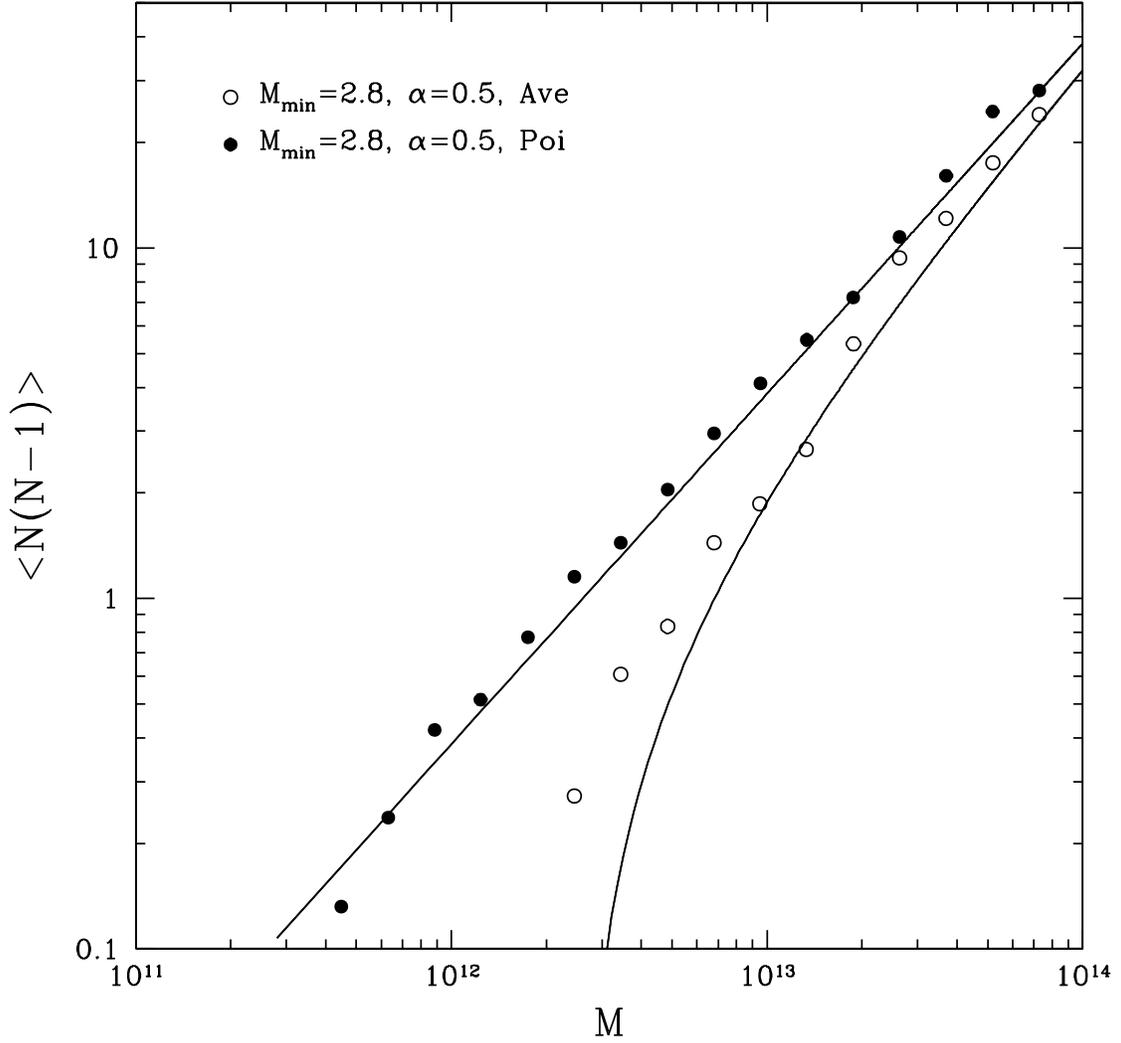}
}
\caption{Recovery of $\NN_M$ from $\xih$ using eq.~(\ref{eqn:emp2}).
The solid and open circles show the recovered $\NN_M$ for Poisson
and Average $\PNNavg$ HOD models, respectively.  Also shown for
comparison, are the ``true'' $\NN_M$ relations that were used as inputs 
for the HOD models (lines).
}
\label{fig:21}
\end{figure}

Figure~\ref{fig:21} illustrates the performance of equation~(\ref{eqn:emp2}).
Despite the rather crude, linear ramp approximation for $F(r/2\Rvir)$, the
estimates of $\NN_M$ derived from $\xih$ recover the true values shown by the
solid lines for both of these HOD models, which have
$\Mmin=2.8\times 10^{11}\hMsun$, $\alpha=0.5$, and Poisson and Average
$\PNNavg$, respectively.  In practice, one could only apply this method to
halos with virial diameters $2\Rvir \lesssim 0.5x_m^{-1}\hmpc \sim 0.8\hmpc$
(mass $M \lesssim 5\times 10^{12}\hMsun$ for $\Omegam=0.3$), since $\xig$
departs from $\xih$ at $r \gtrsim 0.5\hmpc$ (see Fig.~\ref{fig:7}).  However,
one could extend the range somewhat by calculating the 2-halo contribution to
$\xig$ and subtracting it to obtain $\xih$; alternatively, a numerical method
that uses equation~(\ref{eqn:xi7}) directly instead of the linear ramp
approximation would not be difficult to implement.  The important lesson of
equation~(\ref{eqn:emp2}) and Figure~\ref{fig:21} is that $\xih$ provides
fairly well localized information about $\NN_M$, in just the mass range where
the group matching method breaks down.

The small scale correlation function depends on $\NN_M$ rather than $\NavgM$,
and it has an additional dependence on the assumed galaxy profile (though
Fig.~\ref{fig:6} shows that this dependence is not strong).  However, it
is clear from the discussion above that matching $\ngrpN$, $\ng$, and $\xig$
already places strong constraints on $\PNM$ for a given cosmology.  A trial
HOD inferred by the techniques outlined above can be tested and refined using
other clustering statistics.  The VPF gives strong guidance on the low mass
end of $\PNM$, especially the value of $\Mmin$.  Higher order correlations
probe the higher moments of $\PNNavg$.  Additional strong tests come from the
mean virial mass as a function of multiplicity and from the pairwise velocity
dispersion, which is sensitive to the mean occupation at high mass and to
$\PNNavg$ at low mass.  Success in matching spatial clustering but failure
with these dynamical measures could be a signature of an incorrect assumption
about the underlying cosmology (e.g., the wrong value of $\Omegam$), or it
could be an indication of velocity bias ($\alpha_v \neq 1$).  The scale
dependence of the pairwise dispersion can help discriminate between these
possibilities, since $\sigma_v^2(r)$ combines terms that depend on $\alpha_v$
with others that do not (see eq.~[\ref{eqn:vdisp3}]).  Lensing measurements of
the galaxy-mass correlation function (or the analogous group-mass correlation
function) provide stronger discrimination by probing halo mass scales
independently of galaxy peculiar velocities.  The 1-halo regime of $\xigm$
also constrains $\NavgM$ in a range where $\xig$ constrains $\NN_M$.

Suppose we find an HOD that reproduces all of these properties of observed
galaxy clustering.  Can we take this success as evidence that the assumed
cosmological model is itself correct, or could there be some other combination
of cosmology and HOD that would yield the same results?  In other words, do
cosmology and HOD bias have degenerate effects on galaxy clustering, or can
sufficiently precise measurements constrain both independently?  We reserve a
detailed investigation of this question for future work, but qualitative
arguments suggest grounds for optimism.

Let ``model 1'' denote our supposed successful combination of cosmological
model and HOD prescription.  If we raise $\Omegam$ but keep the linear theory
matter power spectrum $P_{\mathrm{lin}}(k)$ the same, then the dark matter
halo population will be essentially unchanged except that all masses will grow
by the factor $\Omega_{m,2}/\Omega_{m,1}$, including the characteristic mass
$M_*$ of the halo mass function.  If we maintain the same HOD as a function of
$M/M_*$, then the spatial clustering of galaxies will stay the same, but
dynamical measures like the pairwise velocity dispersion or virial mass vs.
richness relation will not.  Changes to dynamically sensitive statistics could
be partially repaired by introducing velocity bias, with
$\alpha_{v,2} < \alpha_{v,1}$, but this suppression will not disguise changes
in $\sigma_v^2(r)$ or $P^S(k)$ at large scales or changes in $\xigm$ at any
scale.

For a less easily detectable change, we can reduce the amplitude of
$P_{\mathrm{lin}}(k)$ at the same time we increase $\Omegam$, maintaining a
``cluster normalization'' condition
$\sigma_{8,2}\Omega_{m,2}^{0.5} \approx \sigma_{8,1}\Omega_{m,1}^{0.5}$
\citep{white93}.  This change preserves (approximately) the velocity scale of
rich clusters and
large scale flows, but it reduces the length scale of non-linearity in the
matter distribution, by roughly the factor $(\Omega_{m,2}/\Omega_{m,1})^{1/3}$
required to keep $M_*$ constant in physical units.  If we maintain the same
relative galaxy populations as a function of $M/M_*$, then all characteristic
scales in the galaxy clustering drop by the same factor, including the galaxy
correlation length $r_0$.  To return $r_0$ and the space density $\ng$ to
their original values, we would need to raise $\Mmin/M_*$, or change $\alpha$,
or make other alterations to $\PNM$, and these changes would generally have
different impact on different clustering statistics.  Furthermore, the cluster
normalization condition preserves the amplitude of the halo mass function on
cluster scales but changes the overall shape of $dn/dM$ in physical units,
with a steeper fall-off at high masses for models with lower $\sigma_8$
(higher $\Omegam$).  This change could  be partially repaired by altering the
shape of $P_{\mathrm{lin}}(k)$, but that would alter the halo spatial
clustering.  We tentatively conclude that substantial changes to the value of
$\Omegam$ or the normalization or shape of $P_{\mathrm{lin}}(k)$ cannot be
masked by changes in the HOD.  (See \citeauthor{zheng02} [\citeyear{zheng02}]
for further discussion of these issues.)

Given the increasingly tight constraints on cosmological parameters that come
from CMB anisotropies, the Lyman-$\alpha$ forest, gravitational lensing,
Type~Ia supernovae, cluster evolution, and other data, a procedure that
recovers the HOD for an assumed cosmology would be very valuable in itself.
Successful application of this procedure to surveys like the 2dFGRS and SDSS
would tell us essentially everything that local ($z=0$) galaxy clustering has
to say about galaxy formation, assuming that the adopted cosmological model is
correct.  Determinations of the HOD for different galaxy classes would provide
detailed targets for analytic or numerical models of galaxy formation and
insight into the physics that controls galaxy luminosities, colors, and
morphologies.

If, as we have speculated, the effects of cosmology and bias are not
degenerate, then the HOD determination procedure becomes something still more
powerful: a systematic way of testing cosmological models against galaxy
clustering data given only some very general assumptions about galaxy
formation (basically the validity of the HOD framework itself).  Incorrect
cosmological models would fail to reproduce the observations for any choice of
HOD.  Of course, galaxy clustering provides a testing ground for cosmology
even with less detailed assumptions about galaxy formation, since the effects
of ``local'' bias on the power spectrum, higher order moments, redshift-space
distortions, and galaxy density and velocity fields become relatively simple
on sufficiently large scales (see, e.g., \citealt{coles93}; \citealt{fry93};
\citealt{mann98}; \citealt{scherrer98}; \citealt{narayanan00};
\citealt{berlind01}).  However, by incorporating HOD determination into
analyses of galaxy clustering, one can use information from the highly
nonlinear scales of individual halos to calculate effective ``bias factors''
for a variety of clustering statistics in the intermediate, moderately
nonlinear regime, where the impact of bias may not be simple.  This approach
sharpens the sensitivity of clustering studies to small departures from the
``minimal'' cosmological model, which could be produced by light but not massless
neutrinos, by a non-standard relativistic background, by non-scale invariant
inflation, or even by a small contribution from isocurvature or non-Gaussian
fluctuations.  Analysis of separate galaxy classes with different HODs would
provide redundant checks of the results, since the properties of the
underlying mass clustering should be independent of the tracer population used
to infer them.

The monotonic matching method proposed above is similar to the technique
suggested by \citet{peacock00}, except that they use the ``galaxy systems
luminosity function'' in place of the multiplicity function and extend the
assumption of a monotonic relation between halo mass and luminosity into the
single-galaxy regime.  Applying the technique to the data of Moore et al. (1993)
and the $\Lambda$CDM halo mass function, \citet{peacock00} recover an $\NavgM$
relation that is shallow at low mass and steepens towards higher masses,
qualitatively similar to the predictions of semi-analytic models and
hydrodynamic simulations (\citealt{benson00}; Berlind et al., in preparation)
and to the broken power-law model whose correlation function is illustrated by
the solid curve in Figure~\ref{fig:9}.  As the first determination of the HOD
from empirical data, this recovery of a relation that resembles theoretical
predictions and yields approximately the observed galaxy power spectrum
(\citealt{peacock00}, figure~8) is very encouraging.  Using similar methodology,
\citet{marinoni01} have recently obtained a similar $\NavgM$ relation for a
new sample of optically selected galaxies.
With a different, model fitting approach, \cite{jing98b}, \cite{jing98c},
and \cite{jing02} conclude that suppression of the fraction
of galaxies in high mass halos can help explain the observed form
and amplitude of the galaxy two-point and three-point correlation
functions and pairwise velocity dispersion.  Determinations of the
HOD from the 2dF and SDSS redshift surveys will yield a far more detailed
picture of the relation between galaxies and dark matter and among different
galaxy types themselves.  The improved constraints on dark matter clustering
have the potential to reveal effects that are quantitatively subtle but have
profound physical implications.

\section{Summary}

In the HOD framework, the bias of a population of galaxies is fully defined by
the conditional probability $\PNM$ that a halo of virial mass $M$ contains $N$
galaxies, together with prescriptions that specify the relative spatial and
velocity distributions of galaxies and dark matter within virialized dark
matter halos.  This framework provides an informative basis for
connecting observations of galaxy clustering to the physics of galaxy
formation, and it opens a route to the empirical determination of bias from
galaxy redshift surveys.  We have investigated the sensitivity of galaxy
clustering statistics to features of the HOD, focusing on models with mean
occupation $\NavgM \propto M^{\alpha}$ for $M>\Mmin$, and found the following
results:

1) The amplitude of the galaxy correlation function on large scales is
determined by the relative numbers of galaxies in high mass and low mass
halos, and is thus sensitive to $\alpha$ and, to a lesser extent, $\Mmin$.
On small scales, $\xig$ receives the largest contribution from galaxy pairs
in halos of virial radius $\Rvir \sim r$, so different separations probe the
occupation at different masses.  The amplitude and shape of $\xig$ are highly
sensitive to $\alpha$, $\Mmin$, and $\PNNavg$, though each feature affects
$\xig$ in a different way.  The spatial distribution of galaxies within halos
has a relatively modest impact that is confined to small spatial scales.

2) In HOD models, a power-law correlation function emerges from a balance of
several competing effects, and it therefore requires rather specific combinations
of parameters.  The sensitivity of $\xig$ to HOD parameters implies that its
observed power-law form is a strong constraint on the physics of galaxy
formation, and that the success of semi-analytic models and hydrodynamic
simulations in reproducing this form is entirely non-trivial.

3) The influence of HOD parameters on the galaxy-mass cross-correlation
function is similar to the effect on $\xig$, except that $\xigm$ is
independent of $\PNNavg$ even on small scales.  The direct dependence
of galaxy-galaxy lensing measurements on the mass of halos makes them an
especially valuable test of the background cosmology.

4) The bispectrum contains complementary information to two-point statistics
due to its greater weighting of high mass halos and its dependence on the
third moment of $\PNNavg$.  The dependence of the reduced bispectrum $Q(k)$
on $\alpha$ and $\Mmin$ is fairly complex, and different from the dependencies
found for $\xig$.

5) The void probability function (VPF) is sensitive to the low $M$ regime
of $\PNM$, where halos have a significant probability of containing no galaxies.
In particular, the VPF is most sensitive to the cut-off halo mass $\Mmin$.

6) The pairwise velocity dispersion is particularly sensitive to $\alpha$,
which controls the fraction of galaxies in high dispersion halos, and to the
presence of velocity bias in halos.  In addition, $\sigma^2_v(r)$ is sensitive
to $\Omegam$ and the amplitude of the matter power spectrum.

7) The ratio of the redshift to real-space power spectra is sensitive to
$\NavgM$ and velocity bias, as well as $\Omegam$.  Velocity dispersions within
halos influence the redshift-space power spectrum out to large scales.

8) The group multiplicity function is sensitive to $\Mmin$ and $\alpha$ in
different ways: increases in $\Mmin$ cause a roughly horizontal shift of
$\ngrpN$ towards greater $N$, whereas increases in $\alpha$ cause a drastic
change in the slope of $\ngrpN$.

9) The richness - virial mass relation is a direct probe of $\PNM$.  However,
it is subject to biases that make it difficult to extract $\NavgM$.  The mean
halo mass per multiplicity, $M_{\mathrm{avg}}(N)$, can be extracted much more
robustly, though even this measure can be affected by velocity bias.

The different sensitivities of these clustering statistics to different features
of the HOD suggest that the HOD can be empirically determined from
observations of galaxy clustering.  We have outlined a strategy for measuring
the HOD from a galaxy redshift survey, which involves obtaining a first guess
at $\PNM$ and then refining that guess with additional clustering statistics.
In its present form, our strategy assumes that the cosmological model is known
{\it a priori}.  However, the effects of changing the HOD should generally be
different from the effects of changing cosmological parameters, so an
incorrect cosmological model may be unable to reproduce the full suite of
galaxy clustering statistics for any choice of HOD.  At the very least,
determinations of the HOD for different classes of galaxies from the 2dF and
SDSS redshift surveys will provide considerable insight into the physics of
galaxy formation and the origin of galaxy properties.  If the degeneracy
between cosmology and bias can indeed be broken, then determinations of the
HOD will also sharpen tests of the standard cosmological model, perhaps
revealing the imprint of new physics on the large scale structure of the
universe.

\acknowledgments
We are indebted to Vijay Narayanan for many helpful discussions on the 
subject of bias and for supplying some of the software used in this project.
We thank Carlton Baugh, Andrew Benson, James Bullock, Chris Burke, Shaun Cole, 
Carlos Frenk, Andrey Kravtsov, and Roman Scoccimarro for helpful discussions, 
input, and comments, and Carlton Baugh for making the APM correlation function 
available to us in electronic format.  We also thank the Virgo consortium for 
making their simulations publicly available.  This work was supported by NSF 
grants AST-9802568 and AST-0098584.  AAB received additional support from a 
Presidential Fellowship from the Graduate School of The Ohio State University.


\clearpage

\begin{thebibliography}{}

\bibitem[Baugh(1996)]{baugh96}
Baugh, C. 1996, \mnras, 280, 267

\bibitem[Benson et al.(2000)]{benson00}
Benson, A. J., Cole, S., Frenk, C. S., Baugh, C. M., \& Lacey, C. G. 2000,
\mnras, 311, 793

\bibitem[Benson(2001)]{benson01}
Benson, A. J. 2001, \mnras, 325, 1039

\bibitem[Berlind et al.(2001)]{berlind01}
Berlind, A. A., Narayanan, V. K., \& Weinberg, D. H. 2001, \apj, 549, 688

\bibitem[Blanton et al.(1999)]{blanton99}
Blanton, M., Cen, R., Ostriker, J. P., \& Strauss, M. A. 1999,
\apj, 522, 590

\bibitem[Bond et al.(1991)]{bond91}
Bond, J. R., Cole, S., Efstathiou, G., \& Kaiser, N. 1991, \apj, 379, 440

\bibitem[Brainerd, Blandford, \& Smail(1996)]{brainerd96}
Brainerd, T. G., Blandford, R. D., \& Smail, I. 1996, \apj, 466, 623

\bibitem[Bullock et al.(2001)]{bullock01}
Bullock, J. S., Kolatt, T. S., Sigad, Y., Somerville, R. S.,
Kravtsov, A. V., Klypin, A. A., Primack, J. R., \& Dekel, A. 2001,
\mnras, 321, 559

\bibitem[Bullock et al.(2002)]{bullock02}
Bullock, J. S., Weschsler, R. H., \& Somerville, R. S. 2001, \mnras, 329, 246

\bibitem[Casas-Miranda et al.(2001)]{casas01}
Casas-Miranda, R., Mo, H. J., Sheth, R. K., \& B\"orner, G. 2001,
\mnras, in press

\bibitem[Cen \& Ostriker(2000)]{cen00}
Cen, R., \& Ostriker, J. P. 2000, \apj, 538, 83

\bibitem[Cole et al.(1994)]{cole94}
Cole, S., Fisher, K. B., \& Weinberg, D. H.  1994, \apj, 267, 785

\bibitem[Cole et al.(1995)]{cole95}
Cole, S., Fisher, K. B., \& Weinberg, D. H.  1995, \apj, 275, 515

\bibitem[Coles(1993)]{coles93}
Coles, P. 1993, \mnras, 262, 1065

\bibitem[Coles, Melott, \& Munshi(1999)]{coles99}
Coles, P., Melott, A. L., \& Munshi, D. 1999, \apj, 521, L5

\bibitem[Col\'{\i}n et al.(1999)]{colin99}
Col\'{\i}n, P., Klypin, A. A., Kravtsov, A. V., \& Khokhlov, A. M. 1999,
\apj, 523, 32

\bibitem[Colless et al.(2001)]{colless01}
Colless, M. et al. 2001, \mnras, 328, 1039

\bibitem[Connolly et al.(2001)]{connolly01}
Connolly, A. J. et al. 2001, \apj, submitted (astro-ph/0107417)

\bibitem[Dav\'e et al.(2000)]{dave00}
Dav\'e, R., Hernquist, L., Katz, N., \& Weinberg, D. H. 2000, in Proc.
Rencontres Internationales de l'IGRAP, "Clustering at High Redshift",
eds. A. Mazure, O. Le Fevre, \& V. Le Brun (San Francisco:
ASP Conference Series v.200), p.402

\bibitem[Davis \& Peebles(1983)]{davis83}
Davis, M., \& Peebles, P. J. E. 1983, \apj, 267, 465

\bibitem[Davis et al.(1985)]{davis85}
Davis, M., Efstathiou, G., Frenk, C. S., \& White, S. D. M. 1985, \apj,
292, 371

\bibitem[de Filippis et al.(1999)]{defilippis99}
De Filippis, E., Longo, G., Andreon, S., Scaramella, R., Testa V.,
de Carvalho, R. R., \& Djorgovski, S. G. 1999, MmSAI, in press, (astro-ph/9909368)

\bibitem[Dekel \& Lahav(1999)]{dekel99}
Dekel, A., \& Lahav, O. 1999, \apj, 520, 24

\bibitem[dell'Antonio \& Tyson(1996)]{dellantonio96}
dell'Antonio, I. P., \& Tyson, J. A. 1996, \apj, 473, 17L,

\bibitem[Efstathiou, Bond, \& White(1992)]{efstathiou92}
Efstathiou, G., Bond, J. R., \& White, S. D. M. 1992, \mnras, 258, 1

\bibitem[Eke, Cole, \& Frenk(1996)]{eke96}
Eke, V. R., Cole, S., \& Frenk, C. S. 1996, \mnras, 282, 263

\bibitem[Evrard(1987)]{evrard87}
Evrard, A. E. 1987, \apj, 316, 36

\bibitem[Fischer et al.(2000)]{fischer00}
Fischer, P. et al. 2000, \aj, 120, 1198

\bibitem[Fry(1984)]{fry84}
Fry, J. 1984, \apj, 279, 499

\bibitem[Fry \& Gazta\~naga(1993)]{fry93}
Fry, J. N., \& Gazta\~naga, E. 1993, \apj, 413, 447

\bibitem[Fry(1994)]{fry94}
Fry, J. N. 1994, Phys. Rev. Lett., 73, 215L

\bibitem[Gazta\~naga \& Juszkiewicz(2001)]{gaztanaga01}
Gazta\~naga, E., \& Juszkiewicz, R. 2001, \apj, 558, 1L

\bibitem[Gott \& Turner(1977)]{gott77}
Gott, J. R., \& Turner, E. L. 1977, \apj, 216, 357

\bibitem[Griffiths et al.(1996)]{griffiths96}
Griffiths, R. E., Casertano, S., Im, M., Ratnatunga, K. U. 1996,
\mnras, 282, 1159

\bibitem[Guzik, \& Seljak(2001)]{guzik01}
Guzik, J., \& Seljak, U. 2001, \mnras, 321, 439

\bibitem[Guzzo et al.(1997)]{guzzo97}
Guzzo, L., Strauss, M. A., Fisher, K. B., Giovanelli, R., \& Haynes, M. P.
1997, \apj, 489, 37

\bibitem[Hamilton(1998)]{hamilton98}
Hamilton, A. J. S. 1998, in The Evolving Universe, ed. Hamilton, D.,
(Kluwer Academic, Dordrecht) p. 185 (astro-ph/9708102)

\bibitem[Hamilton et al.(1991)]{hamilton91}
Hamilton, A. J. S., Matthews, A., Kumar, P., \& Lu, E. 1991, \apj, 374, L1

\bibitem[Hatton \& Cole(1999)]{hatton99}
Hatton, S. J., \& Cole, S.  1999, \mnras, 310, 113

\bibitem[Heisler, Tremaine, \& Bahcall(1985)]{heisler85}
Heisler, J., Tremaine, S., \& Bahcall, J. N. 1985, \apj, 298, 8

\bibitem[Hudson et al.(1998)]{hudson98}
Hudson, M. J., Gwyn, S. D. J., Dahle, H., Kaiser, N. 1998, \apj, 503, 531

\bibitem[Jenkins et al.(1998)]{jenkins98}
Jenkins, A., Frenk, C. S., Pearce, F. R., Thomas, P. A., Colberg, J. M.,
White, S. D. M., Couchman, H. M. P., Peacock, J. A., Efstathiou, G., \&
Nelson, A. H. 1998, \apj, 499, 20

\bibitem[Jing's(1998)]{jing98a}
Jing, Y. P. 1998, \apj, 503, 9

\bibitem[Jing \& B\"orner(1998)]{jing98c}
Jing, Y. P., \& B\"orner, G. 1998, \apj, 503, 37

\bibitem[Jing, Mo, \& B\"orner(1998)]{jing98b}
Jing, Y. P., Mo, H. J., \& B\"orner, G. 1998, \apj, 494, 1

\bibitem[Jing, B\"orner \& Suto(2002)]{jing02}
Jing, Y. P., B\"orner, G., \& Suto, Y. 2002, \apj, 564, 15

\bibitem[Juszkiewicz et al.(1995)]{juszkiewicz95}
Juszkiewicz, R., Weinberg, D. H., Amsterdamski, P., Chodorowski, M.,
\& Bouchet, F. 1995, \apj, 442, 39

\bibitem[Kaiser(1987)]{kaiser87}
Kaiser, N.  1987 \mnras, 227, 1

\bibitem[Kauffmann et al.(1999)]{kauffmann99}
Kauffmann, G., Colberg, J. M., Diaferio, A., \& White, S. D. M. 1999,
\mnras, 303, 188

\bibitem[Kauffmann, Nusser, \& Steinmetz(1997)]{kauffmann97}
Kauffmann, G., Nusser, A., \& Steinmetz, M. 1997, \mnras, 286, 795

\bibitem[Lacey \& Cole(1993)]{lacey93}
Lacey, C. G., \& Cole, S. 1993, \mnras, 262, 627

\bibitem[Lemson \& Kauffmann(1999)]{lemson99}
Lemson, G. \& Kauffmann, G. 1999, \mnras, 302, 111

\bibitem[Limber(1954)]{limber54}
Limber, D. 1954, \apj, 119, 655

\bibitem[Little \& Weinberg(1994)]{little94}
Little, B., \& Weinberg, D. H. 1994, \mnras, 267, 605

\bibitem[Ma \& Fry(2000)]{ma00}
Ma, C., \& Fry, J. N. 2000, \apj, 543, 503

\bibitem[Maddox et al.(1990)]{maddox90}
Maddox, S. J., Efstathiou, G., Sutherland, W. J., \& Loveday, J. 1990
\mnras, 242, 43

\bibitem[Maia, da Costa, \& Latham(1989)]{maia89}
Maia, M. A. G., da Costa, \& L. N., Latham, D. W. 1989, ApJS, 69, 809

\bibitem[Mann, Peacock, \& Heavens(1998)]{mann98}
Mann, R. G., Peacock, J. A., \& Heavens, A. F. 1998, \mnras, 293, 209

\bibitem[Marinoni \& Hudson(2001)]{marinoni01}
Marinoni, C., \& Hudson, M. J. 2001, \apj, submitted (astro-ph/0109134)

\bibitem[McClelland \& Silk(1977)]{mcclelland77}
McLelland, J., \& Silk, J. 1977, \apj, 217, 331

\bibitem[McKay et al.(2001)]{mckay01}
McKay, T. et al. 2001, \apj, submitted, (astro-ph/0108013)

\bibitem[Mo \& White(1996)]{mo96}
Mo, H. J., \& White, S. D. M. 1996, \mnras, 282, 347

\bibitem[Moore, Frenk, \& White(1993)]{moore93}
Moore, B.; Frenk, C. S.; White, S. D. M 1993, \mnras, 261, 827

\bibitem[Narayanan et al.(2000)]{narayanan00}
Narayanan, V. K., Berlind, A. A., \& Weinberg, D. H. 2000, \apj, 528, 1

\bibitem[Navarro, Frenk, \& White(1997)]{navarro97}
Navarro, J. F., Frenk, C. S., \& White, S. D. M. 1997, \apj, 490, 493

\bibitem[Neyman \& Scott(1952)]{neyman52}
Neyman, J., \& Scott, E. L. 1952, \apj, 116, 144

\bibitem[Nolthenius, Klypin, \& Primack(1994)]{nolthenius94}
Nolthenius, R., Klypin, A., Primack, J. R. 1994, \apj, 422, 45L

\bibitem[Norberg et al.(2001)]{norberg01}
Norberg, P. et al. 2001, \mnras, 328, 64

\bibitem[Norberg et al.(2002)]{norberg02}
Norberg, P., et al.\ 2002, \mnras, in press, astro-ph/0112043

\bibitem[Peacock \& Dodds(1994)]{peacock94}
Peacock, J. A., \& Dodds, S. J. 1994, \mnras, 267, 1020

\bibitem[Peacock \& Dodds(1996)]{peacock96}
Peacock, J. A., \& Dodds, S. J. 1996, \mnras, 280, 19

\bibitem[Peacock \& Smith(2000)]{peacock00}
Peacock, J. A., \& Smith, R. E. 2000, \mnras, 318, 1144

\bibitem[Peacock et al.(2001)]{peacock01}
Peacock, J. A. et al. 2001, Nature, 410, 169

\bibitem[Pearce et al.(1999)]{pearce99}
Pearce, F. R., Jenkins, A., Frenk, C. S., Colberg, J. M., White, S. D. M.,
Thomas, P. A., Couchman, H. M. P., Peacock, J. A., \& Efstathiou, G. 1999
\apj, 521L, 99

\bibitem[Peebles(1974)]{peebles74}
Peebles, P. J. E. 1974, \aap, 32, 197

\bibitem[Porciani \& Giavalisco(2001)]{porciani01}
Porciani, C., \& Giavalisco, M. 2001, \apj, 565, 24

\bibitem[Press-Schechter(1974)]{press74}
Press, W. H., \& Schechter, P. 1974, \apj, 187, 425

\bibitem[Ramella, Geller, \& Huchra(1989)]{ramella89}
Ramella, M., Geller, M. J., \& Huchra, J. P. 1989, \apj, 344, 57

\bibitem[Ramella et al.(1999)]{ramella99}
Ramella, M. et al. 1999, \aap, 342, 1

\bibitem[Sargent \& Turner(1977)]{sargent77}
Sargent, W. L. W., \& Turner, E. L. 1977, \apj, 212, 3L

\bibitem[Scherrer \& Bertschinger(1991)]{scherrer91}
Scherrer, R. J., \& Bertschinger, E. 1991, \apj, 381, 349

\bibitem[Scherrer \& Weinberg(1998)]{scherrer98}
Scherrer, R. J., \& Weinberg, D. H. 1998, \apj, 504, 607

\bibitem[Scoccimarro et al.(2000)]{scoccimarro00}
Scoccimarro, R., Sheth, R. K., Hui, L., \& Jain, B. 2000, \apj, 546, 20

\bibitem[Seljak(2000)]{seljak00}
Seljak, U. 2000, \mnras, 318, 203

\bibitem[Seljak(2001)]{seljak01}
Seljak, U. 2001, \mnras, 325, 1359

\bibitem[Sheth et al.(2001)]{sheth01}
Sheth, R. K., Hui, L., Diaferio, A., \& Scoccimarro, R. 2001, \mnras, 325, 1288

\bibitem[Smith et al.(2001)]{smith01}
Smith, D. R., Bernstein, G. M., Fischer, P., Jarvis, M. 2001, \apj, 551, 643

\bibitem[Somerville et al.(2001)]{somerville01}
Somerville, R. S., Lemson, G., Sigad, Y., Dekel, A., Kauffmann, G., \&
White, S. D. M. 2001, \mnras, 320, 289

\bibitem[Tully-Fisher(1977)]{tully77}
Tully, R. B., \& Fisher, J. R. 1977, \aap, 54, 661

\bibitem[Vogeley et al.(1994)]{vogeley94}
Vogeley, M. S., Geller, M. J., Park, C., \& Huchra, J. P 1994, \aj, 108, 745

\bibitem[Wang, Tegmark, \& Zaldarriaga(2001)]{wang01}
Wang, X., Tegmark, M., \& Zaldarriaga, M. 2001, Phys. Rev. D., submitted
(astro-ph/0105091)

\bibitem[Weinberg(1995)]{weinberg95}
Weinberg, D. H. 1995, in Wide-Field Spectroscopy and the Distant Universe,
eds. S. J. Maddox and A. Arag\'on-Salamanca
(Singapore: World Scientific), 129

\bibitem[White(2001)]{white01a}
White, M. 2001, \mnras, 321, 1

\bibitem[White, Hernquist, \& Springel(2001)]{white01b}
White, M., Hernquist, L, \& Springel, V. 2001, \apj, 550, L129

\bibitem[White, Hernquist, \& Springel(2001)]{white01c}
White, M., Hernquist, L, \& Springel, V. 2001, \apj, submitted
(astro-ph/0107023)

\bibitem[White, Efstathiou, \& Frenk(1993)]{white93}
White, S. D. M., Efstathiou, G. P., \& Frenk, C. S. 1993, \mnras, 262, 1023

\bibitem[Wilson, Kaiser, \& Luppino(2001)]{wilson01}
Wilson, G., Kaiser, N., \& Luppino, G. A. 2001, \apj, 556, 601

\bibitem[Wu et al.(1998)]{wu98}
Wu, X.,  Chiueh, T., Fang, L., \& Xue, Y. 1998, \mnras, 301, 861

\bibitem[York et al.(2000)]{york00}
York, D. et al. 2000, \aj, 120, 1579

\bibitem[Yoshikawa et al.(2001)]{yoshikawa01}
Yoshikawa, K., Taruya, A., Jing, Y. P., \& Suto, Y. 2001, \apj, 558, 520

\bibitem[Zehavi et al.(2001)]{zehavi01}
Zehavi, I. et al. 2001, \apj, submitted (astro-ph/0106476)

\bibitem[Zheng et al.(2002)]{zheng02}
Zheng, Z., Tinker, J. L., Weinberg, D. H., \& Berlind, A. A. 2002,
\apj, submitted, astro-ph/0202358

\end{thebibliography}
\end{document}